\documentclass{aa}

\usepackage{graphicx}
\usepackage{txfonts}
\usepackage{xcolor}
\usepackage{booktabs}
\usepackage{multirow}
\usepackage{multicol}

\newcommand{\feh}{[Fe/H]}
\newcommand{\ofe}{[O/Fe]}
\newcommand{\mgfe}{[Mg/Fe]}
\newcommand{\mgh}{[Mg/H]}
\newcommand{\sife}{[Si/Fe]}
\newcommand{\cafe}{[Ca/Fe]}

\newcommand{\alphafe}{[$\alpha$/Fe]}
\newcommand{\rgui}{$\langle R_{\rm{g}} \rangle$}
\newcommand{\rapo}{$\langle R_{\rm{apo}} \rangle$}
\newcommand{\rperi}{$\langle R_{\rm{peri}} \rangle$}
\newcommand{\rbirth}{$\langle R_{\rm{b}} \rangle$}
\newcommand{\eccentricity}{$\langle e \rangle$}
\newcommand{\zmax}{$\langle Z_{\rm{max}} \rangle$}
\newcommand{\etot}{$\langle E_t \rangle$}
\newcommand{\lphi}{$\langle L_{\phi} \rangle$}
\newcommand{\lr}{$\langle L_r \rangle$}

\newcommand{\lz}{$\langle L_z \rangle$}
\newcommand{\jr}{$\langle J_r \rangle$}
\newcommand{\jp}{$\langle J_{\phi} \rangle$}
\newcommand{\jz}{$\langle J_z \rangle$}

\definecolor{codeblue}{RGB}{31, 119, 180}
\definecolor{codeorange}{RGB}{255, 127, 14}
\definecolor{codegreen}{RGB}{44, 160, 44}
\definecolor{codepurple}{RGB}{148, 103, 189}
\definecolor{codegray}{RGB}{128, 128, 128} % Grey color

\begin{document}

    \title{Probing the origins}

    \subtitle{I. Generalised additive model inference of birth radii for Milky Way stars in the solar vicinity}

    \titlerunning{Origins I -- From birth to orbit: Dissecting stellar motion in the Galactic thin disc}

    \author{M.~L.~L.~Dantas\inst{\ref{ia_puc}, \ref{astroing_puc}, \ref{cft},  \ref{camk}}
           \and
           R.~Smiljanic\inst{\ref{camk}}
           \and
           R.~S.~de Souza\inst{\ref{herts},\ref{iag},\ref{unc}}
           \and
           P.~B.~Tissera\inst{\ref{ia_puc}, \ref{astroing_puc}}
           \and
           L.~Magrini\inst{\ref{inaf_oaa}}
           }

    \institute{
    Instituto de Astrofísica, Pontificia Universidad Católica de Chile, Av. Vicuña Mackenna 4860, Santiago, Chile \label{ia_puc}
    \and
    Centro de Astro-Ingeniería, Pontificia Universidad Católica de Chile, Av. Vicuña Mackenna 4860, Santiago, Chile \label{astroing_puc}
    \and 
    Center for Theoretical Physics, Polish Academy of Sciences, al. Lotników 32/46, 02-668, Warsaw, Poland \label{cft}
    \and
    Nicolaus Copernicus Astronomical Center, Polish Academy of Sciences, ul. Bartycka 18, 00-716, Warsaw, Poland \label{camk} \\
    \email{mlldantas@protonmail.com}
    \and
    Centre for Astrophysics Research, University of Hertfordshire, College Lane, Hatfield, AL10~9AB, UK \label{herts}
    \and 
    Instituto de Astronomia, Geofísica e Ciências Atmosféricas, Universidade de São Paulo, Rua do Matão 1226, 05508-090, São Paulo, Brazil \label{iag}
    \and 
    Department of Physics \& Astronomy, University of North Carolina at Chapel Hill, NC 27599-3255, USA \label{unc}
    \and
    INAF -- Osservatorio Astrofisico di Arcetri, Largo E. Fermi, 5, 50125 Firenze, Italy \label{inaf_oaa}
    }

    \date{Received XXX; accepted XXX}

    % \abstract{}{}{}{}{} 
    % 5 {} token are mandatory
    \abstract
    {As stars traverse the Galaxy, interactions with structures such as the bar and spiral arms can alter their orbits, leading either to `churning', where changes in angular momentum shift their guiding radii, or `blurring', where angular momentum is preserved. Churning is what is commonly known as radial migration.}
    {Here, we probe the orbital characteristics of a diverse set of stars in the thin disc observed by the \textit{Gaia}-ESO survey. We aim to discern whether their orbits are predominantly influenced by churning or if they keep their orbital birth radii (i.e. were blurred or remained undisturbed).}
    {We employed a generalised additive model (GAM) to address the limitations inherent in radial metallicity gradients predicted by chemical evolution models, thereby facilitating estimation of the birth radii for the thin disc stars in our sample based on their age and chemical composition. We then juxtaposed the birth radius predictions derived from the GAM with the calculated guiding radii, among other dynamic parameters. This comparison was performed within distinct groups of our dataset, categorised through hierarchical clustering (HC) based on 21 chemical abundances spanning 18 species.}
    {Our results indicate that groups of stars with different chemical abundances exhibit distinct orbital behaviours. Metal-rich stars, formed in the inner regions of the Milky Way, seem to be predominantly churned outward. Their metal-poor counterparts, formed in the outer thin disc, exhibit the opposite behaviour. Also, the proportion of blurred/undisturbed stars generally increases with decreasing metallicity when compared to their churned counterparts. Approximately three-fourths of the sample has been affected by (inward or outward) churning, while the remaining part of the sample ($\sim 1/4$) has either been influenced by blurring or remained undisturbed. These percentages vary considerably across different metallicity-stratified groups. Additionally, we identified a large age gap between churned and blurred/undisturbed sub-samples within each HC-based group: the outward-churned stars were systematically the oldest, inward-churned stars the youngest, and blurred/undisturbed stars at intermediate ages. Yet, given that our sample mostly comprises old stars, we suspect that those classified as blurred/undisturbed may have primarily undergone blurring due to their extended interactions with Galactic structures, considering that their median ages are $\sim$ 6.61 Gyr. We also detected significant differences in angular momenta in the $z$ component for stars that have either churned inward or outward when compared to their blurred/undisturbed counterparts. The action components also provide interesting insights into the orbital history of our different metallicity- and motion-stratified groups. Additionally, we observed the potential effects of the pericentric passage of the Sagittarius dwarf galaxy in our most metal-poor subset of stars formed in the outer disc. Finally, we estimate that the Sun's most probable birth radius is $7.08 \pm 0.24$ kpc, with a 3$\sigma$ range spanning from 6.46 to 7.81 kpc, which is in agreement with previous studies.}
    {}
 

    \maketitle
%
%-------------------------------------------------------------------

\section{Introduction} \label{sec:intro}

The Milky Way (MW) is a complex combination of stars, gas, dust, and dark matter \citep[e.g.][and references therein]{BinneyVasiliev2024}. Detailed studies of the MW are important for contextualising its role as the cradle of our Solar System \citep{Gonzalez2001, Stojkovic2019} and as a fundamental archetype for unravelling galactic structures and evolutionary processes, which makes the MW a reference for studying other galaxies \citep[e.g.][and references therein]{KobayashiTaylor2023}.

Our Galaxy is a barred spiral with several substructures -- thin and thick discs, the halo, the (box/peanut-shaped) bulge, spiral arms, and the bar -- all interacting intricately \citep{Bland-HawthornGerhard2016}. As new stars form and evolve, they are influenced by these structures and their gravitational tugs, which can perturb their orbits \citep[e.g.][]{SellwoodBinney2002, Lepine2003, Roskar2008, Brunetti2011, MartinezBautista2021, Carr2022, Lu2022, Fujimoto2023, Iles2024, Nepal2024}. Some stars may have been formed near their current Galactocentric orbital radii, while others may have migrated, a result of being disturbed by the complex interplay of forces within the MW. Additionally, there is evidence that satellite galaxies could also induce radial mixing in the outer disc of MW-like galaxies \citep[e.g.][]{Quillen2009}. By studying the stars in the solar vicinity, one can learn about the various Galactic stellar populations and gain insight into their histories and origins.

In this context, the dynamic process most commonly referred to as radial migration is known as churning. It involves changes in a star's angular momentum that cause a shift in its guiding radius without necessarily affecting its orbital eccentricity. A related process often discussed alongside churning is blurring, which is a process where stars undergo epicyclic motion around their guiding radii, maintaining their angular momentum, and typically orbiting at high eccentricities \citep[see, for instance,][and references therein]{Sellwood2014, Halle2015, Frankel2020, Wozniak2020}.

Many previous investigations have detected stars in the solar vicinity that exhibit characteristics typical of systems that have radially migrated from the inner regions of the Galaxy \citep[e.g.][to mention a few]{Castro1997, Pompeia2002, Trevisan2011, Chen2019, Zhang2021, Lehmann2024}. Additionally, there is evidence suggesting that the Sun itself has relocated from the inner Galaxy, most likely from the proximity of the bulge \citep[see, for instance,][in which the authors analyse the potential imprints on the Earth's geological history]{Tsujimoto2020}.

Indeed, in a previous work \citep{Dantas2023}, we explored a set of super-metal-rich stars currently inhabiting the solar vicinity that were observed by the \textit{Gaia}-ESO public spectroscopic survey \citep{Gilmore2012, Gilmore2022, Randich2013, Randich2022}. These stars are old (typical median ages of $\sim$ 8 Gyr), have near-circular orbits (median eccentricities around 0.2), and can reach maximum Galactic heights between 0.5-1.5 kpc. These are all characteristics of stars with perturbed orbits due to the interactions with major Galactic structures, that is, the bar and/or spiral arms. In addition to migrating metal-rich stars, there is probably a fraction of metal-poor thin-disc stars that are also migrators but originate from the outer regions of the disc \citep{Haywood2008}.

To try to quantify how much a star has been dislocated from its original birth radius, it is usually necessary to make assumptions about how the radial metallicity gradient of the interstellar medium changed with time \citep[see, for instance,][]{Frankel2018, Minchev2018, Feltzing2020, Ratcliffe2023}. However, how the metallicity gradient changes with time is itself affected by the radial motions of gas and stars \citep[see, for instance,][]{Kubryk2015_01, Kubryk2015_02}. Observational constraints on the temporal variations of the metallicity gradient are important ingredients for models of Galactic chemical evolution. Studies in that direction have been done with samples of stars \citep[e.g.][]{Xiang2015, Anders2017, Willett2023} and open clusters \citep{ChenZhao2020, Spina2021, Magrini2023} with precise ages. Young components of the Galaxy, such as Cepheids and some open clusters, are very useful, as they have hardly had enough time to interact with the Galaxy's structures and therefore have probably not (yet) been subject to relocation \citep[e.g.][]{daSilvaDorazi2023, ViscasillasVazquez2023}.

Such chemical evolution models play a crucial role by mapping the spatial and temporal evolution of the MW. They are intrinsically connected to the idea of the `inside-out' formation and growth of the Galaxy. As per this conceptual framework, the initial phase of Galactic evolution entails the progressive consolidation of the central region, where star formation is intense and fast and quickly results in material of high metallicity. The outer regions of the disc are progressively consolidated but with an outward gradient in star formation intensity, resulting in temporal differences in their chemical enrichment \citep[e.g.][]{Chiappini2001, Kobayashi2006, Magrini2009, Bergemann2014, Andrews2017, Schonrich2017, Hu2023, Magrini2023}. Additionally, there is myriad evidence suggesting that the inside-out formation scenario also happens in other galaxies, not necessarily only in spiral ones \citep[e.g.][]{Bezanson2009, AvilaReese2023}.

Previous studies have estimated stellar birth radii ($R_{\rm b}$) using various datasets and methodologies. For instance, \citet{Feltzing2020} and \citet{Ratcliffe2023} used data from the Apache Point Observatory Galactic Evolution Experiment \citep[APOGEE;][]{Majewski2017}, while \citet{Feltzing2020} relied on various distinct chemical enrichment models to derive $R_{\rm b}$, aiming to constrain and broaden the range of plausible $R_{\rm b}$. \citet{Ratcliffe2023} employed the empirical approach developed in \citet{Lu2024}. Similarly, \citet{Minchev2018} adopted a semi-empirical method, applying it to a sample of stars observed with the High Accuracy Radial velocity Planet Searcher instrument \citep[HARPS;][]{Mayor2003}. These diverse approaches are valuable, as they provide important benchmarks for assessing and comparing new techniques and methodologies.

In this paper, we propose an alternative approach for deriving $R_{\rm b}$ that serves as an additional framework that can complement and be compared to previous studies. Specifically, we employ generalised additive models \citep[GAMs; introduced by][]{HastieTibshirani} to expand theoretical chemical evolution models for the thin disc (in this case, those of \citealt{Magrini2009}) of the MW (which are characterised by widely spaced bins in age and radius) to investigate and differentiate the motion of thin disc stars (i.e. churn and blur/lack of interaction). Our GAM approach incorporates two independent variables consisting of chemical abundances (\feh\ or \mgh; the analysis with \mgh\ is in the Appendix) and $t$, the age of the Universe at the time the star was formed (roughly 13.8 Gyr\footnote{We adopted the estimated age of the Universe as 13.8 Gyr \citep{PlanckCollaboration2020}.} minus the current estimated age of a star). The Galactocentric distance ($R$) serves as the response variable (i.e. the output of our regression model). Using GAM, we can effectively model the complex relationship between the chemical abundances and $t$ (independent variables) and $R$ without imposing rigid assumptions regarding the functional forms of these relationships.

This method enables us to refine and extend existing chemical evolution models, offering a statistically robust and flexible approach to estimating $R_{\rm b}$ directly from stellar samples. Unlike previous studies, which rely on distinct methods such as empirical calibrations or semi-empirical techniques, our approach makes use of the flexibility of GAMs to account for non-linear relationships between chemical abundances, stellar ages, and Galactocentric distances. By avoiding the rigid assumptions often inherent in other methodologies, GAM provides a framework that ensures both statistical reliability and adaptability. While this method complements earlier works by providing an independent line of inquiry, it also introduces a powerful tool that can be applied to any chemical enrichment model, facilitating broader and more nuanced explorations of the MW’s chemo-dynamical evolution.

Furthermore, while there is evidence that the thin and thick discs formed somewhat concurrently \citep[e.g.][]{BeraldoSilva2021}, chemical evolution models for the MW have predominantly focused on the thin disc. This focus is justified by the thin disc’s relatively smooth chemical and dynamical trends, which make it more straightforward to model. Moreover, the thin disc dominates the stellar population in the solar neighbourhood, providing a wealth of observational data to constrain theoretical models.

In contrast, the thick disc -- despite its distinct \alphafe\ enhancements, older stellar populations, and kinematically hot nature \citep[see e.g.][]{Soubiran2003, Bensby2011, RecioBlanco2014, Bland-HawthornGerhard2016} -- presents greater complexities that remain less well understood. Structures such as the halo and bulge pose even greater challenges for chemical evolution models, as their diverse formation mechanisms result in extreme variations in metallicity and dynamics. For instance, the halo reflects the earliest stages of Galactic chemical evolution and accretion events, being predominantly metal-poor \citep[see e.g.][]{Gonzalez-Jara2025}. In contrast, the bulge, characterised by its old stellar populations and strikingly wide metallicity variations \citep[$-1.5 \lesssim \rm{\feh} \lesssim +0.5$; see][for a comprehensive review]{Barbuy2018}, encapsulates the star formation history of the Galactic centre. Despite its importance, providing a self-consistent empirical chemo-dynamical model for the Galactic bulge remains a significant challenge due to its complexity \citep[e.g.][]{Grieco2012, Barbuy2018}.

Insights from numerical simulations, particularly regarding metallicity gradients and dynamical evolution across these Galactic components, provide valuable context for understanding their formation and evolution \citep[e.g.][]{Grand2017, Tissera2022, Jara-Ferreira2024}. However, the theoretical predictions must be compared with observations to ensure consistency with the MW's history and evolution. Consequently, while the thin disc offers a tractable and data-rich framework for advancing our understanding of chemical evolution, expanding these models to incorporate other Galactic components -- guided by both simulations and observational constraints -- remains an essential and ongoing objective \citep[see e.g.][for a study combining chemical enrichment models of the MW's disc with numerical simulations of galactic discs]{Minchev2013}.

This work forms the foundation of a broader investigation. Here, we focus on analysing results derived from the GAM and providing an in-depth exploration of our observed \textit{Gaia}-ESO star sample. Our goal is to examine other aspects of these findings in future studies in order to better understand the causes and implications of the stellar motion trends revealed here.

This paper is organised as follows: Section \ref{sec:data_method} details the dataset and methodologies used in this study, thus providing the foundation for the subsequent analysis. The heart of the paper lies in Section \ref{sec:analysis}, where we delve into the analysis and discuss the implications of our findings. Concluding the paper, Section \ref{sec:conclusions} synthesises our insights and summarises the key takeaways and conclusions drawn from this phase of our investigation. Finally, we provide the details of our catalogue publicly available in the CDS in Section \ref{sec:data_availability}.

%-------------------------------------------------------------------
\section{Dataset and methodology}  \label{sec:data_method}

% -----------------
\subsection{Data description} \label{subsec:data_description}

We used the same comprehensive chemical abundance dataset for 1460 stars discussed in \citet{Dantas2023}. The spectra of these stars were observed in high resolution with UVES (resolving power at 47 000), within the range of 4800--6800\AA, by the \textit{Gaia}-ESO survey. The details of the processing of the spectra were presented in \citet{Sacco2014}. Methods for estimating atmospheric parameters and abundances were originally described in \citet{Smiljanic2014} with updates given in \citet{Worley2024}. The survey homogenisation process was discussed in \citet{Hourihane2023}. The data is part of the internal data release 6 (iDR6) of \textit{Gaia}-ESO, which is equivalent to the final public data release\footnote{\url{https://www.eso.org/sci/publications/announcements/sciann17584.html}}. Unlike \citet{Dantas2022, Dantas2023}, which focused on a subset of metal-rich and super-metal-rich stars, our present analysis encompasses the entire dataset.

\begin{figure*}
    \centering
    \includegraphics[width=\linewidth]{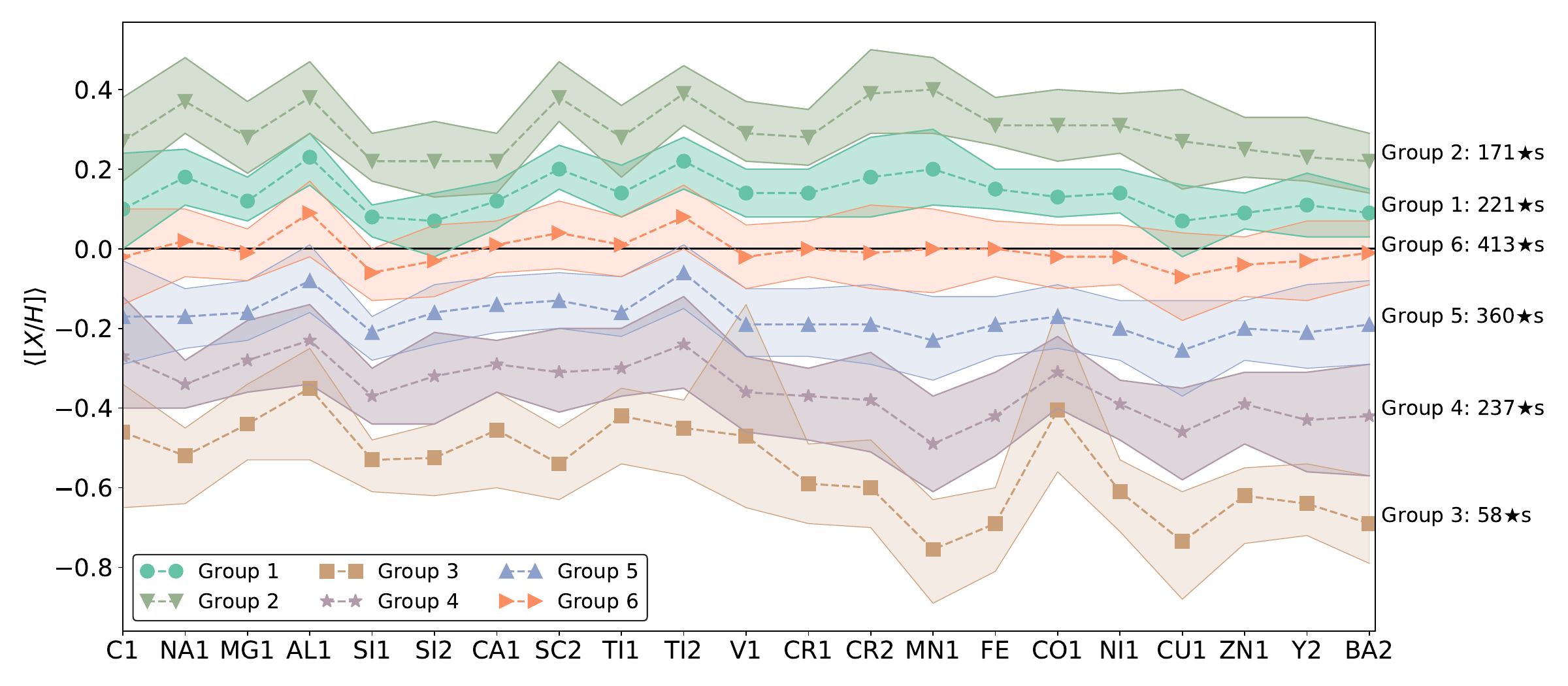}
    \caption{Chemical abundances, arranged in ascending order of atomic number ($x$-axis), of the six main stellar groups ($y$-axis; in terms of solar abundance) classified using an HC algorithm. On the right side of the figure, each group is labelled with its corresponding number, along with the total count of stars in each group. $\langle \rm{[X/H]} \rangle = 0$ is depicted by the black horizontal to represent the solar abundances. It is important to note that the lines in the image are not indicative of regression but are rather utilised to visually represent the relative increase or decrease in element abundance compared to the preceding element. The shaded areas depict the 1$\sigma$ (16-84\%) confidence interval after bootstrapping the data of each group.}
    \label{fig:hc_groups}
\end{figure*}

Following up on the ideas presented in \citet{BoessoRochaPinto2018}, we classified the stars in this sample employing 21 species of 18 elements as parameters and utilise a non-parametric technique for this purpose, namely a hierarchical clustering (HC) algorithm \citep{Murtagh&Contreras2012, Murtagh2014}. The 21 abundances used are: \ion{C}{i}, \ion{Na}{i}, \ion{Mg}{i}, \ion{Al}{i}, \ion{Si}{i}, \ion{Si}{ii}, \ion{Ca}{i}, \ion{Sc}{ii}, \ion{Ti}{i}, \ion{Ti}{ii}, \ion{V}{i}, \ion{Cr}{i}, \ion{Cr}{ii}, \ion{Mn}{i}, Fe,\footnote{The abundance of Fe was estimated via the [Fe/H] provided by iDR6 of \textit{Gaia}-ESO. Hence, there is no ionisation level associated with this abundance.} \ion{Co}{i}, \ion{Ni}{i}, \ion{Cu}{i}, \ion{Zn}{i}, \ion{Y}{ii}, and \ion{Ba}{ii}. The description of this methodology is very thorough in \citet{Dantas2023}. Therefore, in this section we provide only a broad overview of the particularities most relevant to the current paper. This classification allowed us to categorise the stars into different groups and subgroups according to their chemical composition. These stellar groups based on chemical abundances can be seen in Fig. \ref{fig:hc_groups}; note that the super-metal-rich group was analysed \citet{Dantas2022, Dantas2023} is in greyish-green with abundances in inverted triangles (Group 2). For consistency, the numbers of the HC groups are kept the same as in the aforementioned papers. Figure \ref{fig:hc_groups} also showcases the number of stars in each group retrieved from the HC, as well as the confidence interval corresponding to 1$\sigma$ (16-84\%) which has been estimated via bootstrap of the sample in each group.  

Age estimation was performed by employing \textsc{unidam} \citep{Mints2017, Mints2018} alongside PARSEC isochrones \citep{Bressan2012}. We computed stellar orbits with \textsc{galpy} \citep{Bovy2015}, adopting the MW potential model proposed by \citet{McMillan2017}, and using as input parallaxes and proper motions from \textit{Gaia}-EDR3 \citep{Gaia2016, GaiaEDR3}. All dynamic parameters presented in this study are expressed in their median form, such as \rgui~representing the median guiding radius. This approach is adopted due to our error and confidence interval estimation procedure, which involves resampling the observed parameters from all stars. For details of the parameterisation methodologies of both \textsc{unidam} and \textsc{galpy}, as well as the error estimation strategy, we refer the interested reader to \citet[][Sect. 2 therein]{Dantas2023}.

\begin{figure*}
    \centering
    \includegraphics[width=\linewidth]{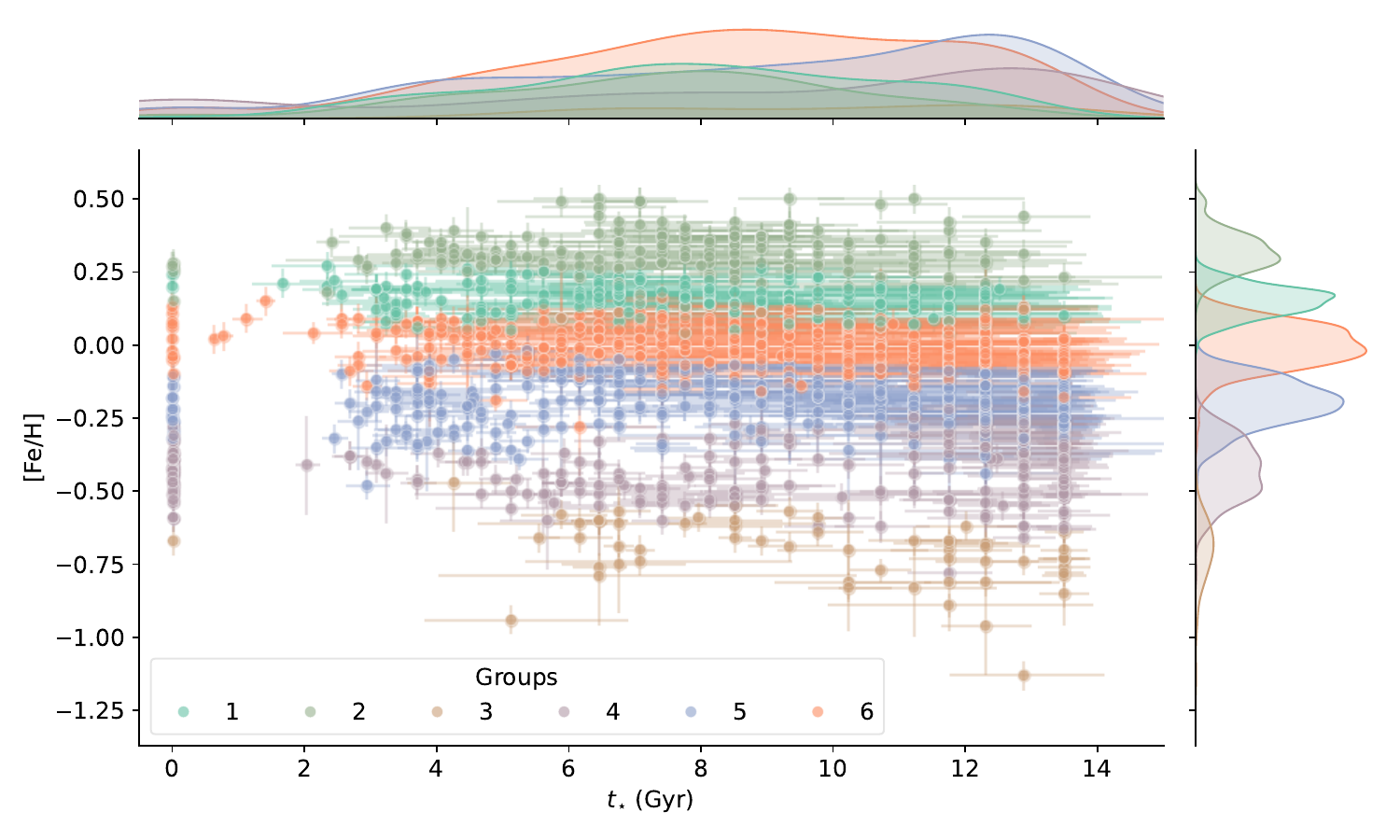}
    \caption{Scatter plot displaying the relationship between metallicity, \feh, and stellar ages, $t_{\star}$, using \textsc{unidam} estimates for the entire sample. Data points are categorised by HC groups. On top, 1D-Gaussian kernel densities show age distribution; on the right, a similar view presents \feh~distribution for each HC group. The window width for the kernel density estimation is computed using Scott’s method, with a default bandwidth adjustment of 1, which is the standard configuration of the \textsc{seaborn} package in \textsc{python} \citep{Waskom2021}; this configuration is consistently applied to all figures with Gaussian kernel densities throughout this paper. The data points and densities are colour-coded and stratified according to the different HC groups, similarly to Fig. \ref{fig:hc_groups}.}
    \label{fig:feh_age}
\end{figure*}

Figure \ref{fig:feh_age} illustrates the variations in \feh~and the stellar age ($t_{\star}$) across the entire sample. The distributions of these parameters are visualised on the adjacent axes in shape 1D-Gaussian kernel density plots. Notably, the influence of the HC is distinctly observable in the \feh~distributions for each respective group.

% -----------------
\subsection{The Galactic chemical evolution models of Magrini et al.} \label{subsec:chemoevo_models}

In this Section, we briefly present the Galactic chemical evolution model described by \citet{Magrini2009}, which we use to estimate the birth radii of our sample of stars. The models delineate radial gradients of various elemental abundances, specifically \ofe, \sife, \mgfe, \cafe, and \feh, across different epochs (i.e. five age bins: 2.1, 3.3, 8.0, 11.0, and 13.7 Gyr).

Elements such as O, Si, Mg, and Ca are classified as $\alpha$-elements and predominantly synthesised within massive stars (with $M_{\star} \geq 8 M_{\odot}$). They are ejected into the interstellar medium (ISM) primarily through core-collapse supernovae (Type II SNe). Conversely, Fe is produced in part by Type II SNe and also, in larger amounts, by Type Ia SNe, which are thermonuclear explosions of white dwarfs in binary systems. Consequently, the $\alpha$-elements to Fe ratios are instrumental in tracing the contributions of both Type II and Type Ia SNe to Galactic chemical enrichment. These ratios are pivotal in delineating the star formation history and chemical evolution of galaxies, including the MW, providing insight into the temporal dynamics of their elemental composition \citep[e.g.][]{Tinsley1979, Venn2004, Blancato2019}. It is worth noting that, as an end product of stellar nucleosynthesis, Fe's relative abundance to $\alpha$-elements serves as a critical indicator, mapping the rate of ISM enrichment with these elements. Iron's usual role as a chemical clock arises from its contribution from both SN II and SN Ia on short and long timescales \citep{MatteucciGreggio1986, MatteucciRecchi2001, Palicio2023}.

The adopted model, \citet{Magrini2009}, integrates high-resolution spectroscopy of open clusters across a range of ages and Galactocentric distances with the theoretical multiphase approach described by \citet{Ferrini1992, Ferrini1994}. This model posits that the Galactic disc was assembled through the accretion of gas from both the halo and the interstellar medium (ISM). The primary gas infall scenario adheres to an exponentially declining law, consistent with an inside-out formation process, where the inner regions of the disc evolve and consolidate more rapidly than those at greater Galactocentric distances. Additionally, \citet{Magrini2007, Magrini2009} propose an alternative parameterisation that combines a constant gas inflow per unit area with the exponential decline. This variation proves useful in reproducing the metallicity gradient, albeit with distinct implications for star formation in the outer disc. For further details on these models, we direct the reader to \citet{Magrini2007, Magrini2009}.

\begin{figure}
    \centering
    \includegraphics[width=\linewidth]{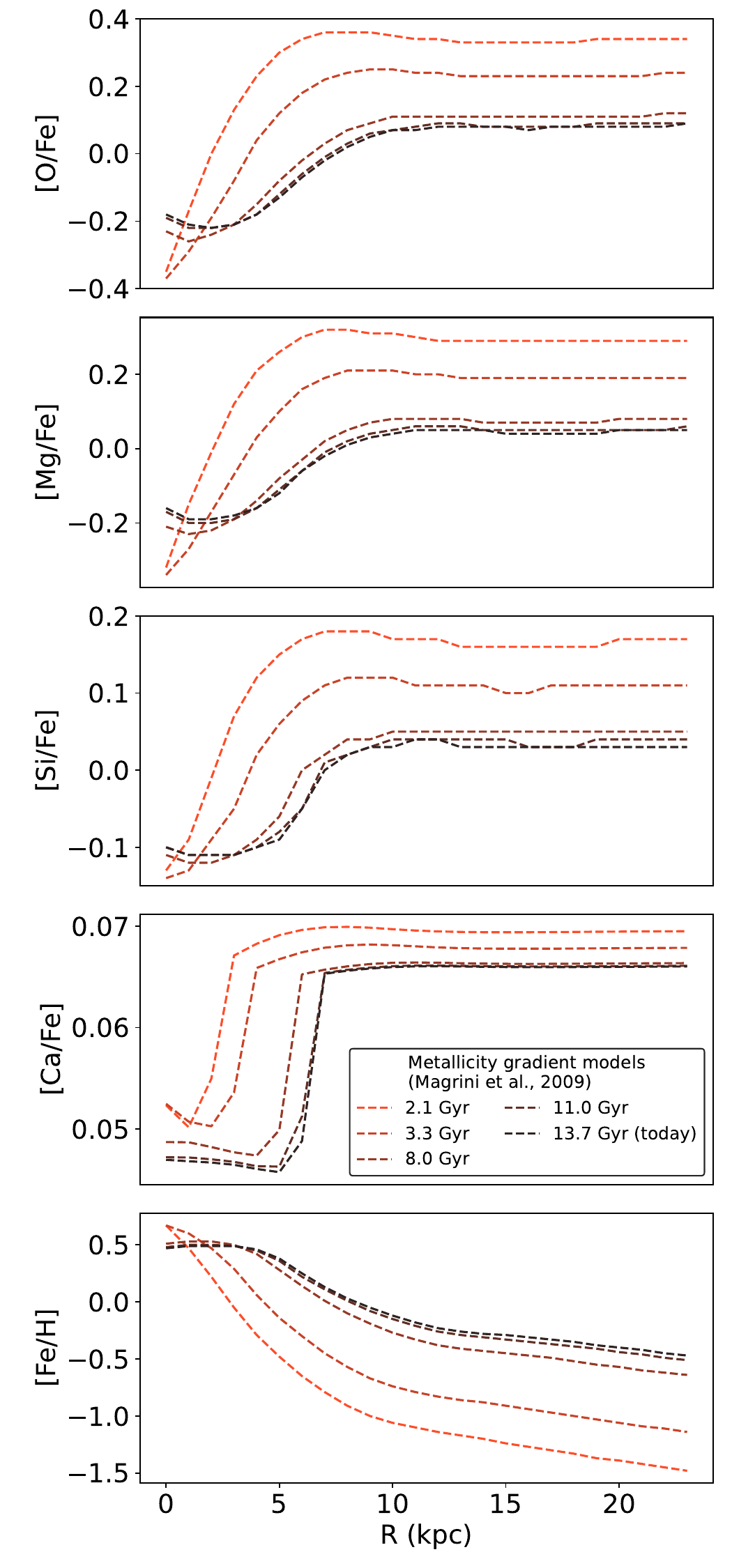}
    \caption{Chemical enrichment models of the MW as described by \citet{Magrini2009} illustrating the relative abundances of key elements in comparison to iron (Fe) across various Galactocentric radii ($R$). The elements, arranged sequentially by atomic number relative to Fe, include \ofe\ at the top panel, followed by \mgfe, \sife, \cafe, and culminating with \feh\ on the bottom panel. The colour gradation, ranging from a lively orange to black, represents the ages of the Universe at the time of star formation within the MW, starting at $t$=2.1 Gyr and extending to $t$=13.7 Gyr. This visual spectrum effectively portrays the chronological progression of the Galaxy's chemical composition. The legend, for clearer visualisation, is incorporated within the \cafe\ subplot.}
    \label{fig:MagriniModels}
\end{figure}

\begin{figure}
    \centering
    \includegraphics[width=\linewidth]{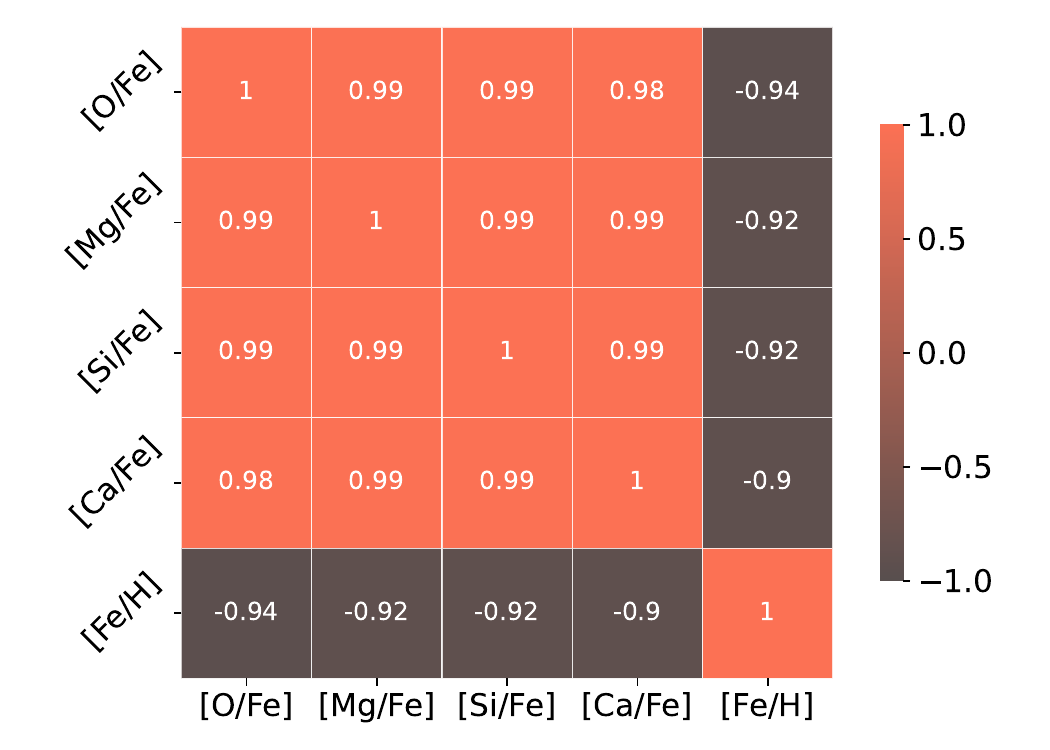}
    \caption{Heatmap visualisation of the correlation matrix showing the pairwise correlation coefficients between elemental abundances relative to iron (\ofe, \mgfe, \sife, \cafe) and the iron abundance relative to hydrogen (\feh). Orange indicates a strong positive correlation, while brown signifies a strong negative correlation, demonstrating the tight coupling between these elemental abundances in the context of \citeauthor{Magrini2009}'s model. The correlation matrix was estimated via Spearman's rank correlation coefficient \citep{Spearman1904}.}
    \label{fig:model_correlations}
\end{figure}

In Fig. \ref{fig:MagriniModels} the radial profiles of the \citet{Magrini2009} models are shown. It is noticeable that the bottom panel, depicting \feh, shows the smoothest curves (potentially the best estimated of the profiles), whereas all the other ones have constant or near-constant abundances after a certain $R$. This is an important feature to have in mind since this lack of smoothness can be detrimental to the implementation of the GAM. This is discussed further in Sect. \ref{subsec:gam}. Moreover, there is a clear pattern of strong positive or negative correlations between the abundances in the models, as evidenced by the heatmap in Fig. \ref{fig:model_correlations}.

In the context of this paper, we have limited our statistical model to \feh~and $t$ as the independent variables, with $R$ serving as the dependent variable. To demonstrate that a single chemical abundance can suffice to estimate $R$, we have included an analysis incorporating \mgfe~in the Appendix \ref{appendix:comparison_feh_mgh}. The decision to select \feh~or \mgfe~for our examination is substantiated by both physical and statistical considerations, as discussed in Sect. \ref{subsec:gam}. Nonetheless, we underscore that relying on a solitary chemical abundance is inadequate for capturing the full spectrum of stellar characteristics; this is corroborated by our approach of utilising 21 abundances to categorise and collectively analyse the stars in our sample. The choice to focus on a single chemical abundance in this instance is informed by the observed correlations within \citeauthor{Magrini2009}'s model. However, we stress that a similar analysis considering other chemical abundances, such as those of neutron-capture elements, can be considered, especially if there is a lower correlation between these parameters.

% -----------------
\subsection{Generalised additive models} \label{subsec:gam}

% -----------------
% -----------------
\subsubsection{Why we adopt generalised additive models}

Astrophysics has long relied on traditional statistical methods, such as linear regression models \citep{Isobe1990}, where the expected value of the response variable is assumed to be linearly dependent on its coefficients. While these techniques have been integral to the field and remain highly effective in many contexts, they can encounter challenges when applied to more complex, non-linear models.

Recently, there has been increasing recognition of the need for more flexible statistical frameworks that can better exploit the information present in both observational and simulated data. Recent trends highlight a growing intersection between Astronomy and fields such as Statistics and Computer Science. For instance, \citet{Veneri2022} show that papers in Astronomy have increasingly cited these fields between 2010 and 2020. As a case in point, Generalised Linear Models (GLMs) have been successfully applied to astrophysical problems, with notable examples from the COsmostatistics INitiative \citep[COIN\footnote{\url{https://cosmostatistics-initiative.org/}}; ][]{Elliott2015, deSouza2015AC, deSouza2015MNRAS, deSouza2016}. A GLM has also been applied to estimate the occurrence of early-type galaxies with ultraviolet excess as a function of redshift, stellar mass, and emission lines \citep{Dantas2020}, not to mention the numerous applications of GLMs in astrophysics illustrated in the textbook by \citet*{Hilbe2017}.

Generalised additive models \citep[introduced by][]{HastieTibshirani}, an extension of GLMs, offer even greater flexibility by allowing for non-linear relationships between predictors and the response variable, using smooth functions. Despite their potential, to our knowledge, GAMs have been previously used fewer times in astrophysics, such as in photometric redshift estimation \citep{Beck2017}, in the analysis of ionising photon escape fractions from dark haloes \citep[][]{Hattab2019}, and more recently in the examination of the relationship between galaxy structure and star formation rate \citep{Stephenson2024}.

In recent years, astrophysics seems to have skipped over these `intermediate' techniques, with many researchers jumping directly from classical methods to highly complex approaches such as Gaussian processes and deep learning. While these advanced techniques are undoubtedly valuable, methods such as GLMs and GAMs offer elegant and interpretable solutions that are more straightforward to implement and understand. This accessibility makes them highly valuable in a range of disciplines, shown by its use in ecology \citep[e.g.][]{Simpson2018, Pedersen2019, Kosicki2020}, linguistics \citep[e.g.][]{Wieling2016, Tomaschek2018}, medicine \citep[e.g.][]{deSouza2019}, to mention a few.

% -----------------
% -----------------
\subsubsection{The general form of a generalised additive model}

The general form of a GAM can be expressed as follows:

\begin{equation} \label{eq:gams}
    g(\mu) = \beta_0 + s_1(x_1) + s_2(x_2) + \ldots + s_p(x_p) + \epsilon.
\end{equation}

\noindent Here, $g$ is a specified link function that relates the linear predictor to the expected value of the response variable, $\mu$ represents the expected value of the response variable, $\beta_0$ is the intercept term, $s_i$ represents the smooth function associated with the $i$th predictor variable, and $x_i$ denotes the $i$th predictor variable.

In this framework, the smooth functions $s_i(x_i)$ are estimated from the data, allowing for flexible modelling of the predictor-response relationships. The specific form of these smooth functions can vary depending on the chosen smoothing technique, such as spline-based methods or kernel smoothing. The smooth functions $s_i(x_i$) can be represented as a sum of basis functions:

\begin{equation} \label{eq:basisfunc}
    s_i(x_i) = \sum_{j=1}^{m_i} \beta_{ij} B_{ij}(x_i),
\end{equation}

\noindent where $m_i$ is the number of basis functions, $\beta_{ij}$ are the coefficients, and $B_{ij}(x_i)$ are the basis functions associated with the $i$th predictor variable.

In Sect. \ref{subsubsec:gam_profile}, we describe how we applied GAM to our specific problem, and in Appendix \ref{appendix:model} we provide the \textsc{r} code for our model. There, the reader will also find a more in-depth discussion on the practical choices of the parameters, as well as the choice of the smooth parameters used in our model. Other technical implications of our model are also discussed in Appendix \ref{appendix:model}.

% -----------------
% -----------------
\subsubsection{A generalised additive model applied to radial metallicity profiles} \label{subsubsec:gam_profile}

\begin{figure*}[t] 
    \centering
    \includegraphics[width=\linewidth]{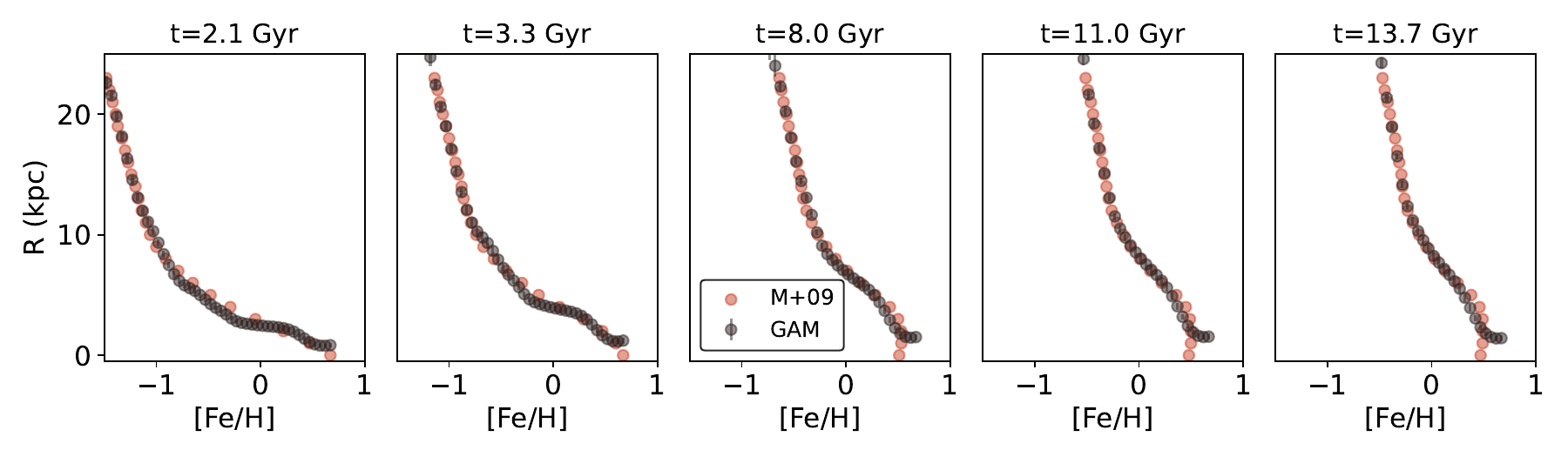}
    \caption{Comparison between the original theoretical models and the GAM estimated curves for the same $t$ bins used as the GAM input (namely 2.1, 3.3, 8.0, 11.0, and 13.7 Gyr, respectively). The original models are depicted by the orange markers, whereas the new estimated GAM curves are represented by the black markers.}
    \label{fig:comparison_feh}
\end{figure*}

We employed a GAM to extend existing theoretical models that map the radial metallicity profile of the Galaxy \citep[such as the ones presented by][]{Magrini2009}, specifically considering the Galactocentric distance ($R$) as the response variable. Our objective was to accommodate stars of diverse ages and evaluate whether they currently reside at their birth radii. To achieve this, we utilised the \texttt{mgcv} package in \textsc{r} to build the GAM, which is formulated as

\begin{equation} \label{eq:our_gam}
    \log(\mathbb{E}[R]) = \beta_0 + s_1(\text{\feh}) + s_2(t) + \text{ti}(\text{\feh}, t) + \epsilon.
\end{equation}

\noindent In Equation \ref{eq:our_gam}, the predictor variables are the metallicity (\feh) and $t$, which represents the age of the Universe when the star was formed (13.7 Gyr minus the current estimated age of the star\footnote{We note that here we are using 13.7 Gyr because this is the age of the Universe used in \citet{Magrini2009}. For other situations, we adopt a more recent estimate by the \citeauthor{PlanckCollaboration2020}, as previously described.}). The function $s(\cdot$) corresponds to smoothing splines, enabling the model to capture potential complex and non-linear relationships between the response and predictors. Specifically, $s_1(\text{\feh})$ captures the effect \feh, while $s_2(t$) captures the age of the Universe (when a given star is formed), $t$, and $\text{ti}(\text{\feh}, t)$ captures the interaction of both variables. Also, we note that we assumed the response variable $R$ follows a Gaussian distribution. This choice is motivated by the continuous and positive nature of the stellar parameter, making the Gaussian distribution a suitable approximation. However, to ensure the model's predictions remain positive, we employ a $\log$ link function. This log link function effectively mimics the behaviour of a LogNormal distribution, which is commonly used for modelling positive continuous variables. The term $\beta_0$ denotes the intercept, and $\epsilon$ represents the error term accounting for unexplained variability. Also, the model was fitted to the data using the restricted maximum likelihood (REML) method (\texttt{method="REML"}), and smoothing parameters were automatically selected based on generalised cross-validation (\texttt{select=TRUE}). For more details on the specific parameters of this GAM, we refer the reader to Appendix \ref{appendix:model}.

The fitting of the functions and parameters in Equation \ref{eq:our_gam} were done with the data points from the model by \citet{Magrini2009}, described above. Through this approach to model specification and parameter estimation, we aim to address the limitations of the sparsely binned theoretical models and provide a flexible and data-driven approach to understanding the metallicity profile of the Galaxy as a function of the age of the Universe when its constituent stars were formed. The use of smoothing splines empowers the model to adaptively estimate the underlying relationships, thereby enhancing our ability to discern potential intricate patterns within the data.

Our selection of \feh\ as the singular chemical tracer in our GAM is driven by a strategic emphasis on simplicity and analytical focus. This choice is substantiated by the pronounced correlation between \feh\ and the \alphafe\ ratios, a relationship delineated in Fig. \ref{fig:model_correlations}. The relationship between \feh\ and the \alphafe\ ratios, as underscored not only by the theoretical models but also by the observed data from the MW \citep[i.e. the Tinsley-Wallerstein diagram;][]{Wallerstein1962, Tinsley1979}, establishes \feh\ as a highly representative tracer that effectively captures most of the intricacies of these elemental correlations. Moreover, the \feh\ radial gradient profiles demonstrate superior smoothness compared to other elemental profiles, enhancing the clarity and interpretability of the model. This approach not only avoids the complications of multiple, highly correlated variables but also bolsters the robustness of our analysis in capturing the nuances of Galactic chemical evolution with the information we have at hand.

Although our methodology differs in several aspects, it is noteworthy that our use of \feh\ and $t$ as primary parameters aligns conceptually with the approach of \citet{Ratcliffe2023}, who also relied on these two variables to derive $R_{\rm b}$ from their stellar sample. This similarity underscores the importance of these parameters in tracing stellar birth radii and highlights how different methods can complement each other.

In Appendix \ref{appendix:comparison_feh_mgh}, we provide evidence to support our decision to use \feh~as the primary chemical feature in our GAM. We achieve this by presenting an analysis of the GAM that exclusively considers \mgh, thereby isolating its effects from those of Fe. Additionally, the appendix includes a comparative study between two GAMs: one using \feh~and the other \mgh. This comparison aims to show how the choice of chemical feature impacts the estimated Galactocentric birth distances of stars. The main body of our paper focuses on the model described in Eq. \ref{eq:our_gam}, which uses \feh.

Figure \ref{fig:comparison_feh} shows the comparison between the original theoretical models of \citet[][in orange]{Magrini2009} and the new curves calculated using the GAM (in black). Some imperfections can be noted in the GAM estimations when compared to the original curves, but overall the comparison is very satisfactory. One of them is the flattening of the metallicity gradient in the inner radii at high values of $t$ (last $\sim 5-6$ Gyr). To understand the imperfections, it is crucial to note that the original models contain just five sparse age bins. This limited granularity presents a significant hurdle for the GAM, hindering its ability to achieve a smoother and more comprehensive representation of the original models. Additionally, the use of a \texttt{log}-link function, which is highly effective for the overall model, introduces limitations in accurately estimating $R_{\rm b}$ at very small radii (below 2.5 kpc). For a more detailed discussion of these imperfections and their implications, we direct the reader to Sect. \ref{subsec:challenges} and Appendix \ref{appendix:model}.

\begin{figure*}
    \centering
    \includegraphics[width=\linewidth]{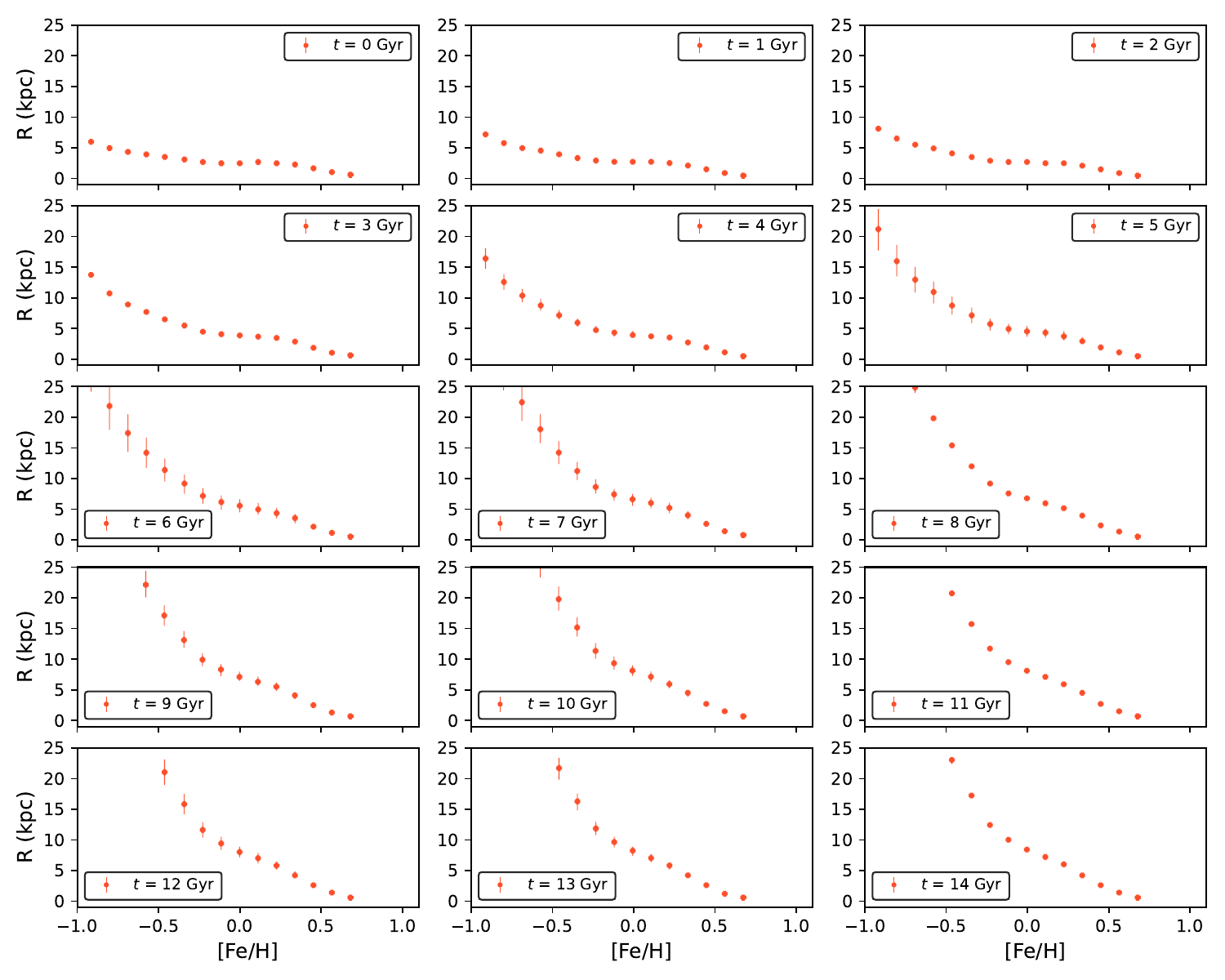}
    \caption{New metallicity profile grids generated using the GAM. Each curve illustrates the radial metallicity profile of stars formed at different epochs in the Universe's history, ranging from 0 to $t$ Gyr (age of the Universe). It is also worth mentioning that each subplot has its $y$-axis ($R$) truncated up to 25 kpc. Also, it is worth mentioning that the GAM does not limit the radii it can reach in a regression. Therefore, we chose to display up to physically meaningful radii.}
    \label{fig:new_grids_split_feh}
\end{figure*}

Figure \ref{fig:new_grids_split_feh} illustrates the radial metallicity profiles estimated from the GAM across a more dense grid of ages spaced at a 1 Gyr interval. For the $y$-axis, even though the GAM is not restricted by physical limitations, we display the response variable $R$ within the meaningful Galactocentric range of 0 to 25 kpc. This value encompasses the upper estimates for the MW's Galactocentric radius, which is generally thought to range from 17 to 25 kpc [see e.g. \citet[][]{Huang2016} for an estimated radius based on the rotation curve of the MW and \citet{Lian2024} for a smaller estimation of 17 kpc]. Additionally, the open clusters used to create the models described by \citet{Magrini2009}, reach up to 22 kpc. Thus, our choice of 25 kpc is a safe choice, encompassing even the largest estimates for the MW.

The sequence of radial gradients displayed in Fig. \ref{fig:new_grids_split_feh} indicates that, as the Universe ages, the formation of very metal-poor stars increasingly occurs farther away from the Galactic centre. It is also important to note the standard error provided by the calculations performed with the \textsc{mgcv} package, depicted by the error bars in both figures; these errors increase with increasing radii. For \feh, this is because the original theoretical models provide the radial profiles for low values of \feh~ only when the Universe was very young (age bin of 2.1 Gyr). For all other ages, particularly older ones, the theoretical models do not predict the formation of stars with very low metallicity, leading to greater uncertainty in the GAM's predictions.

% -----------------
% -----------------
\subsubsection{A generalised additive model applied to a Gaia-ESO sample of stars} \label{subsubsec:gam_gaia_eso}

The model described in Eq. \ref{eq:our_gam}, with functions fitted to the \citet{Magrini2009} models, was then applied to the 1460 stars in our sample (described in Sect. \ref{subsec:data_description}). The regression is performed directly using the stars' parameters: \feh~and $t$, where $t$ is the age of the Universe when the star was formed, calculated as follows:

\begin{equation} 
    \label{eq:ages}
    t = t_{\text{u}} - \overline{t}_{\star}.
\end{equation} 

\noindent Here, $t_{\text{u}}$ is the current estimated age of the Universe \citep[adopted value: 13.8 Gyr; see][]{PlanckCollaboration2020}, and $\overline{t}_{\star}$ is the median age of the star estimated using \textsc{unidam} \citep{Mints2017, Mints2018}, as described in Sect. \ref{sec:data_method} \citep[and in more depth by][]{Dantas2023}.

The GAM is then useful not only for providing finer grids -- as described previously -- but also for directly estimating the most likely birth radius ($R_{\rm b}$) for a star with these parameters. To estimate the uncertainties for $R_{\rm b}$, we applied the GAM after using a bootstrap method, resampling each star's parameters 1000 times. The parameters used for the bootstrap were \feh~and $t$, along with their respective uncertainties. The bootstrapping procedure closely follows the methodology outlined in Sect. 2.2 of \citet{Dantas2023}, and therefore we omit the detailed steps here for simplicity. This approach not only provides $R_{\rm b}$ for each star, but also yields the associated quantiles, standard deviation, and other statistical measures.

Therefore, whenever we use $R_{\rm b}$, we refer to the birth radius estimation derived from the GAM. Meanwhile, \rbirth~ denotes the median birth radius for each star, as estimated through the bootstrapping method and the GAM. 

%-------------------------------------------------------------------
\section{Analysis and results}  \label{sec:analysis}

\begin{figure*}
    \centering
    \includegraphics[width=\linewidth]{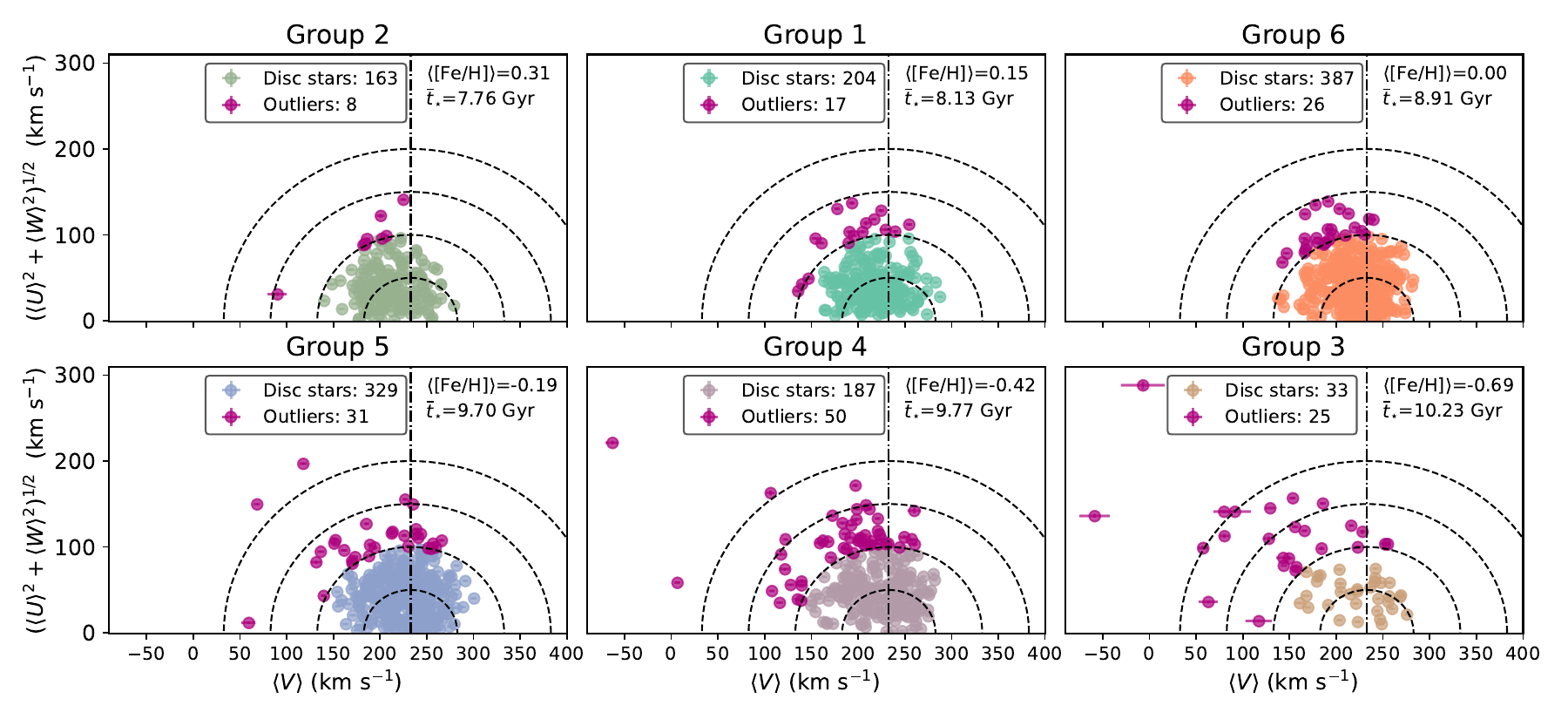}
    \caption{Toomre diagram illustrating the six groups extracted from the HC, arranged in descending order of median metallicity, $\langle \text{\feh} \rangle$. The vertical dot-dashed line marks the Sun's velocity at 232.8 km s$^{-1}$, as estimated by \citet{McMillan2017}. Concentric dashed circles represent constant total velocities of 50, 100, 150, and 200 km s$^{-1}$, centred around the Sun's velocity in this diagram. Velocities are corrected for the solar motion using \{$U_{\odot}$, $V_{\odot}$, $W_{\odot}$\}= \{8.6$\pm$0.9, 13.9$\pm$1.0, 7.1$\pm$1.0\} km s$^{-1}$, following the values reported by \citet[][Table 2]{McMillan2017}. A consistent range on both axes is maintained to facilitate comparison across groups and highlight outliers, particularly those with lower metallicities. The colour scheme corresponds to that used in previous figures: stars with combined velocities exceeding 100 km s$^{-1}$ are designated as outliers and marked in magenta. Subplot labels indicate the number of disc stars and outliers for each group, with \( \langle \text{\feh} \rangle \) and stellar ages, \( \overline{t}_{\star} \), presented. The number of stars likely to belong to the disc as well as the number of outliers are shown in the top legend box.}
    \label{fig:toomre}
\end{figure*}

This Section begins with an examination of the chemo-dynamic properties of our sample. This analysis is crucial, as it helps identify stars that may not belong to the Galactic thin disc and therefore cannot have their birth radii estimated using our selected chemical evolution models. This analysis is performed in Sect. \ref{subsec:chemodynamics} and we briefly describe the final sample in Sect. \ref{subsubsec:final_sample_post_assessment}, after the removal of halo and thick disc intruders.

Following this initial assessment, we dive into our main goal and compare the birth radii (\rbirth\footnote{We refer the reader to Sect. \ref{subsubsec:gam_gaia_eso} for details on the usage of \rbirth.}, the response variable predicted by the GAM) and the current guiding radii (\rgui) and other dynamic parameters for the stars of our sample, to quantify potential shifts in their Galactocentric distance due to either lack of disturbance/blurring, and/or churning. The \rbirth\ used in this Section is the one estimated in Eq. \ref{eq:our_gam}, hence using \feh~and $t$.

% -----------------
\subsection{Preliminary chemo-dynamic analysis} \label{subsec:chemodynamics}

% -----------------
% -----------------
\subsubsection{Distinguishing between disc and halo stars} \label{subsubsec:disc_or_halo}

\begin{figure*}
    \centering
    \includegraphics[width=\linewidth]{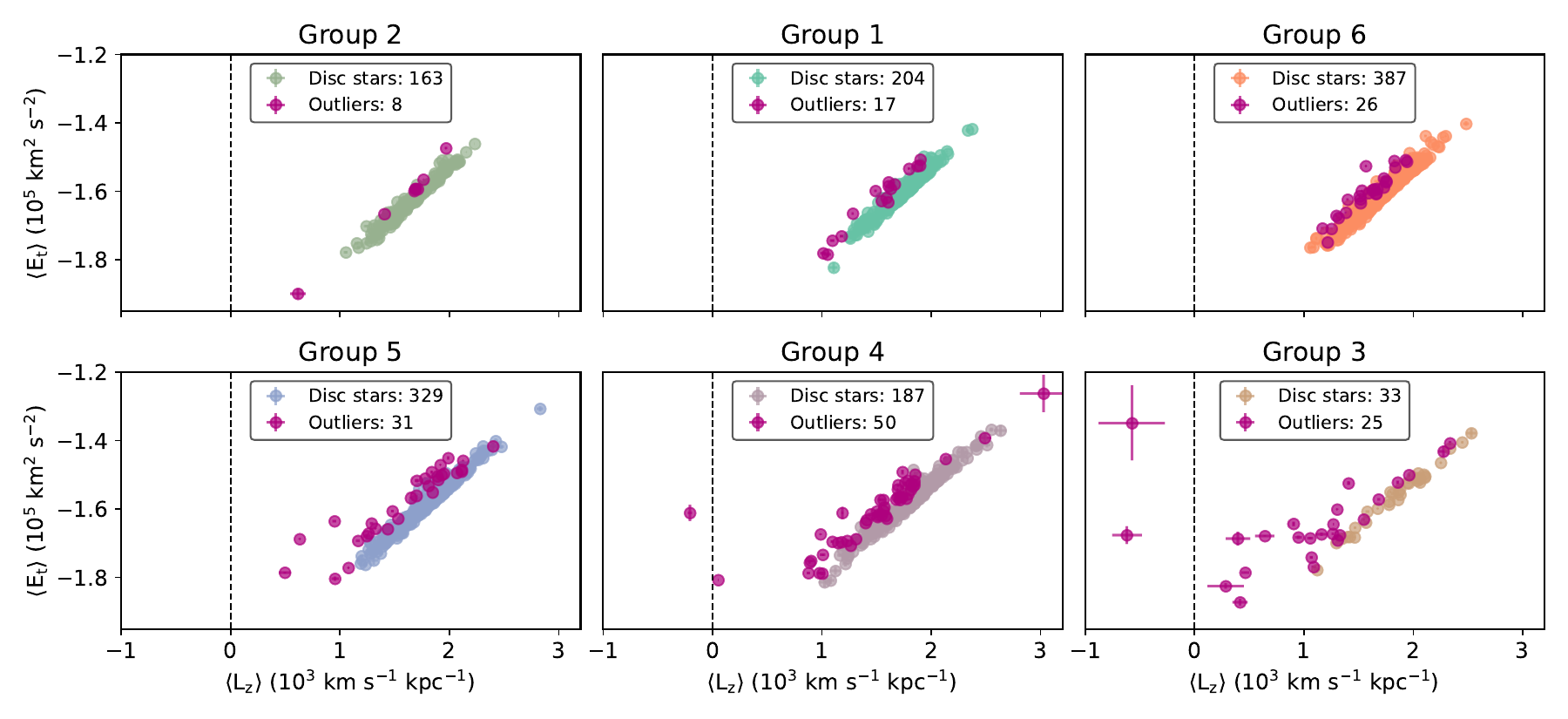}
    \caption{Lindblad diagram for all six groups retrieved from the HC, ordered by descending median metallicity $\langle \rm{\feh} \rangle$. A vertical dashed line is depicted at \lz=0, which denotes the stars with retrograde orbits around the Galactic centre. Stars with \lz~between 1 and 3 show stars that are probably well constrained within the MW's disc, whereas those with 0 $\leq$ \lz $<1$ have a higher probability of not being part of the disc, and those with \lz $<0$ probably come from the MW's halo. A uniform range is maintained on both the $x$ and $y$-axes, serving to direct the reader's attention towards outliers manifesting within lower metallicities. The colour scheme remains consistent with the preceding figures.}
    \label{fig:lindblad}
\end{figure*}

In an attempt to remove stars that potentially do not belong to the Galactic disc, we used as a threshold the relative absolute velocity to the Sun up to 100 km s$^{-1}$, as shown in the Toomre diagram displayed in Fig. \ref{fig:toomre}. The stars above this 100 km s$^{-1}$ concentric circle are a potential contamination of stars from the Galaxy's halo and should not be considered in our discussion of radial migration. The stars are also shown in the Lindblad diagram in Fig. \ref{fig:lindblad} where the same outliers identified in the Toomre diagram are highlighted. For easier reference, we also display these numbers in Table \ref{tab:disc}. We provide a complementary supporting figure which depicts the action map of our sample in Appendix \ref{appendix_subsec:prelim_chemodynamic} (Fig. \ref{fig:action_map}).

These outliers exhibit consistent characteristics across all plots. Notably, as metallicity decreases, a proportionally higher number of outliers are identified. Stars in groups 4 and 3, in particular, are the most metal-poor and have the largest median ages. In addition to their low metallicities and high ages, some of the outliers in these groups possess \lz~and \etot~values consistent with remnants of accretion and/or merger events. These characteristics align with expectations from the Gaia-Enceladus merger (GE; \citealt{Belokurov2018, Helmi2018}; and see \citealt{GiribaldiSmiljanic2023} for a precise time estimate of the GE merger, suggesting that it was completed $9.6 \pm 0.2$ Gyr ago). Since the focus of this paper is not to explore the various merger or accretion scenarios that have influenced the MW's evolution, we did not include these outlier stars in our analysis.

We acknowledge, however, that the Toomre diagram is not a one-size-fits-all solution for classifying stellar populations. In this study, we employed it as a straightforward means to exclude halo stars. Nevertheless, some disc stars, particularly those with lower $U$ or $W$ velocities, may be misclassified as halo stars. Appendix \ref{appendix_subsec:disc_or_halo} addresses this issue in detail, showing that only about 1.2\% of the total sample (17 out of 1460 stars) may be subject to this misclassification, with an upper estimate of 3.4\% (50 out of 1460 stars); therefore, we regard these cases as having a minor impact.  In total, only around 12\% of the total sample is excluded as halo stars (Table \ref{tab:disc}). Additionally, we also apply a thin-thick disc separation criterion in the Appendix, as outlined in Sect. \ref{subsubsec:thin_thick_disc}.

\begin{table}
    \caption{Classification of stars based on their membership in the Galactic disc.}
    \centering
    \begin{tabular}{c c r}
    \toprule
    Cluster Group & Outside Disc? & {Total Stars} \\
    \midrule
    \midrule
    \multirow{2}{*}{2}     & False  & 163 \\
                           & ~True  &  8  \\
    \midrule
    \multirow{2}{*}{1}     & False  & 204 \\
                           & ~True  &  17 \\
    \midrule
    \multirow{2}{*}{6}     & False  & 387 \\
                           & ~True  &  26 \\
    \midrule
    \multirow{2}{*}{5}     & False  & 329 \\
                           & ~True  &  31 \\
    \midrule
    \multirow{2}{*}{4}     & False  & 187 \\
                           & ~True  &  50 \\
    \midrule
    \multirow{2}{*}{3}     & False  & 33  \\
                           & ~True  &  25 \\
    \midrule
    \multirow{2}{*}{Total} & False & ~1303 \\
                           & True  & ~~157 \\
    \bottomrule
    \end{tabular}
    \tablefoot{
    This table shows the total number of stars (`Total Stars') for each `Cluster Group', categorised by their location relative to the Galactic disc (`Outside Disc?' = True for stars outside the disc and False for stars within the disc). The classification is based on criteria derived from the Toomre Diagram (Fig. \ref{fig:toomre}), as detailed in the current section.}
    \label{tab:disc}
\end{table}

% -----------------
% -----------------
\subsubsection{Assessing thin and thick disc membership} 
\label{subsubsec:thin_thick_disc}

After analysing which stars are unlikely to belong to the Galactic disc, we proceed to distinguish the classification of our stars in relation to the thin or thick disc. The radial metallicity models we adopted here focus on the temporal chemical enrichment of the thin disc, which renders the estimation of \rbirth\ for stars in the thick disc less certain. To evaluate whether the stars in our sample predominantly belong to the thin or thick disc, we adopted a variation of the prescription used by \citet{RecioBlanco2014}, and decided to differentiate between the two discs as follows:

\begin{itemize}
    \item For $-0.1 \leq \text{\feh} \leq +0.5$, \text{\mgfe} = +0.05.
    \item For $-0.5 \leq \text{\feh} \leq -0.1$, \text{\mgfe} extends from +0.05 to +0.15.
    \item For $-1.0 \leq \text{\feh} \leq -0.5$, \text{\mgfe} progresses from +0.15 to +0.25.
\end{itemize}

\noindent In our adaptation of this model, we employed a second-order spline to fit these lines, promoting a smoother transition between the thin and thick discs. To assess the classification probabilities of stars in our sample, we estimated the total error by combining the uncertainties for both \mgfe\ and \feh:

\begin{equation} \label{eq:sigma_tot_met}
    \sigma = (\sigma_{\rm \mgfe}^{2} + \sigma_{\rm \feh}^{2})^{1/2}.
\end{equation}

\begin{figure*}
    \centering
    \includegraphics[width=0.49\linewidth]{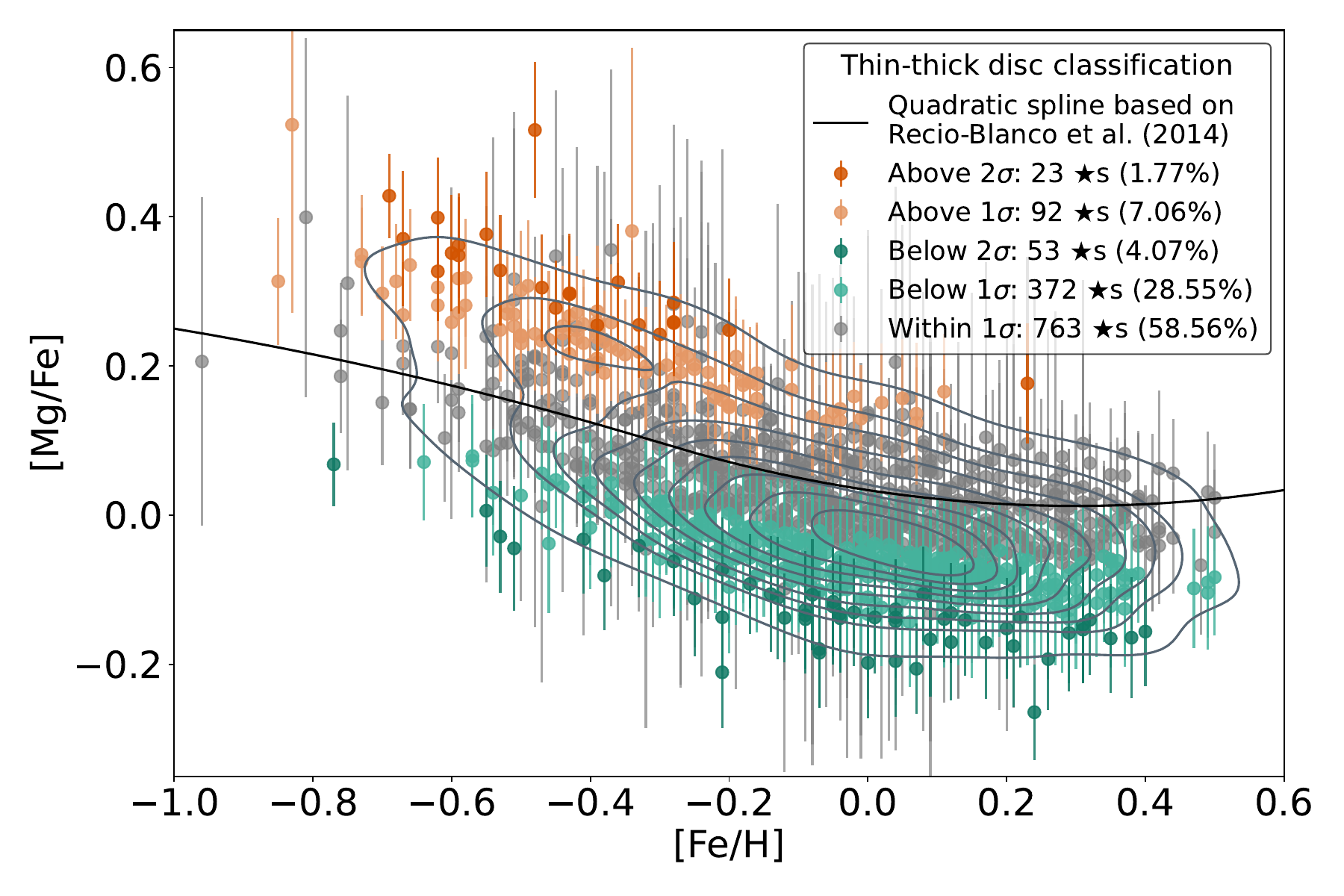}
    \includegraphics[width=0.49\linewidth]{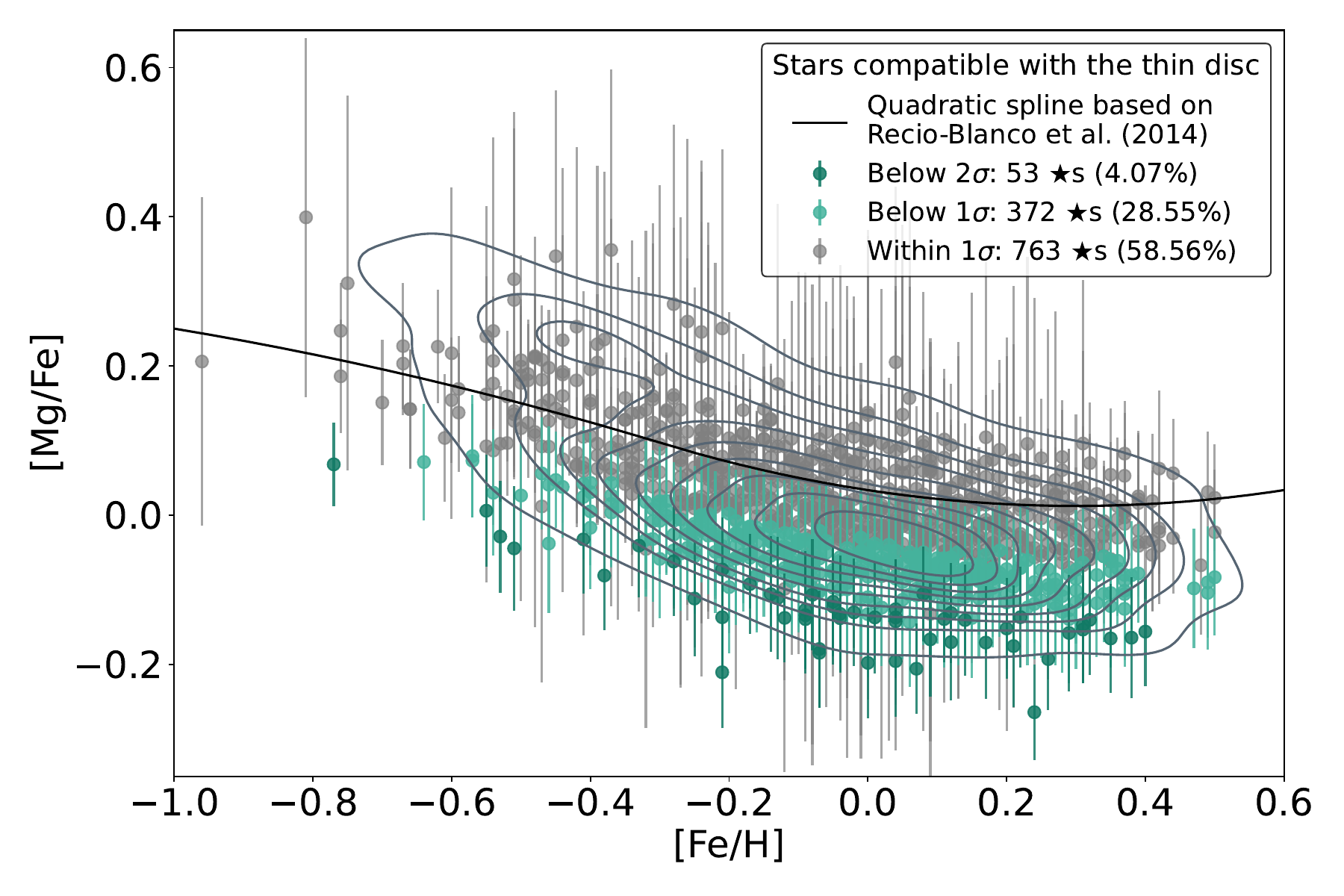}
    \caption{\mgfe\ against \feh\ for all disc stars in our sample (left panel) and for the stars compatible with the thin disc (right panel). Grey markers indicate stars that are compatible within 1$\sigma$ of the quadratic spline threshold line based on \citet{RecioBlanco2014}. Stars classified as belonging to the thin disc are shown in shades of cyan, while those associated with the thick disc are depicted in shades of orange. The varying shades represent classifications based on whether the stars fall within 1 or 2$\sigma$. It is important to note that stars classified within 1$\sigma$ in the figure exclude those classified within 2$\sigma$ to avoid double counting in the legend. In other words, the actual number of stars classified within 1$\sigma$ is the sum of those shown in both the 1 and 2$\sigma$ categories, meaning 115 and 425 stars for those above and below threshold respectively. Additionally, dark grey 2D Gaussian kernel densities are overlaid, revealing a double peak in star density: a prominent peak below the thin disc threshold and a secondary peak at low metallicity, highlighting the likely location of most potential interlopers from the thick disc.}
    \label{fig:thin_thick_disc}
\end{figure*}

\noindent The classification results are presented in Fig. \ref{fig:thin_thick_disc} and Table \ref{tab:thin_thick}. This method aligns well with current literature, as the Galactic thick disc is generally recognised for its enhanced $\alpha$-element abundance relative to the thin disc (e.g. \citealt[][]{Soubiran2003, Bensby2011, Trevisan2011, Cheng2012, Bland-HawthornGerhard2016}; but see also \citealt{Adibekyan2011} for a discussion on a metal-rich $\alpha$-enhanced stellar population originated from the inner disc). We estimate that contamination by thick disc stars constitutes between 1.77\% and 8.83\%, corresponding to 2$\sigma$ and 1$\sigma$ levels, respectively, with most contamination concentrated in the metal-poor regime. The remaining stars either fall below the classification threshold, confirming their status as thin disc members, or are consistent with it within acceptable uncertainties, thus remaining compatible with a thin disc classification.

To enhance our thin disc sample’s purity, while trying to maintain a good level of completeness, we excluded only the stars marked in orange, representing almost 9\% of the disc stars. These stars exceed the 1$\sigma$ threshold ($\sim 68\%$ confidence interval), indicating a high likelihood of thick disc membership. We refrained from additional exclusions to retain a more comprehensive thin disc representation, acknowledging that a few thick disc interlopers may persist.

\begin{table}
    \caption{Table expressing the number of stars classified in the thin or thick disc according to the  prescription adopted here for each metallicity-stratified group (`G').}
    \centering
    \begin{tabular}{c|rrr|rrr|r}
    \toprule
    \multirow{2}{*}{G} & \multicolumn{3}{c|}{Thick disc} & \multicolumn{3}{c|}{Thin disc} & Within \\
          & $1\sigma$ & $2\sigma$ & $1\sigma_{+}$ & $1\sigma_{\rm}$ & $ 2\sigma$ & $1\sigma_{+}$ & $1\sigma$\\
    \midrule
    \midrule
    2     &  0 &  1 &   1 &  46 &  7 &  53 & 109 \\
    1     &  4 &  0 &   4 &  63 & 10 &  73 & 127 \\
    6     & 13 &  0 &  13 & 116 & 14 & 130 & 244 \\
    5     & 26 &  8 &  34 & 100 & 15 & 115 & 180 \\
    4     & 40 & 11 &  51 &  44 &  5 &  49 &  87 \\
    3     &  9 &  3 &  12 &   3 &  2 &   5 &  16 \\
    \midrule
    Total & 92 & 23 & 115 & 372 & 53 & 425 & 763 \\
    \%    & 7.06 & 1.77 & 8.83 & 28.55 & 4.07 & 32.62 & 58.56 
    \end{tabular}
    \tablefoot{The number of stars classified as belonging to either the thick or thin discs is based on their combined errors (described in Eq. \ref{eq:sigma_tot_met}). Classifications are determined by whether these errors are above (thick) or below (thin) the threshold defined by the quadratic spline adaptation to \citeauthor{RecioBlanco2014}'s prescription. The columns marked as $1\sigma_{+}$ sum the two preceding columns, as all stars classified above/below the $2\sigma$ threshold are also classified above/below the $1\sigma$ threshold. In other words, these columns reflect the total count of stars represented by both the orange and cyan shades in Fig. \ref{fig:thin_thick_disc} for both discs. The last two columns show the total number of stars in each group and their percentage of the total 1303 stars.}
    \label{tab:thin_thick}
\end{table}

% -----------------
% -----------------
\subsubsection{The final sample post chemo-dynamic assessment}
\label{subsubsec:final_sample_post_assessment}

Combining the removal of stars that either potentially belong to the halo (Sect. \ref{subsubsec:disc_or_halo}) and the Galactic thick disc (Sect. \ref{subsubsec:thin_thick_disc}, our final sample is comprised of 1188 stars.

\begin{figure}
    \centering
    \includegraphics[width=\linewidth]{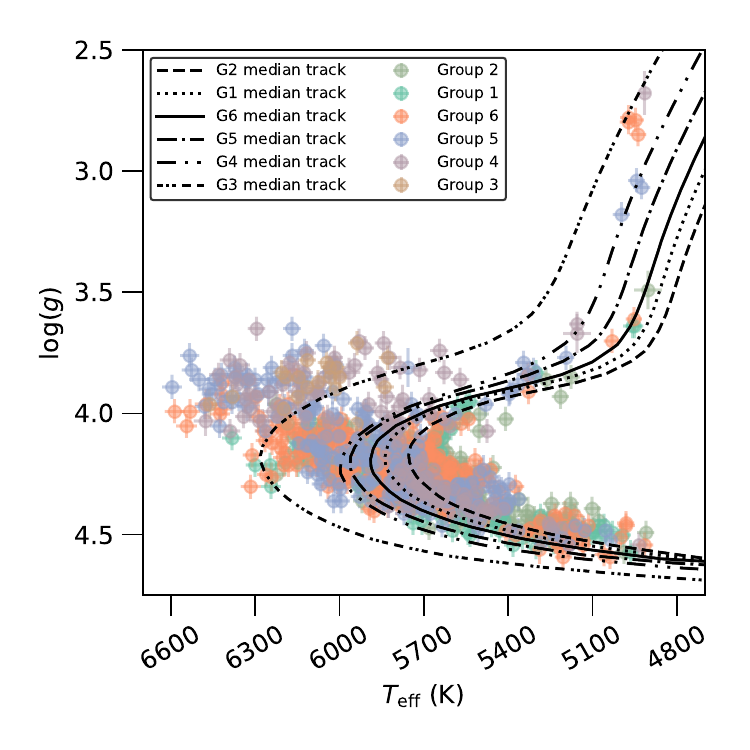}
    \caption{Kiel diagram for all thin disc stars in our final sample. The colour scheme aligns with that used in the preceding figures. Tracks indicating the median ages for each metallicity-stratified group are displayed with unique patterned line styles to facilitate group distinction.}
    \label{fig:disc_stars_kiel}
\end{figure}

\begin{figure}
    \centering
    \includegraphics[width=\linewidth]{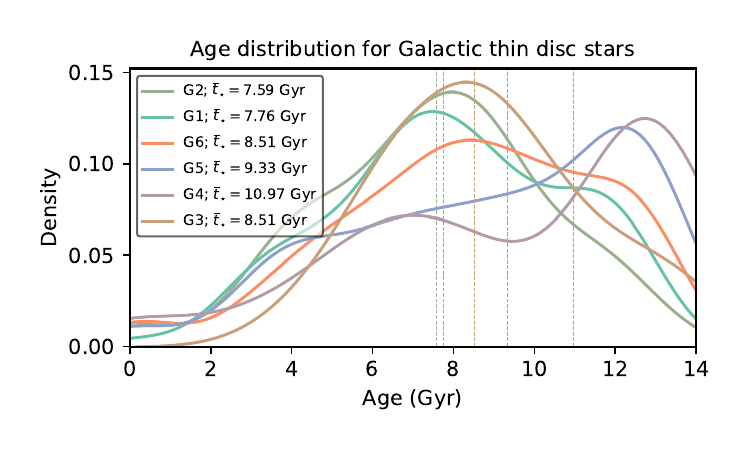}
    \caption{Age distribution for the stars belonging to the Galactic disc illustrated in the shape of Gaussian kernel densities for easier view (instead of histograms). The $x$-axis has been truncated to represent a physically meaningful range of stellar ages. The vertical dashed lines in corresponding colours show the computed median ages for each subgroup in our disc stellar sample. The colour scheme is consistent with the preceding figures.}
    \label{fig:disc_stars_ages}
\end{figure}

It is noteworthy that, after excluding stars potentially belonging to the Galactic halo and those that seem to belong to the thick disc, the median age, $\overline{t}_{\star}$, for the groups changes, and the variation becomes smaller. The difference between the youngest and oldest groups is reduced to $\sim$ 2 Gyr, which is within the expected age uncertainty \citep[$\approx$ 1-2 Gyr; see][]{Mints2018}. These are median values, and we discuss other effects of age and kinematics in Sect. \ref{subsec:churn_x_blurr}. Additionally, Group 4, which is the second most metal-poor, is now the oldest when comparing their age distributions. We display both the Kiel diagram for entire final sample and the distribution of these ages in Figs. \ref{fig:disc_stars_kiel} and \ref{fig:disc_stars_ages}, respectively. The tracks shown in Fig. \ref{fig:disc_stars_kiel} are PARSEC isochrones retrieved from the CMD web interface.\footnote{\url{http://stev.oapd.inaf.it/cgi-bin/cmd}} Regarding Fig. \ref{fig:disc_stars_ages}, we note that this distribution is slightly different from the one in the adjacent upper axis of Fig. \ref{fig:feh_age}, which contains the full sample, not only the stars in the Galactic thin disc.

For reference, we also provide the heliocentric positions for all the stars in our sample in Appendix \ref{appendix_subsec:helio_dist}.

% -----------------
\subsection{Birth radii estimation versus current Galactocentric distances} \label{subsec:birthradii}

\begin{figure*}
    \centering
    \includegraphics[width=\linewidth]{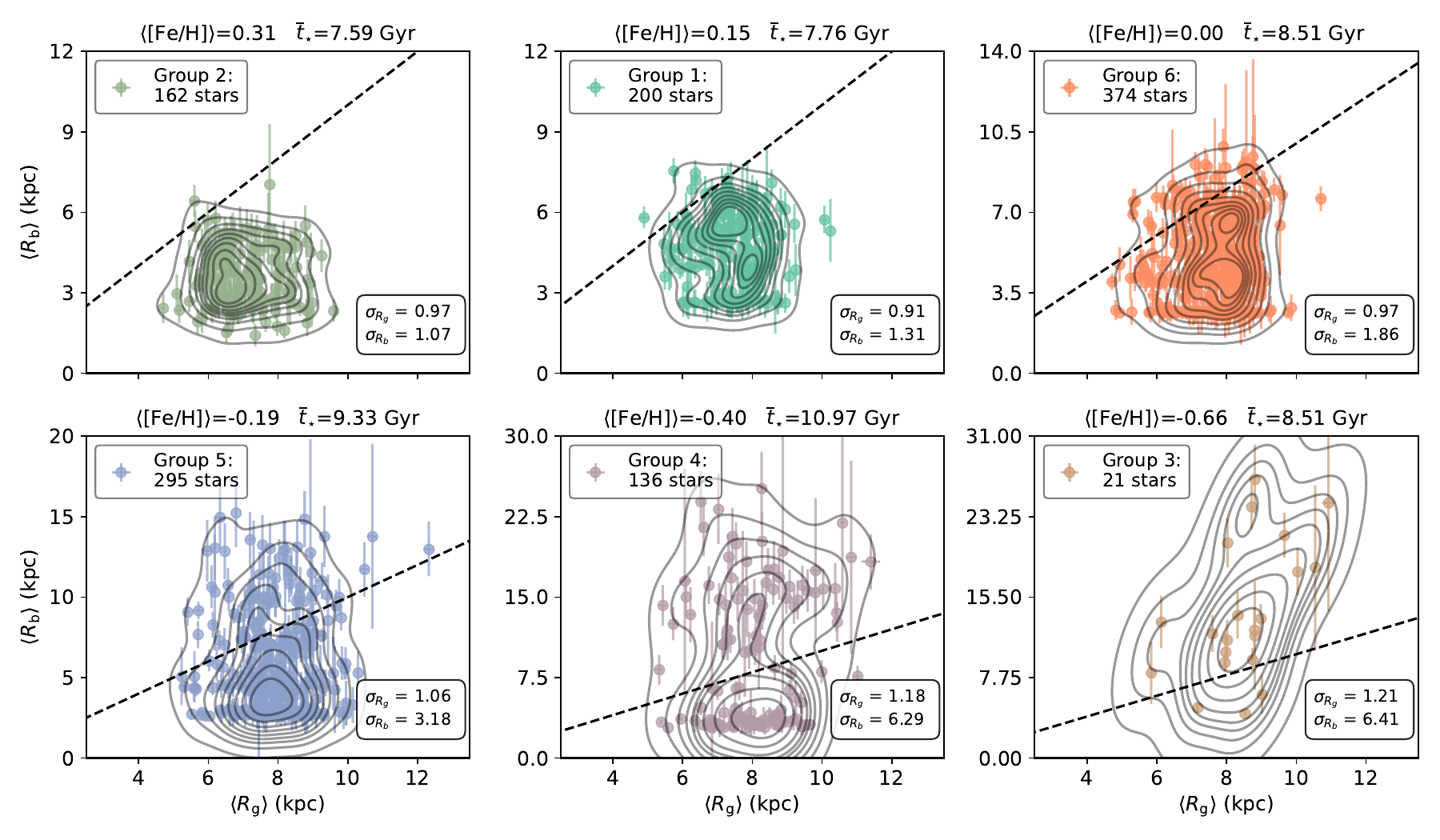}
    \caption{Estimated Galactocentric birth radii, \rbirth, in contrast to the present guiding radii, \rgui. Both radii are measured in kiloparsecs and encompass all stars in our sample that are part of the disc, according to Fig. \ref{fig:toomre}. Each distinct subplot delineates a specific star group, classified using hierarchical clustering (as illustrated in Fig. \ref{fig:hc_groups}). The subplots are arranged in descending order of the median iron abundance, $\langle \rm{\feh} \rangle$. At the upper part of each subplot, $\langle \rm{\feh} \rangle$ and $\overline{t}_{\star}$ are displayed. Furthermore, the overall standard deviations ($\sigma$) for \rbirth~and \rgui~are exhibited at the lower right corner of each subplot. We overlay the 2D-Gaussian kernel densities to illustrate the concentration patterns of the stars in each subplot. A reference black dashed line is incorporated, signifying the point of parity between both radii (\rbirth=\rgui). The colour scheme remains consistent with the preceding figures.}
    \label{fig:rb_rg_groups}
\end{figure*}

Figure \ref{fig:rb_rg_groups} presents a comparison between the estimated \rbirth, derived from the GAM, and the \rgui\ values calculated using \textsc{galpy}. It is evident that the most metal-rich groups tend to have \rgui\ values predominantly larger than their expected \rbirth. This trend gradually shifts as metallicity decreases, resulting in a greater number of stars residing within their original \rbirth, or even exhibiting smaller \rgui\ compared to their original \rbirth. This is an important finding, which we discuss further in Sect. \ref{subsubsec:migration_patterns} and \ref{subsec:churn_x_blurr}.

In several groups shown in Fig. \ref{fig:rb_rg_groups}, a notable limitation arises at the lower bound of the inferred \rbirth. Specifically, the GAM struggles to accurately derive \rbirth\ for stars with very small radii ($\sim$0 to 3 kpc) across all metallicity-stratified groups, except for Group 2. This limitation stems from the resolution constraints of the original chemical enrichment models, as discussed in Sect. \ref{subsec:challenges} and detailed in Appendix \ref{appendix:model}. Most importantly, these models do not predict the formation of non-super-metal-rich stars in the innermost regions of the MW. Notably, this effect suggests that these stars could have likely formed at even smaller radii than predicted, further supporting our conclusion of significant outward migration. This phenomenon is explored further in Appendix \ref{appendix_subsec:rb_other_rs}, with additional context provided by Fig. \ref{fig:rb_age}. It is also important to note that at these Galactocentric distances, we are probing regions very close to the bulge itself, rather than the (thin) disc. Consequently, our analysis and conclusions remain consistent.

Additional supporting figures illustrating \rperi\ and \rapo\ are provided in Figs. \ref{fig:rb_rperi_groups} and \ref{fig:rb_rpapo_groups}, respectively, in Appendix \ref{appendix:additionalmaterial}. In Figs. \ref{fig:rb_rg_groups}, \ref{fig:rb_rperi_groups} and \ref{fig:rb_rpapo_groups} we also include a black dashed line where \rbirth=\rgui, representing the location a star would occupy if it remained at its original estimated birth radius. Additionally, we show $\langle \rm{\feh} \rangle$ and $\overline{t}_{\star}$ at the top of each subplot. Although Fig. \ref{fig:rb_rg_groups} presents 2D Gaussian kernel density estimates to indicate where the combined distributions of \rgui\ and \rbirth\ are concentrated for each metallicity-stratified group, we also provide the separate distributions of \rbirth\ and \rgui\ in Fig. \ref{fig:rg_rb_kdes} in the same Appendix. This allows for a clearer view of the shifts in orbital radii.

% -----------------
% -----------------
\subsubsection{Challenges in predicting birth radii: Limitations of the chemical enrichment models and the generalised additive model} \label{subsec:challenges}

Upon examining the groups depicted in Fig. \ref{fig:rb_rg_groups}, particularly those with sub-solar metallicities (i.e. groups 5, 4, and 3), we observe that some stars exhibit relatively large birth radii (\rbirth~$\sim$ 28 kpc), exceeding the estimated radius of the MW, which is approximately 25 kpc (as previously discussed in Sect. \ref{subsubsec:gam_profile}). One possible explanation is contamination from thick disc stars, which poses challenges for elimination without significantly affecting the completeness of the thin disc population. However, if these stars predominantly belong to the thin disc, as seems more likely for the majority, alternative explanations may apply. Rather than interpreting this as evidence of stars originating outside the Galactic disc, it more likely reflects limitations in the chemical evolution models' ability to accurately account for the age-metallicity relationship in certain stars. Estimating accurate stellar ages, particularly for field stars, remains a significant challenge. We must also consider potential mismatches between model predictions and the actual metallicity evolution in the outer regions of the Galactic thin disc. These combined limitations lead to predictions that some stars have birth radii larger than the known extent of the MW. While the precise birth radii for these stars may be inaccurate, the broader conclusion -- that they formed in the outer disc and migrated inward -- remains robust.

Indeed, the ongoing improvements in chemical enrichment models highlight the evolving nature of this challenge. Periods of metal-poor gas infall, such as those described by \citet{Chiappini1997}, \citet{Micali2013}, and \citet{Spitoni2020}, complicate the task of accurately predicting stellar birth radii by altering the metallicity gradients in the Galaxy. The recent work by \citet{Palla2024}, which discusses a third recent gas infall episode in the MW, further illustrates how these events reshape metallicity gradients and explain the observed abundance patterns. Consequently, the challenge of estimating birth radii for stars, particularly those with low metallicities and high ages, is heightened by these chemical evolution processes, which introduce complexities that models cannot always capture accurately.

Furthermore, we recognise that there may be inherent metallicity variations in a given radius of the Galaxy at any time, leading to natural scatter around the $R_{\rm b}$ estimated by our models in conjunction with the GAM. This is of course challenging to fully capture with models that assume homogeneity of the chemical distribution with radius. This is a point that could be addressed with more complex chemical evolution models in future work.

A complementary explanation for the discrepancies in \rbirth\ predictions could lie in the limitations of the GAM in predicting \rbirth\ for stars with unusual combinations of characteristics, such as high age despite low metallicity, as seen in groups 5, 4, and 3. This issue is evident when considering the original models and the estimated GAM curves shown in Fig. \ref{fig:comparison_feh}. There is a notable gap between the Universe’s age of 3.3 and 8 Gyr, corresponding to the formation period of metal-poor stars. During this time, the curves fail to extend to lower metallicity ranges (\feh$\lesssim-0.5$) while maintaining a reasonable $R$. Consequently, while the GAM represents an improvement over prior models, it remains constrained by their limitations. Thus, any inaccuracies in \rbirth\ predictions may not reflect shortcomings in the GAM itself but rather the limitations imposed by the original chemical evolution models.

% -----------------
% -----------------
\subsubsection{Galactocentric migration patterns across different stellar groups} \label{subsubsec:migration_patterns}

Figure \ref{fig:rb_rg_groups} shows that $\overline{t}_{\star}$ decreases as $\langle \rm{\feh} \rangle$ increases (except for Group 3), accompanied by a reduction in the overall variance of both \rbirth~and \rgui. In other words, as we look at groups with larger metallicities, the stars seem to cluster and become more tightly grouped within a narrower range of both \rbirth~and \rgui~(we note in particular that while the range of \rgui~values is the same in all panels, the range of \rbirth~values increases from the top left to the bottom right). The clustering in \rbirth~for the stars with high metallicity can be easily understood with a glance at the gradients displayed in Figs. \ref{fig:comparison_feh} and \ref{fig:new_grids_split_feh}. Old stars of high metallicity can only form in the inner regions of the Galaxy \citep[as we discussed previously in][]{Dantas2023}. In addition, we note that a decrease in the number density of metal-rich stars towards the outer disc can be seen in the sample of other surveys as well \citep[e.g.][]{Hayden2015, RecioBlanco2023}.

Another important feature seen in Fig. \ref{fig:rb_rg_groups} is that some of our groups of stars seem to be concentrated mostly above the black dashed line (e.g. groups 3 and 4), whereas others seem to be mostly below it (e.g. groups 1 and 2). Furthermore, this trend appears to be intricately linked with metallicity and age. It is noteworthy that not only do the most metal-rich stars appear to have originated in the inner regions of the MW, subsequently migrating to larger orbital radii, but also some of the most metal-poor stars currently reside at smaller Galactocentric distances in contrast to their initial birth radii. Stars exhibiting intermediate metallicities exhibit a more balanced distribution along this dashed line. This migration pattern is likely the most important finding of the paper, offering critical insights into the dynamic evolution of the Galaxy.

Most studies investigating radial migration focus on the metal-rich stars likely to come from the inner Galaxy \citep[e.g.][]{Trevisan2011, Chen2019, Zhang2021}, but a smaller set of studies also show that there is a population of metal-poor stars in the solar vicinity that is likely to have originated from the outer disc \citep[e.g.][]{Haywood2008}. The results of this paper show the migration pattern observed in the solar vicinity that encompasses both the metal-rich and the metal-poor regimes. These results are generally in agreement with other studies, such as the predictions from the simulated models by \citet{Martinez-Medina2016}, in which a mixing from both the inner and the outer disc is detected in the solar vicinity, as well as the larger amount of stars with metallicities near-solar (as seen in Sect. \ref{subsec:churn_x_blurr}).

Later, in Sect. \ref{subsec:churn_x_blurr}, we evaluate in more detail whether each star has \rgui~consistent with its \rbirth~within a certain threshold and evaluate the behaviour of those stars that show dissonant values of \rgui~and \rbirth.

% -----------------
% -----------------
\subsubsection{Other signs of migration} \label{subsubsec:other_signs_migration}

\begin{figure*}
    \centering
    \includegraphics[width=\linewidth]{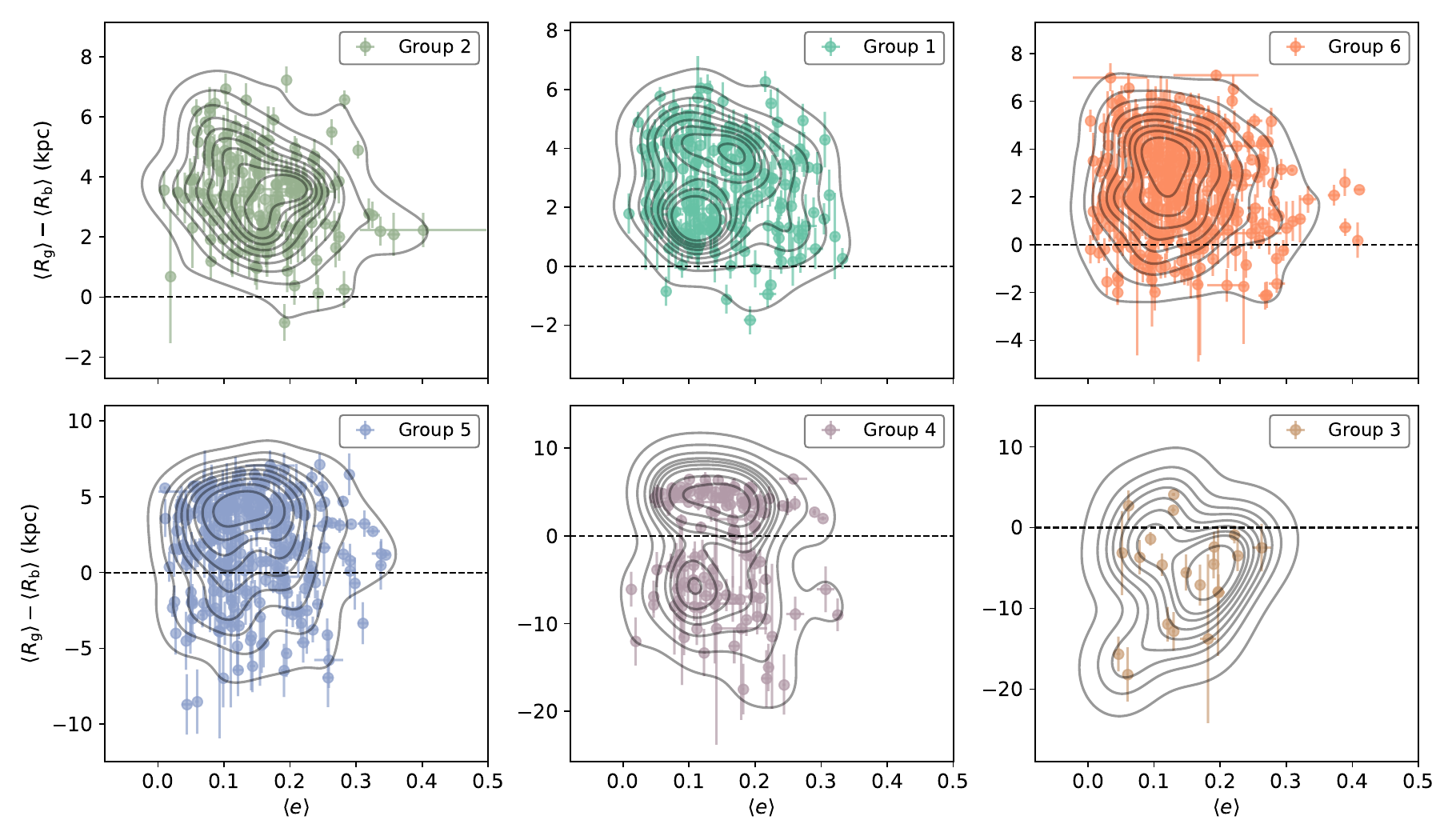}
    \caption{Variation of Galactocentric distance versus median eccentricity (\rgui-\rbirth~ $\times$ \eccentricity) across stellar groups, arranged in descending order of median metallicity $\langle \rm{\feh} \rangle$. The range of eccentricity (\eccentricity) on the $x$-axis remains consistent across all subplots, while the $y$-axis retains an unconstrained scale, enhancing the visualisation of the pronounced \rgui-\rbirth~dispersion. Corresponding to Fig. \ref{fig:rb_rg_groups}, a horizontal dashed line is positioned at \rgui-\rbirth=0 to accentuate the region where Galactocentric distances coincide. To illuminate the 2D distribution of stars within the designated parameter space, 2D-Gaussian kernel densities are overlaid upon the scatter markers. Consistent with prior visuals, the colours in this figure maintain the same scheme.}
    \label{fig:deltar_ecc}
\end{figure*}

\begin{figure*}
    \centering
    \includegraphics[width=1\linewidth]{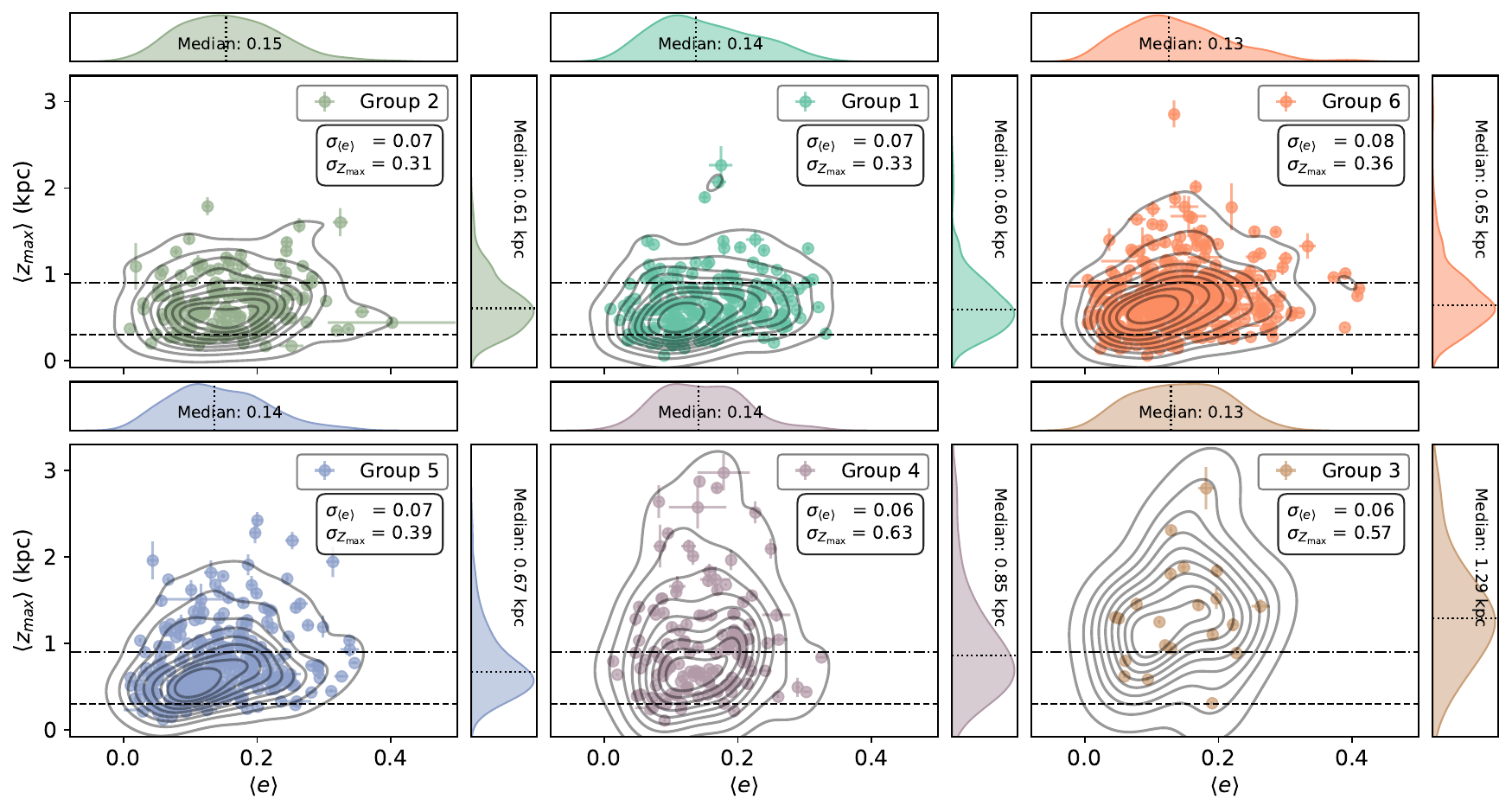}
    \caption{Median maximum Galactic azimuthal height, \zmax, versus the median eccentricity, \eccentricity, for all disc stars in the six groups, ordered by decreasing metallicity from top left to bottom right. The 1D-Gaussian kernel density distributions for \zmax~and \eccentricity~are shown adjacent to the main subplots, with medians marked by dotted lines and annotated values. The main subplots display all disc stars, with thin (300 pc) and thick disc (900 pc) scale heights \citep{McMillan2017} marked by dashed and dot-dashed lines, respectively, and a 2D Gaussian kernel density plot indicating higher density regions. Consistent $x$- and $y$-axis ranges highlight the shift in distributions across metallicity groups, with the proportion of stars with larger \zmax~increasing as metallicity decreases.}
    \label{fig:eccentricity_zmax_kdes}
\end{figure*}

Beyond estimating the \rbirth~of the stars in our sample, other dynamic features can assist us in understanding the nature of their movement. For instance, orbital eccentricity, \eccentricity, appears to play a significant role in determining whether a star has experienced churning. Stars with lower eccentricities seem to be more susceptible to the constant influence of the Galactic structures (such as the bar and spiral arms), which forces them to migrate to outer radii. Yet, there is no consensus on the role of eccentricity when investigating churning and blurring. Some believe that churning happens due to a change in angular momentum but without a change in eccentricity \citep[e.g.][]{VC2014, Martinez-Medina2016}. Conversely, other studies suggest that the orbits of stars affected by churning tend to become increasingly circular over time, indicating that their eccentricities decrease \citep[see][where the authors discuss the influence of the bar in reducing the eccentricity when stars are churned]{Khoperskov2020a}. There is also some evidence that eccentricity can also slightly increase for some churned stars \citep{Roskar2012} potentially caused by the transient spiral arms. Although the causal relationship between eccentricity and the migration process -- i.e. churning -- remains unclear, it is imperative to consider eccentricity as a critical variable when tracking the evolving trajectories of stars.

Another noteworthy parameter implicated in churning is the maximum vertical distance, \zmax, that stars can attain from the Galactic plane. \citet{Roskar2013} show that the Galactic disc tends to thicken as stars undergo outward radial migration. Additionally, evidence suggests that azimuthal variations in metallicity among old stars may indicate strong bar activity, leading to radial migration, specifically churning \citep{DiMatteo2013}. However, this interpretation is debated: \citet{Debattista2025} argue that these azimuthal metallicity variations may instead arise from different stellar populations reacting in distinct ways to spiral arm perturbations. Furthermore, recent studies by \citet{ViscasillasVazquez2023, ViscasillasVazquez2024} show that older stellar clusters exhibit more disturbed orbits compared to their younger counterparts, with higher Galactic heights and eccentricities. These studies serve as evidence that supports the idea that Galactic structures introduce substantial azimuthal perturbations to stellar orbits. Given that our sample of thin disc stars has $\overline{t}_{\star}$ ranging from $\simeq$ 7.6 to $\simeq$ 11 Gyr (considering the different metallicity-stratified groups), we consider them sufficiently old to have been significantly influenced by such Galactic structures (e.g. the central bar, spiral arms, dense molecular clouds) over time.

Therefore, we analyse our stars' \eccentricity\ and \zmax\ to explore whether these parameters are connected to migration patterns. Figure \ref{fig:deltar_ecc} presents the variations in radii, \rgui-\rbirth\ (or simply, $\Delta R$), against \eccentricity\ for all groups in our sample. This figure provides alternative an view of the results shown in Fig. \ref{fig:rb_rg_groups} by incorporating \eccentricity. In Fig. \ref{fig:deltar_ecc}, the eccentricities of stars across all groups appear constrained to $\sim$0.4, though Group 6 contains a few outliers with higher eccentricities. As we move towards metal-poor groups, the variation in $\Delta R$ increases, but \eccentricity\ remains largely unchanged.

It is worth noting that \citet{Kordopatis2015} also report very low eccentricities for super-metal-rich stars in the solar vicinity, which have migrated from the inner Galaxy. This observation is consistent with our findings for Group 2, and as discussed further in \citet{Dantas2023}, it reinforces the idea that churning dominates in this range of metallicities.

Another way to visualise the behaviour of the stars in our sample is by plotting \zmax\ versus \eccentricity, which is displayed in Fig. \ref{fig:eccentricity_zmax_kdes}. The subplots in Fig. \ref{fig:eccentricity_zmax_kdes} are organised by decreasing metallicity, from top left to bottom right. The scale heights of the thin and thick discs are indicated in the figure by dashed and dot-dashed lines at 0.3 and 0.9 kpc, respectively \citep[estimates from][]{McMillan2017}. The consistent axis ranges across all subplots allowed us to observe the shifting distributions as metallicity decreases, notably the increasing proportion of stars with larger \zmax\ values as metallicity decreases, especially for Groups 4 and 3. In fact, the median \zmax\ for Group 4 is approximately at the scale height of the thick disc, and significantly above it for Group 5. The standard deviations for both \eccentricity\ and \zmax\ are displayed in each subplot, demonstrating that the variance in \zmax\ increases as metallicity decreases, while the variance in \eccentricity\ remains pretty constant. As for \eccentricity, it is noticeable that the medians are all $<0.2$, suggesting that most stars have near-circular orbits, but all have somewhat low \eccentricity, almost all having \eccentricity\ $<0.4$; few outliers can be seen in Groups 2 and 6.

As metallicity decreases, a greater number of stars appear to reach larger \zmax\ values. This could be attributed to residual interlopers from the thick disc not previously identified in Sect. \ref{subsubsec:thin_thick_disc}. However, if most of these stars indeed belong to the Galactic thin disc, as expected, this trend could be potentially explained by two factors. (i) Stars in the most metal-poor groups tend to concentrate at older ages (as is discussed in Sect. \ref{subsubsec:dynamics_churn_blur_overall} and detailed in Table \ref{tab:churning_blurring}), giving them more time to interact with Galactic structures (e.g. giant molecular clouds), which may contribute to disc thickening, as shown by \citet{Roskar2013}. However, this applies mainly to stars churned outward, and as we discuss in Sect. \ref{subsec:churn_x_blurr}, the proportion of such stars in these metal-poor groups is low. (ii) Alternatively, this trend can be due to the interaction between the outer disc and the first pericentric passage of Sagittarius, which occurred around 6 Gyr ago. According to \citet{Das2024}, stars older than 6 Gyr are significantly kinematically hotter than their younger counterparts. Indeed, our observations indicate that the stars in the most metal-poor groups -- those likely formed at greater Galactocentric distances -- have higher \zmax\ values. This is consistent with the hypothesis that stars formed at larger radii are more susceptible to the influence of interactions with Sagittarius, leading to dynamic heating. Furthermore, as we show and discuss in Sect. \ref{subsec:churn_x_blurr}, a substantial proportion of metal-poor stars appear to be migrating inward from the outskirts of the MW. 

\begin{figure*}
    \centering
    \includegraphics[width=1\linewidth]{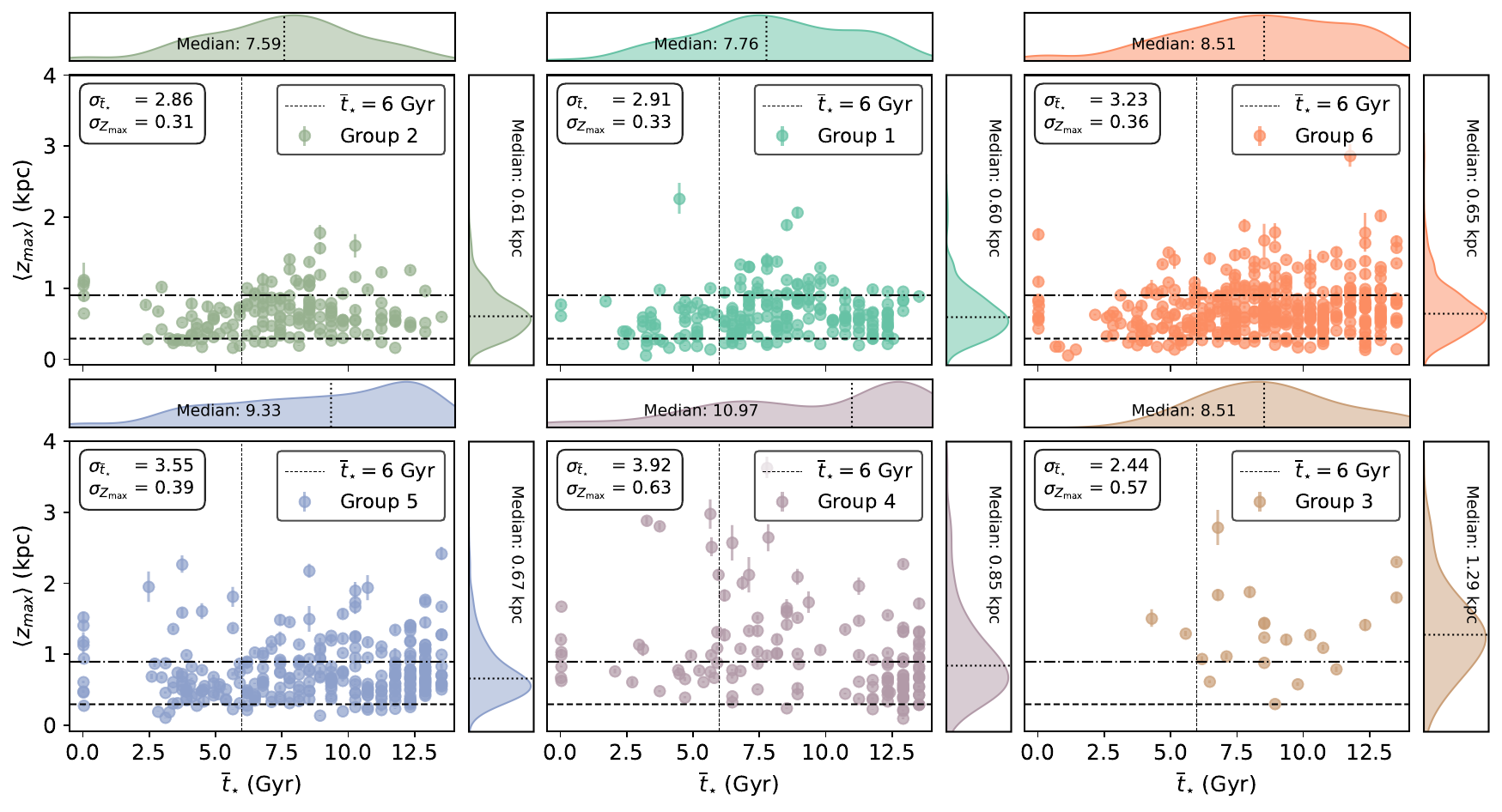}
    \caption{Median maximum Galactic height, \zmax, versus the median stellar age, $\overline{t}_{\star}$, for all disc stars in the six groups, ordered by decreasing metallicity from top left to bottom right. This image is similar to Fig. \ref{fig:eccentricity_zmax_kdes}, with adjacent 1D-Gaussian kernel distributions helping the visualisation of the distribution of both parameters and horizontal lines indicating the scale heights for the thin and thick discs (dashed and dot-dashed lines respectively). A vertical line is added at $t=6$ Gyr to flag the time of the first pericentric passage of Sagittarius. Age uncertainties are not displayed to improve visualisation, but typical errors are of 2 Gyr \citep{Mints2018}. It is noticeable the increase in the proportion of stars with larger \zmax~with decreasing metallicities (i.e. groups 5, 4, and 3), and thus we have included the standard deviation of both parameters ($\sigma_{\overline{t}_{\star}}$ and $\sigma_{Z_{\rm max}}$) for each group in the top left corner. Both standard deviations tend to increase as $\langle \rm{\feh} \rangle$ decreases.}
    \label{fig:zmax_age}
\end{figure*}

To further explore this hypothesis, we depict \zmax\ versus $\overline{t}_{\star}$ in Fig. \ref{fig:zmax_age}, stratified by the HC groups. It is evident in Fig. \ref{fig:zmax_age} that stars aged 6 Gyr or older exhibit a higher proportion of larger \zmax~values, especially in the more metal-poor groups, such as Groups 5, 4, and 3. Overall, if the majority of these stars are not interlopers from the thick disc, we propose that this increased variation in \zmax\ is likely due to a combination of outward churning and the dynamic influence of the pericentric passage of Sagittarius.

\begin{figure*}
    \centering
    \includegraphics[width=0.49\linewidth]{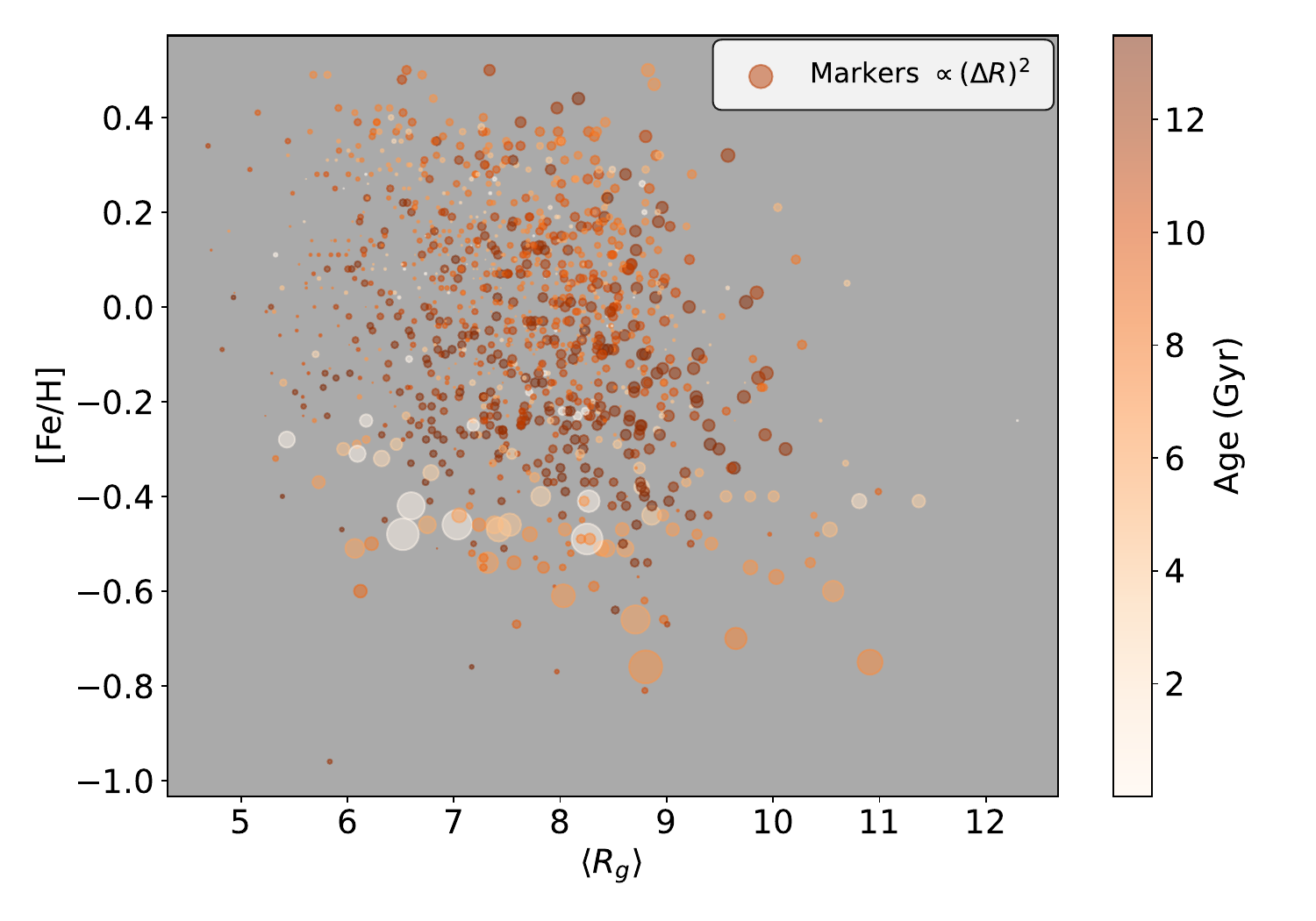}
    \includegraphics[width=0.49\linewidth]{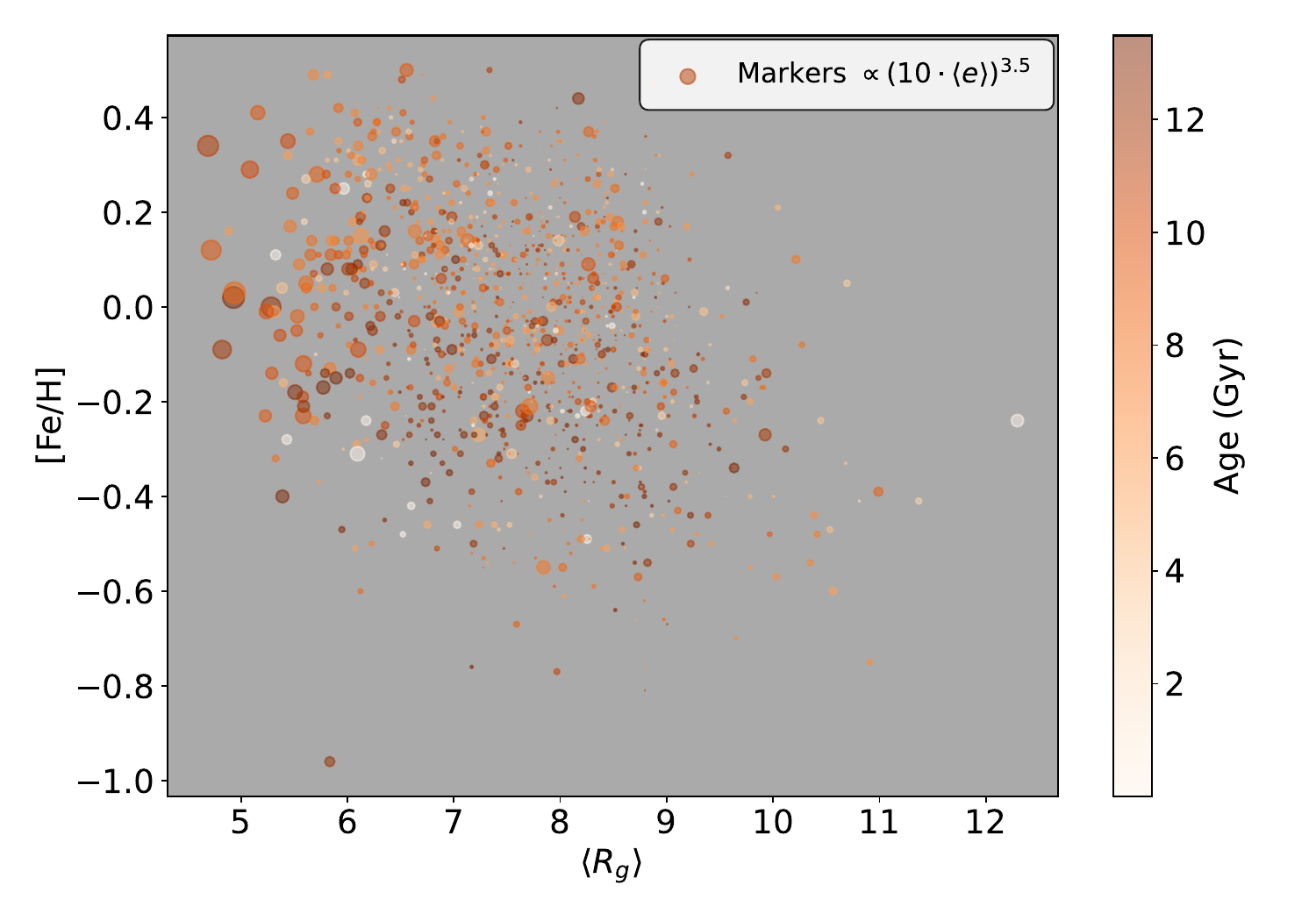}
    \caption{Scatterplot delineating the distribution of metallicity (\feh) against the median guiding radius (\rgui) for our sample of disc stars. The colour gradient encodes the age in Gyr. Left panel: the marker size is modulated as a function of $(\Delta R)^{2}$ [i.e. (\rgui-\rbirth)$^2$], as the $y$-axis of Fig. \ref{fig:deltar_ecc}. Right panel: The marker size is proportional to $(10 \cdot \langle e \rangle)^{3.5}$, where the minimum and maximum \eccentricity\ values are 0.003 and 0.410 respectively. The marker size in each panel is arbitrary and slightly exaggerated to highlight each feature related to their movement, $\Delta R$ or $\langle e \rangle$. We do not attribute physical meaning to the sizes of the markers, but focus on showcasing where the most important differences lie. By doing this, we qualitatively verify that stars with lower metallicity tend to have higher $\Delta R$, whereas stars with smaller \rgui\ seem to have a slightly higher $\langle e\rangle$.}
    \label{fig:cmap}
\end{figure*}

The subsequent image, Fig. \ref{fig:cmap}, condenses a lot of information from our sample into two simple panels. It shows two colour maps of \feh\ versus \rgui\ for our sample of 1188 thin disc stars, with the colour-map representing stellar ages. The key difference between these panels is the basis for marker size. In the left panel, marker sizes are proportional to $(\Delta R)^2$, where $\Delta R$ is the difference between \rgui~and \rbirth. In contrast, the right panel uses marker sizes proportional to eccentricity, specifically $(10 \cdot \langle e \rangle)^{3.5}$. The marker sizes are exaggerated to better highlight stars with the largest $\Delta R$ and \eccentricity.

In the right panel, we greatly enlarged the marker sizes to make the high \eccentricity\ region more prominent. As a result, stars with the largest eccentricities — least likely to be affected by churning — cluster on the left side of the plot \citep[e.g.][]{Khoperskov2020a}. These stars show a wider range of metallicities, smaller \rgui\ values (closer to the Galactic centre), and tend to be older. Conversely, in the left panel, stars with the largest radial differences, $\Delta R$, span a range of \rgui~values but are predominantly metal-poor, with ages ranging from young to intermediate (about 0 to 4–5 Gyr). Although the marker size indicates stars with larger variation in their past and current orbital radii, which can be a sign of radial migration, it does not necessarily confirm it, since we are not accounting for errors-in-measurements here. This is discussed further in Sect. \ref{subsec:churn_x_blurr}.

We revisit these dynamic features later in this paper, after classifying the stars based on whether they underwent churning, blurring, or are undisturbed, in the upcoming sections.

% -----------------
\subsection{Churned versus blurred/undisturbed stars}
\label{subsec:churn_x_blurr}

Using this refined analysis, we quantified which stars have undergone churning, blurring or are undisturbed based on the differences between their \rbirth\ and current \rgui. We applied a straightforward criterion to classify the motion of our stars. Our criterion is as follows:

\begin{enumerate}

    \item We first combined the errors of \rgui\ and \rbirth\ to estimate the total deviation:

    \begin{equation} \label{eq:sigma_tot_radii}
        \sigma_{\rm tot} = (\sigma_{R_b}^2 + \sigma_{R_g}^2)^{1/2}\text{, similarly to Eq. \ref{eq:sigma_tot_met}.}
    \end{equation}

    \item We defined a threshold $\tau$ to determine whether the current \rgui~of a star is compatible with its \rbirth:

    \begin{equation}
        \tau = \eta \times \sigma_{\rm tot},
    \end{equation}
    
    \noindent where $\eta$ is the number of times $\sigma_{\rm tot}$ is considered. We used $\eta=2$, corresponding to approximately a 95\% confidence interval.

    \item We then compared \rbirth\ with the current Galactocentric distance, \rgui, and classified the stars based on their $|\Delta R|$, where $|\Delta R| = |\langle R_b \rangle - \langle R_g \rangle|$:

    \begin{enumerate}[a.]
        \item $|\Delta R|   >  \tau$: Churned;
        \item $|\Delta R| \leq \tau$: Blurred/Undisturbed.
    \end{enumerate}

    \item Finally, we examined the stars classified as being prone to churning and separately counted those with \rgui>\rbirth~and \rgui<\rbirth. We classified those with \rgui>\rbirth~as having moved outward and those with \rgui<\rbirth~as having moved inward.
\end{enumerate}

By definition, stars undergoing churning experience changes in their angular momentum ($L$), while those primarily affected by blurring retain their $L$. A reliable way to distinguish between these processes is by examining variations in orbital radii, which serve as a proxy for changes in $L$. However, undisturbed stars also maintain their angular momentum, making it challenging to distinguish them from blurred stars based solely on this criterion.

To address this, we classify stars with $|\Delta R| \leq \tau$ as either blurred or undisturbed, as both groups exhibit minimal changes in orbital radii. This approach is supported by \citet{Feltzing2020}, who, using different arguments, also link large variations in orbital radii over time to churning, and smaller variations to blurring. Their findings reinforce the use of this distinction in our classification.

Building on this approach, we aim to assess whether our findings align with the expected behaviour of churned and blurred/undisturbed stellar populations. For instance, we can evaluate their properties across several dynamical parameters, such as \eccentricity, \zmax, and most importantly, $L$ (change or lack thereof) -- the latter being the defining characteristic of both blurring and churning. This serves as an independent verification of our observational results, which can later be compared with expectations from simulations.

\renewcommand{\arraystretch}{1.2}

\begin{table*}
    \caption{Comprehensive data summary of stars affected by either churning (C) or a combination of blurring and undisturbed motion (B/U) in the various cluster groups.}
    \centering
    \begin{tabular}{cclccccccccc}
    \toprule
    Group & Movement & Direction & $N_{\star}$ & \% & $\langle \rm{\feh} \rangle$ & $\,\overline{t}_{\star}$ &\,\eccentricity & \zmax  & \lr & \lphi & \lz \\
    & & & & & & Gyr & & kpc &  \multicolumn{3}{c}{$\text{kpc} \cdot \text{km/s}$}\\
    \midrule
    \midrule
    \multirow{3}{*}{2} & \multirow{2}{*}{C} & Inward  &   0 &  0.00 &   N/A &   N/A &  N/A &  N/A &     N/A &     N/A &     N/A \\  
                       &                    & Outward & 148 & 91.36 &  0.32 &  7.76 & 0.15 & 0.60 &   29.38 &   29.12 & 1649.04 \\
                       & B/U                & Equal   &  14 &  8.64 &  0.27 &  3.89 & 0.18 & 0.68 &   58.47 &  -11.51 & 1404.36 \\
    \midrule 
    \multirow{3}{*}{1} & \multirow{2}{*}{C} & Inward  &   3 &  1.47 &  0.08 &  3.72 & 0.19 & 0.54 &   96.16 &  -11.62 & 1314.92 \\
                       &                    & Outward & 153 & 76.50 &  0.16 &  8.51 & 0.13 & 0.64 &   56.04 &   17.92 & 1792.52 \\
                       & B/U                & Equal   &  44 & 21.57 &  0.13 &  6.46 & 0.14 & 0.49 &   39.77 &   12.46 & 1595.18 \\
    \midrule 
    \multirow{3}{*}{6} & \multirow{2}{*}{C} & Inward  &  15 &  3.88 &  0.02 &  3.09 & 0.11 & 0.52 &   63.30 &    9.93 & 1547.70 \\
                       &                    & Outward & 248 & 66.31 &  0.01 & 10.23 & 0.13 & 0.71 &   52.20 &    4.95 & 1831.85 \\
                       & B/U                & Equal   & 111 & 28.68 & -0.01 &  6.17 & 0.12 & 0.57 &  -31.71 &   30.72 & 1741.60 \\
    \midrule 
    \multirow{3}{*}{5} & \multirow{2}{*}{C} & Inward  &  38 & 12.88 & -0.23 &  3.72 & 0.12 & 0.59 &  -87.58 &  -23.39 & 1689.26 \\
                       &                    & Outward & 163 & 55.25 & -0.18 & 11.75 & 0.14 & 0.71 &   74.99 &   20.21 & 1837.53 \\
                       & B/U                & Equal   &  94 & 31.86 & -0.17 &  6.76 & 0.12 & 0.61 &   35.58 &   -9.93 & 1850.79 \\
    \midrule                       
    \multirow{3}{*}{4} & \multirow{2}{*}{C} & Inward  &  43 & 31.62 & -0.46 &  5.78 & 0.14 & 1.00 &  -87.70 &   50.58 & 1828.45 \\
                       &                    & Outward &  70 & 51.47 & -0.34 & 12.88 & 0.14 & 0.67 &   19.07 &   17.86 & 1879.33 \\
                       & B/U                & Equal   &  23 & 16.91 & -0.45 &  8.51 & 0.17 & 1.34 &  158.55 &  -44.37 & 1888.64 \\
    \midrule 
    \multirow{3}{*}{3} & \multirow{2}{*}{C} & Inward  &   9 & 42.86 & -0.66 &  7.08 & 0.13 & 1.25 &  -56.03 &  -88.74 &  2029.65 \\
                       &                    & Outward &   2 &  9.52 & -0.70 & 13.49 & 0.13 & 2.06 &  229.44 &   66.17 &  1826.74 \\
                       & B/U                & Equal   &  10 & 47.62 & -0.65 &  9.55 & 0.19 & 1.25 &   52.70 &   86.62 &  2042.02 \\
    \midrule
    \multirow{3}{*}{\large Total}    
                       & \multirow{2}{*}{C} & Inward  & 108 &  9.09 & -0.23 &  3.72 & 0.13 & 0.59 &  -56.03 &  -11.62 & 1689.26 \\
                       &                    & Outward & 784 & 65.99 & -0.08 & 10.99 & 0.14 & 0.69 &   54.12 &   19.07 & 1829.29 \\ 
                       & B/U                & Equal   & 296 & 24.92 & -0.09 &  6.61 & 0.15 & 0.65 &   46.24 &    1.27 & 1796.19 \\
\bottomrule
\end{tabular}
\tablefoot{This table presents key parameters for stars in different HC groups, categorised by their movement: churning (C) or blurring/undisturbed motion (B/U). Stars with nearly unchanged orbital radii (\rgui = \rbirth~within a 2$\sigma_{\rm{tot}}$ threshold) are considered blurred or undisturbed (B/U). Stars with divergent orbital radii (\rgui~and \rbirth) beyond the same threshold are considered churned (C). Churned stars are further subdivided into those moving inward (\rgui < \rbirth) and outward (\rgui > \rbirth). For each group, the table shows the number of stars ($N_{\star}$), the percentage of stars within the group (\%), median metallicities ($\langle \rm{\feh} \rangle$), median ages ($\overline{t}_{\star}$), median eccentricities (\eccentricity), median maximum Galactic scale-heights (\zmax), and median angular momentum in the $r$, $\phi$, and $z$ directions (\lr, \lphi, \lz).}

\label{tab:churning_blurring}
\end{table*}

\renewcommand{\arraystretch}{1.}

\begin{figure}
    \centering
    \includegraphics[width=\linewidth]{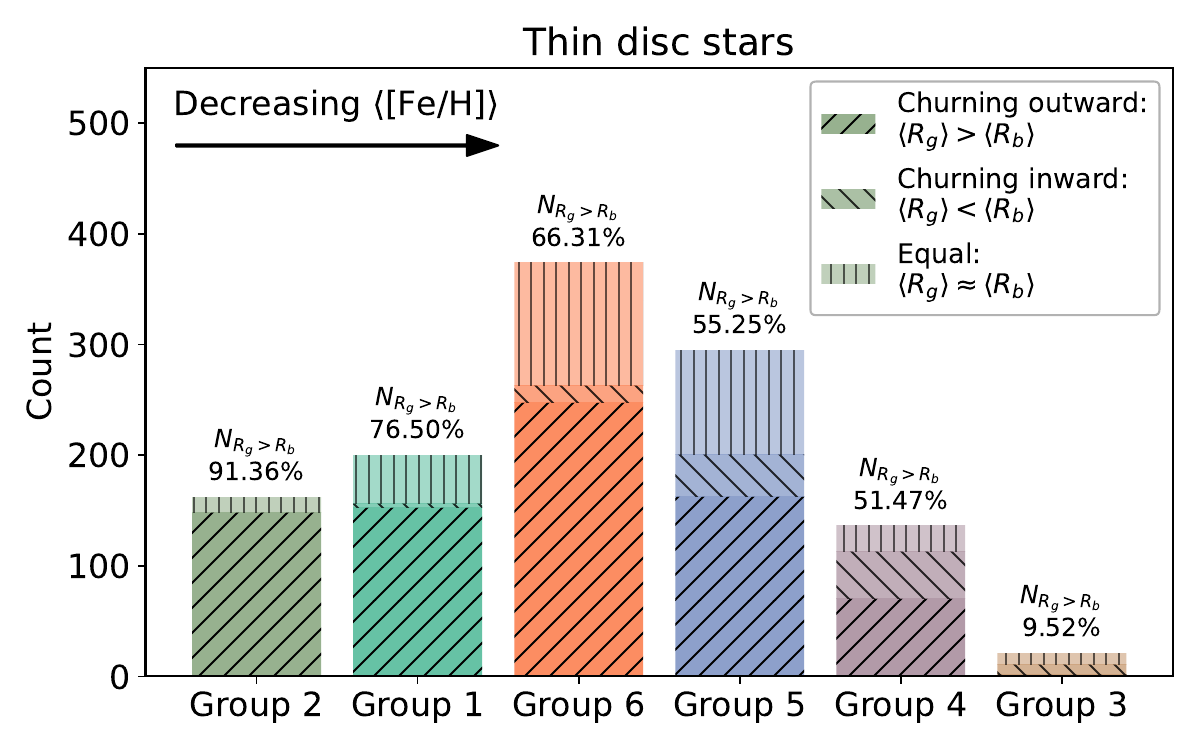}
    \caption{Stacked barplot showing the distribution of stars based on the comparison between their current median guiding radius (\rgui) and birth radius (\rbirth) across various stellar groups, arranged in order of decreasing $\langle \rm{\feh} \rangle$. The lower segments of the bars, shown in darker shades with right-slanted hatching, represent stars with \rgui~greater than \rbirth. The middle segments, in medium shades with left-slanted hatching, indicate stars where \rbirth~is less than \rgui. The top segments, in the lightest shades with vertical hatching, represent stars for which \rgui~and \rbirth~are consistent within the adopted 2$\sigma_{\rm tot}$ threshold. Each group is uniquely coloured according to the scheme used consistently throughout the paper for clarity. Additionally, the percentage of stars with \rgui~$>$~\rbirth~is displayed at the top of each bar.}
    \label{fig:barplot_rg_rb_ges}
\end{figure}

Table \ref{tab:churning_blurring} provides a comprehensive summary of our disc stars, quantifying the effects of the two types of motion according to the criterion described above, marked as those dominated by either churning (C) or a combination of blurring and undisturbed motion (B/U). The stars classified as C are further subdivided into those moving inward or outward. We further depict the proportion of stars that have \rgui~above or below the one-to-one line in Fig. \ref{fig:rb_rg_groups} with a stacked barplot with these quantities in Fig. \ref{fig:barplot_rg_rb_ges}.

The percentages of stars that are either undisturbed/blurred or churned (Table \ref{tab:churning_blurring}) vary across the different metallicity-classified groups. In the most metal-rich group (Group 2), over 90\% of stars exhibit outward radial migration, whereas only about 15\% of stars in Group 3 (the most metal-poor) display this behaviour. The opposite trend is observed for inward radial migration, which is absent in Group 2 (0\%) but plays a significant role in Group 3 ($\sim$ 43\%). This gradual shift in trends follows the decrease in median metallicity across the groups. The fraction of stars without significant changes in orbital radii generally increases as $\langle {\rm \feh} \rangle$ decreases, except for Group 4, which shows a sudden drop in this percentage. Overall, roughly 3/4 of our sample appears to have experienced either inward or outward churning, with the remaining 1/4 being blurred or undisturbed. 

Our findings differ somewhat from those reported by \citet{Feltzing2020}, who found that $\sim$ 50\% of their sample had undergone churning, while $\sim$ 10\% had experienced blurring. They also report that about 5-7\% of stars have not been subject to either churning or blurring. The discrepancies between our results and theirs are likely attributable to methodological differences in how churning and blurring are defined and identified. In this study, we classify stars as churned if their current \rgui\ significantly deviates from our estimate of \rbirth, indicating substantial radial migration. Blurred or undisturbed stars, by definition, are those that have somewhat preserved their initial angular momenta, which should be reflected in relatively unchanged orbital radii (at least in theory; represented by \rgui)\footnote{It is worth recalling that angular momentum is defined as $\mathbf{L} = \mathbf{r} \times \mathbf{p}$, where $\mathbf{r}$ in cylindrical coordinates is $\mathbf{r} = (r \cos \phi, r \sin \phi, z)$, with $r$ representing the orbital radius, and $\mathbf{p}=m \cdot \mathbf{v}$, where $m$ is the stellar mass and $\mathbf{v}$ is the velocity vector. Therefore, for blurred/undisturbed stars, it is reasonable to assume that an unchanged $r$ (or \rgui\ when compared to \rbirth) throughout their orbital histories indicates preserved angular momentum. We do acknowledge, however, that a combination of changes in $r$, $m$, and/or $\mathbf{v}$ could result in a scenario where the angular momentum remains constant despite variations in these parameters. Nonetheless, assessing such a scenario is challenging given the complexity of tracking multiple variables to maintain a stable angular momentum.}, irrespective of other dynamical parameters (as discussed in Sect. \ref{subsubsec:dynamics_churn_blur_ang_mom}). This contrasts with the approach of \citet{Feltzing2020}, who incorporate additional dynamical parameters, such as eccentricity, in their classification.

Indeed, as highlighted in the first paragraph of Sect. \ref{subsubsec:dynamics_churn_blur_overall} and shown in Table \ref{tab:churning_blurring}, \eccentricity\ does not significantly vary among the motion-stratified groups in our sample. By not incorporating these additional parameters, we avoid introducing potential biases into the motion classification of our stars, adhering instead to the theoretical framework of blurred and undisturbed motion, which posits that stars retaining their angular momenta should largely maintain their original birth orbital radii and thus be unaffected by significant radial migration.

These differences in classification highlight the need for consistent and transparent criteria when assessing stellar migration processes. Despite the methodological divergences, the overall proportions between our study and that of \citet{Feltzing2020} remain broadly comparable, suggesting that while the methods differ, the overarching dynamics of radial migration are similarly captured.

In terms of age, the observed trends across the motion-stratified groups are impressive. The age gap between inward and outward churning groups is highly significant, spanning several Gyr. Churned stars moving outward are generally much older ($\overline{t}_{\star} \simeq$ 11.0 Gyr) than those moving inward ($\overline{t}_{\star} \simeq$ 3.7 Gyr). Both $\overline{t}_{\star}$ values tend to increase with decreasing $\langle {\rm \feh} \rangle$, as anticipated by chemical evolution models. Blurred/undisturbed stars have intermediate ages between inward and outward churned stars, with $\overline{t}_{\star}$=6.61 Gyr, also increasing with diminishing $\langle {\rm \feh} \rangle$. These temporal differences suggest that certain Galactic structures require time to influence the movement of these stars significantly. Additionally, since these differences gradually change with varying metallicities and larger \rbirth, these Galactic structures could differentially influence each group of stars.

It is also important to note that, since our sample predominantly consists of old stars, as discussed in the previous paragraph, it is reasonable to assume that most -- if not all -- of the stars classified as blurred/undisturbed may have experienced blurring over time due to prolonged interactions with Galactic structures. This assumption is consistent with the findings of \citet{Feltzing2020}, who report that the fraction of undisturbed stars is small (a maximum of 7\%) and decreases with increasing age (see their Table 2).

Several previous studies have explored the phenomenon of stellar radial migration in the MW \citep[see e.g.][and references therein]{Lepine2003, Minchev2011, Trevisan2011, Chen2019, Wozniak2020, Lian2022, Lu2022, Iles2024, Nepal2024}, and also in other spiral galaxies \citep[e.g.][]{SanchezBlazquez2014}. At first glance, one might question whether the drivers of stellar migration observed here are the same across each HC metallicity-stratified group. It is reasonable to assume that different mechanisms may be at play \citep[e.g.][]{Martinez-Medina2016}; in fact both the bar and spiral arms seem to be able to cause radial migration \citep[e.g.][]{SellwoodBinney2002, Loebman2011, Buck2020, Tsujimoto2020}. The most metal-rich stars, formed in the inner Galaxy, could have been influenced by both the central bar and transient spiral arms \citep[see e.g.][for the effects of the bar on stellar radial migration]{Khoperskov2020a, Iles2024}. The generally old ages of the stars in our sample align with the estimated formation epoch of the bar, around redshift 2 \citep[approximately 10 Gyr ago;][]{Haywood2024, Khoperskov2024}. However, it remains uncertain whether the bar exerted a similarly strong influence on stars formed in the outer disc, where transient spiral arms may have played a more significant role in pulling stars towards the inner regions of the Galaxy \citep[see e.g.][for the influence of spiral arms]{SellwoodBinney2002, Roskar2012, VC2014, Martinez-Medina2016}. We aim at deepening the discussion on the drivers of both outward and inward churning as well as blurring in future works.

The catalogue of our sample, including the relevant dynamical properties of the stars and the derived \rbirth\ values computed using our method and the motion classification is available through the CDS.

% -----------------
% -----------------
\subsubsection{Overview of dynamical properties of churned and blurred/undisturbed stellar populations}
\label{subsubsec:dynamics_churn_blur_overall}

In Sect. \ref{subsubsec:other_signs_migration}, we examined the general characteristics of the stars in our sample that support the thesis of radial migration, focusing on parameters such as \zmax\ and \eccentricity. That section provided a broad overview without distinguishing between stars based on their movement categories, such as blurred/undisturbed or inward/outward churned. To refine this analysis, we stratified these features by motion subgroups within the broader HC groups, as shown in Table \ref{tab:churning_blurring}. Our findings indicate that while \eccentricity~is slightly higher for stars orbiting at smaller \rgui\ (see right panel of Fig. \ref{fig:cmap}), it does not vary significantly across the motion-stratified subgroups, with all displaying low median values below 0.2, suggesting generally near-circular orbits. In contrast, \zmax\ typically increases as metallicity decreases. However, no strong trends emerge when comparing different types of motion within each HC group. Nonetheless, when examining the entire sample (see the last three rows of Table \ref{tab:churning_blurring}), outward-churned stars tend to have a higher \zmax, although this increase is not significantly greater than that observed for blurred/undisturbed or inward-churned stars.

% -----------------
% -----------------
\subsubsection{Angular momentum features of motion-stratified stellar populations} 
\label{subsubsec:dynamics_churn_blur_ang_mom}

\begin{figure*}
    \centering
    \includegraphics[width=\linewidth]{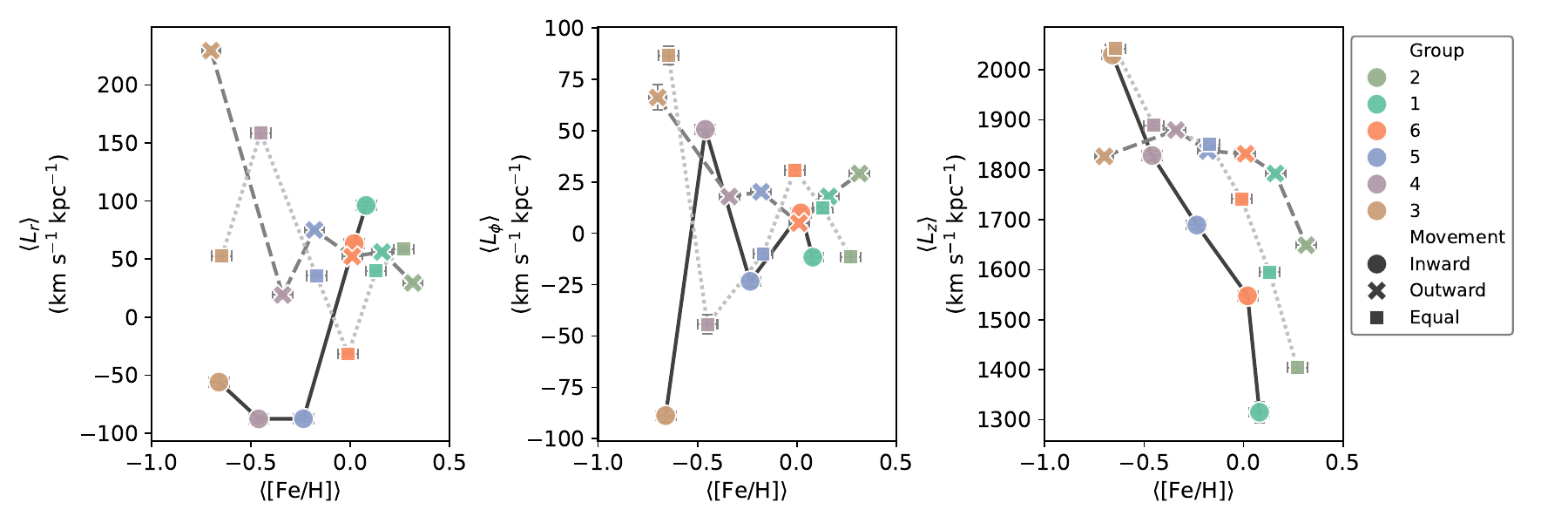}
    \caption{Median angular momenta in the $r$, $\phi$, and $z$ directions -- \lr, \lphi, \lz~-- plotted against the median metallicity, $\langle \rm{[Fe/H]} \rangle$. Each panel presents the angular momentum in a specific direction. Marker colours represent distinct HC groups, consistent with the other figures in this paper, while marker shapes indicate different stellar movements: inward churning, outward churning, and blurring (labelled as `equal'). The lines serve as visual guides rather than regressions, highlighting potential trends: the solid line connects HC groups with inward churning, the dashed line corresponds to outward churning, and the dotted line represents stars with unchanged radii (undisturbed or blurred).}
    \label{fig:ang_mom_move_cyl}
\end{figure*}

Having reviewed these key dynamical parameters, we now turn to the angular momentum ($L$) of our stars, presented in the final three columns of Table \ref{tab:churning_blurring} and illustrated in Fig. \ref{fig:ang_mom_move_cyl}. Initially, we assessed the $L$ components using the Cartesian system ($xyz$) provided by \textsc{galpy}, but as it is less intuitive in this context, we proceeded with the cylindrical coordinate system. For details on our results using the Cartesian system and the transformation process, refer to Appendix \ref{appendix_subsec:ang_moment}.

In Fig. \ref{fig:ang_mom_move_cyl}, we display \lr, \lphi, and \lz\ against $\langle \rm{\feh} \rangle$, \feh\ being a rough proxy for \rbirth\ as illustrated by the radial metallicity gradients shown in this paper. The colours in this figure align with those used throughout this paper, representing the different HC metallicity-stratified groups. Marker shapes indicate the type of movement (inward/outward-churned; and blurred/undisturbed, tagged as `equal'), with distinct lines connecting markers of the same movement to guide the eye towards potential trends (without performing a formal regression analysis).

Figure \ref{fig:ang_mom_move_cyl} highlights the distinct trends in angular momenta across the subgroups stratified by motion. In the left panel, two key patterns emerge for \lr: outward-churned stars and blurred/undisturbed stars do not show a clear trend. Inward-churned stars, on the other hand, show a predominantly increasing trend, except for the most metal-poor HC group. However, as \lr~is not an integral of motion, we refrain from attempting to interpret this trend.

For \lphi, no clear trend is observed, while \lz\ consistently decreases with decreasing $\langle \rm{[Fe/H]} \rangle$ across all motion classes, except in the most metal-poor outward-churned HC groups. These groups (3 and 4) stand out among all motion-stratified subgroups in \lr, \lphi, and \lz, possibly reflecting effects of potential residual interlopers from the thick disc; yet, if indeed these groups are (nearly) free from these intruders, this behaviour can be explained by the disturbance caused by the pericentric passage of Sagittarius in the outer disc, as previously discussed.

The \lz\ trend -- where angular momentum increases with decreasing metallicity -- appears independent of motion type. The trend observed in \lz\ can be interpreted as another view of the radial metallicity gradients for the MW, since we have shown that indeed each motion-stratified group has different $\overline{t}_{\star}$\footnote{Since $L_z = mrv_{\phi}$, $L_z$ is indeed mapping the variations and changes in the Galactic disc.}. Outward-churned HC groups generally exhibit higher $L_z$ compared to blurred/undisturbed and inward-churned counterparts. Notably, within each metallicity-stratified group, outward-churned stars are systematically the oldest (see Table \ref{tab:churning_blurring}). Conversely, inward-churned stars, the youngest in each HC group, consistently have lower $L_z$ than blurred/undisturbed stars, which have intermediate ages. Another interpretation of \lz\ is that that the different motion-stratified groups may interact with the Galaxy's structures differently, gaining or losing momentum. Whether these ages reflect the timescales necessary to produce inward- or outward-churned stars remains unclear, but some studies indicate that they might \citep{Lian2022}. Nonetheless, these observations support the idea that angular momentum components, especially $L_z$, can be considered important tracers of the chemical enrichment of the MW, as well as stellar migration processes and Galactic structure interactions.

% -----------------
% -----------------
\subsubsection{Action components of motion-stratified stellar populations} 
\label{subsubsec:dynamics_churn_blur_action}

\begin{figure*}
    \centering
    \includegraphics[width=\linewidth]{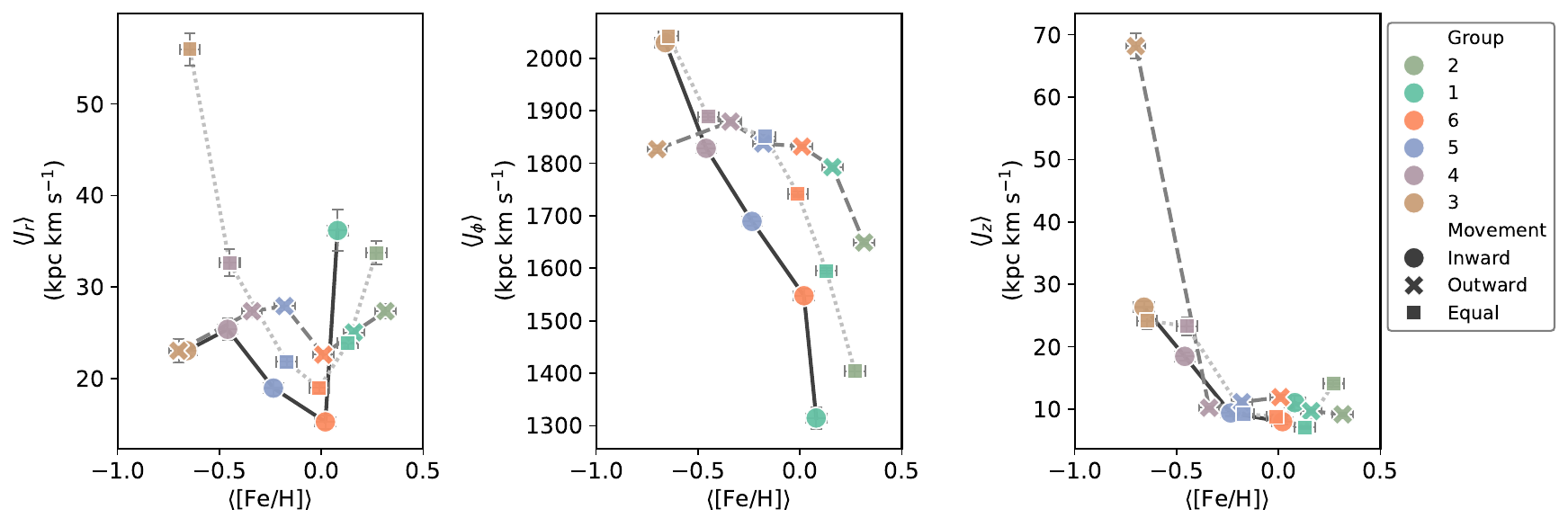}
    \caption{Same as Fig. \ref{fig:ang_mom_move_cyl} but for the action components (\jr, \jp, \jz), instead of angular momentum.}
    \label{fig:action_comp_feh}
\end{figure*}

Action components ($J_r$, $J_{\phi}$, $J_z$), are derived from the integral of angular momentum across an orbit. They provide scalar quantities that summarise radial, azimuthal, and vertical motions, which improves interpretability compared to the vector nature of angular momentum. By capturing motion as distinct scalar values, action variables enable a somewhat more intuitive understanding of orbital characteristics in terms of energy and distance, without the complexity of vector decomposition. Therefore, we expand our analysis by illustrating the median action components in relation to $\langle \rm{\feh} \rangle$ in Fig. \ref{fig:action_comp_feh}, similarly to Fig \ref{fig:ang_mom_move_cyl}.

The radial action, $J_r$, characterises motion occurring within the Galactic plane. For stars in near-circular orbits, a low $J_r$ value indicates minimal radial motion, implying that these stars experience limited fluctuations in their \rgui\ over time. Conversely, for stars with higher eccentricities, $J_r$ quantifies the energy associated with radial motion, reflecting the extent of their radial oscillations. This distinction allows $J_r$ to provide insights into both the stability and the dynamics of stellar orbits in the Galactic plane. Considering this, it is noticeable that Group 6 (depicted in orange; characterised by metallicities near solar levels as shown in Fig. \ref{fig:hc_groups}), exhibits the lowest \jr\ across all motion classes. 

This is an interesting finding, since Group 6 does not have the largest proportion of stars that retain their orbital radii over time; in fact, it is the third highest in this respect ($\sim 29\%$ compared to $\sim 32\%$ for Group 5, the next group in order of decreasing metallicity; and $\sim 48\%$ for Group 3, the most metal-poor group). It may seem a contradiction that the group with lowest \jr\ is not the same as that with the largest proportion of stars that did not suffer radial migration. However, the proportion of stars moving inward, outward, or retaining their radii provides different insights from those captured by \jr, which is a time-integrated quantity.

In fact, \jr\ measures the stability of the stars’ orbits over time, which suggests that the stars in Group 6 may have reached a phase in their orbital evolution where their \rgui\ became relatively stable earlier than those in other groups. A potential explanation for this phenomenon is that the stars in Group 6 (but also those in neighbouring groups, i.e. 1 and 5) were formed closer to the outer Lindblad resonance (OLR) radius, located at approximately 7.2 kpc \citep[see for instance the pioneer work by][]{Dehnen2000}, but see also other estimates on the OLR, such as \citealt{Khoperskov2020a}, with an estimate of $\sim$ 7.4 kpc.

Simulations of MW analogues show that the OLR plays a significant role in stabilising the orbits of stars that reside near it because it is a location where the stars' orbital period matches the pattern speed of the Galactic spiral arms. This resonance reduces the stars' radial motion, meaning they experience less fluctuation in their $R_{\rm g}$ over time. As a result, stars near the OLR are less prone to large radial oscillations, which helps maintain a stable $R_{\rm g}$ \citep[see e.g.][]{Halle2015, Khoperskov2020a}. This proximity could explain why their orbits stabilised earlier, as they may have reached their current \rgui\ more quickly, allowing them to settle into more stable orbits sooner. To probe this hypothesis, we show the \rbirth\ against \rgui\ in Fig. \ref{fig:motion_rb_rg}, where it is noticeable that Groups 6's motion-classified subgroups (in orange) were collectively born closer to the OLR than the other groups.

In contrast, when we assess the other groups, we observe a general increase in \jr\ associated with both increasing and decreasing $\langle \rm{\feh} \rangle$, suggesting that stars in these groups have experienced more substantial and progressive movement over time, which makes sense since they were indeed observed in the solar vicinity, where the local metallicity differs from theirs. These findings are consistent with our earlier discussions.

Turning to $J_{\phi}$, this azimuthal action quantifies angular motion around the Galactic centre. It can be interpreted as a measure of the angular momentum possessed by stars, particularly in circular orbits, which applies to our sample. The trends we observe for \jp\ align with \lz, illustrated in Fig. \ref{fig:ang_mom_move_cyl}, which we have discussed in detail in previous paragraphs. Since our adopted potential for the MW is the one described by \citet[][i.e. does not change in time]{McMillan2017}, these results are expected by definition.

The vertical action, $J_z$, is a measure of oscillations and deviations from the Galactic plane, thus providing insight into the vertical distribution of stellar populations. Our analysis shows that \jz\ is relatively stable across all motion classes, maintaining values around 10 kpc km s$^{-1}$, except within the two most metal-poor groups (Groups 3 and 4). These groups display marked vertical disturbance, with outward-churned stars in Group 3 reaching nearly 70 kpc km s$^{-1}$. This behaviour may suggest a vertical gradient that is more pronounced among metal-poor groups, potentially (but not surely) attributable to residual thick disc interlopers or the influence of the Sagittarius dwarf galaxy during its pericentric passage $\sim$ 6 Gyr ago.

% -----------------
% -----------------
\subsubsection{\rbirth, \rgui, and the OLR} 
\label{subsubsec:rb_rg_olr}

\begin{figure}
    \centering
    \includegraphics[width=\linewidth]{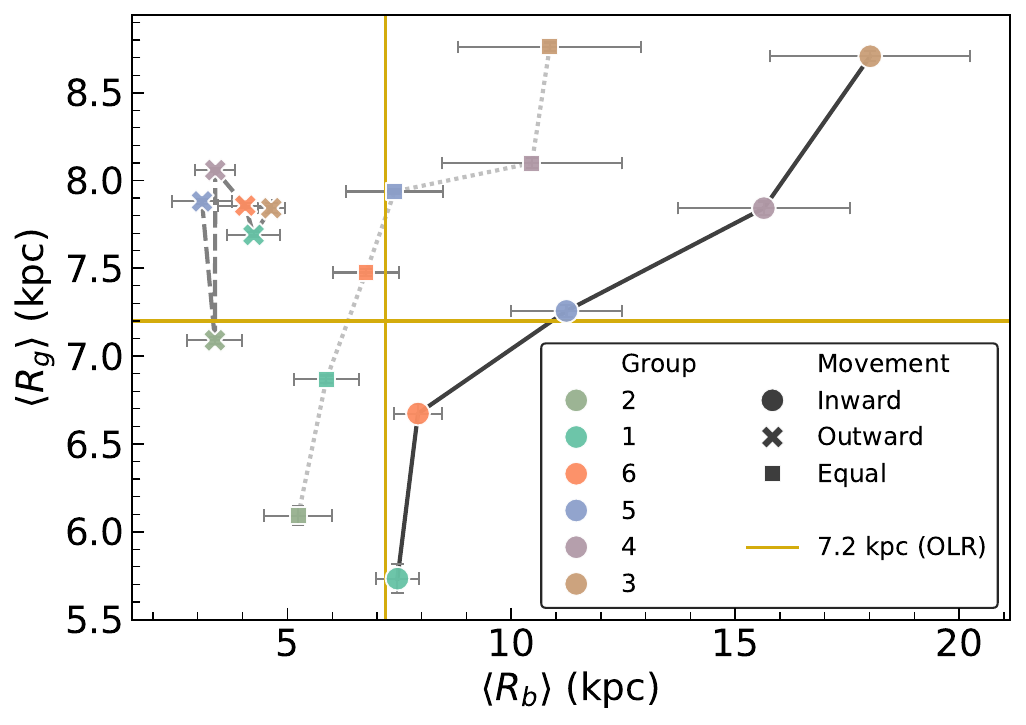}
    \caption{\rbirth\ against \rgui\ for all the metallicity- and motion-stratified groups of our sample. The general features of this plot are consistent with those observed in Figs. \ref{fig:ang_mom_move_cyl} and \ref{fig:action_comp_feh}. The expected outer OLR radius \citep[estimated by][]{Dehnen2000} is marked in yellow.}
    \label{fig:motion_rb_rg}
\end{figure}

Lastly, Fig. \ref{fig:motion_rb_rg}, which we used to investigate the potential reasons behind Group 6's low \jr, also reveals an intriguing pattern across our metallicity- and motion-stratified groups. Specifically, stars moving outward were born in the innermost regions of the MW, whereas stars moving inward show the opposite trend. Groups that retain their radii consistently occupy a position between these two extremes. These findings are independent of metallicity and agree with the gradients observed in \lz\ and \jp. This pattern could suggest that the Galactic factors that cause outward and inward churning are distinct.

Building on the discussion developed right before the beginning of Sect. \ref{subsubsec:dynamics_churn_blur_overall}, the bar structure likely influences stars formed in the inner regions of the MW, whereas stars moving inward may be more sensitive to the transient spiral arms than to the bar. Although there is ongoing debate regarding the length of the MW’s bar and its components \citep[such as the so-called long-bar;][]{LopezCorredoira2007, Wegg2013, Portail2017, Lucey2023}, recent estimates place its length at $\approx$ 3.5 kpc \citep{Lucey2023}, which could constrain the range of the bar's influence, particularly in its proximity. These estimates agree with predictions from \citet{Khoperskov2020a}, which indicate that the bar—and more specifically, its deceleration—contributes to the churning of stars from the inner Galaxy outward towards the OLR.

It is important to note that different models of the MW, especially those concerning its central bar, can shift the position of the OLR \citep[see the discussion in][]{Fragkoudi2019}, explaining why many stars migrate further from the 7.2 kpc estimate of the OLR; not to mention that it is unclear if and how the OLR shifts its position with time. This brings about a much longer discussion, which is beyond the scope of the current paper.

% -----------------
\subsection{Determining the probable birth radius of the Sun} \label{subsec:rbirth_sun}

Numerous studies have provided estimates for the Sun's $R_{\rm b}$, with values varying depending on the methods employed. These estimates range from 7.3 kpc \citep{Minchev2018} to 5.2 kpc \citep{Frankel2018}, and more recently, as low as 4.5 kpc \citep{Lu2024}, among others. \citet{Minchev2013} predicted a range of 4.4-7.7 kpc, which encompasses many of the aforementioned estimates.

Here, we investigate what our GAM estimates as the most probable location of the Sun's $R_{\rm b}$. As input parameters, we adopted the Sun's age, $t_{\odot} = 4.775 \pm 0.039$ Gyr, as provided by \citet{BonannoFrohlich2015}. Specifically, we selected the final estimate from Table 3 of \citet{BonannoFrohlich2015}, which the authors regard as the most reliable. This estimate was chosen based on the highest Bayes factor, derived using the \texttt{Irwin+AdelR+Asplund} model. For the Sun's metallicity, we followed the convention \feh=0 but allowed for an error margin of 0.03 dex, as suggested by the non-local thermal equilibrium estimate presented by \citet{Korn2003} and further discussed in \citet{Asplund2009}.

\begin{figure}
    \centering
    \includegraphics[width=\linewidth]{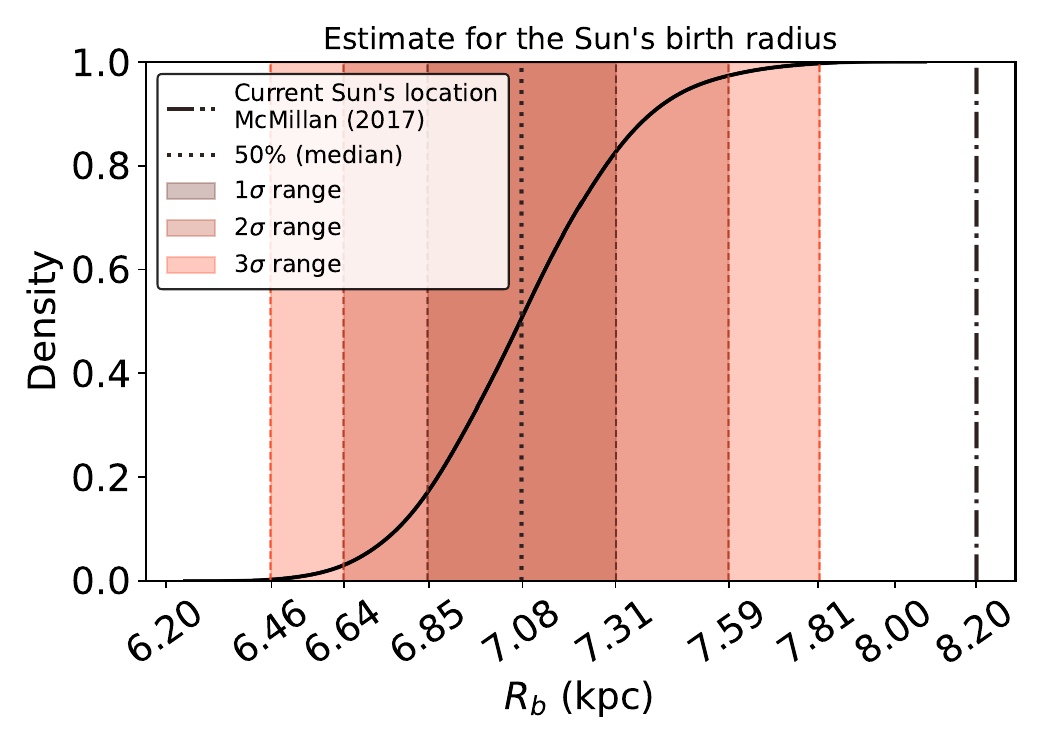}
    \caption{Cumulative distribution function depicting the estimated birth radius of the Sun according to our GAM. The plot illustrates the probability distributions for the 1, 2, and 3$\sigma$ confidence intervals, with different shades of brown and orange used to distinguish these intervals. The values are marked on the tick labels of the $x$-axis. The mean and median values are $7.08 \pm 0.24$ kpc. The current orbital radius of the Sun is shown represented by the dot-dashed vertical line \citep[i.e. 8.20 kpc; see][]{McMillan2017}.}
    \label{fig:sun_birth}
\end{figure}

Using these parameters, we applied a bootstrap resampling procedure (1000 iterations) as described in Sect. \ref{sec:data_method} of this manuscript, which was then fed into our GAM. Our analysis yields a most probable $R_{\rm b}$ for the Sun of $7.08 \pm 0.24$ kpc from the Galactic centre. Additionally, we provide the 1, 2, and 3$\sigma$ ranges for this estimate. Considering the 3$\sigma$ range, we find that the Sun's $R_{\rm b}$ lies between 6.46 and 7.81 kpc. These results are in agreement with the estimates presented earlier in this section and provide further evidence that the Sun likely originated at a smaller distance from the Galactic centre and has since churned outward, in agreement with the motion of other stars in Group 6 (the ones most similar to the Sun). Furthermore, the results demonstrate that the chemical enrichment models of \citet{Magrini2009}, when integrated with the GAM, can recover a similar range for the Sun's $R_{\rm b}$ within very reasonable error margins. The results are shown in Fig. \ref{fig:sun_birth}.

%-------------------------------------------------------------------
\section{Summary and conclusions}  \label{sec:conclusions}

In this study, we have harnessed the power of a GAM to refine the spatial and chemical grids of established chemical enrichment models for the MW \citep{Magrini2009}. Moreover, this GAM framework was adeptly employed to estimate the birth radii of a select group of \textit{Gaia}-ESO stars encompassing a diverse range of metallicities spanning from metal-poor to super-metal-rich previously classified into six metallicity-stratified groups through HC. Detailed elucidation of our data curation and processing methodologies can be found in the comprehensive exposition by \cite{Dantas2023}. Our overarching aim revolved around a quantitative and qualitative exploration of the dynamic effects manifesting within these stars, notably the phenomena of blurring and churning (the latter is also known as radial migration). Our conclusions are as follows:

\begin{enumerate}
    \item The GAM has been proven to be a useful method for directly deriving the Galactocentric birth radii, \rbirth, of the stars in our sample by employing a minimalist approach that uses only \feh~and the age of the Universe at the time of the stars' birth ($t$). This simplicity was intentional, as introducing other available abundances -- such as $\alpha$-elements, which are highly correlated with one another and with \feh\ -- would not necessarily improve GAM's performance. Instead, GAM's effectiveness is more likely to improve if the underlying chemical enrichment models incorporate a broader range of independent information, such as neutron-capture element abundances, or better reflect the complexities of the MW's evolutionary history, including multiple gas infall events. Additionally, models that provide a richer dataset with larger age and metallicity bins as well as smoother radial gradients would allow for more precise predictions. Conversely, including degenerate information does not enhance the performance of the GAM, further justifying our choice of a minimalist approach in this study.
    
    \item We estimated the expected \rbirth\ for the stars of our sample using chemical evolution models by \citet{Magrini2009}, which were used to run a GAM to expand the reach of the original theoretical models. By comparing their \rbirth\ with current guiding radii (\rgui), we classified the stars of our sample according to their motion: inward churning, outward churning, and equal motion (undisturbed or blurred). As a consequence, we unveiled a significant migration pattern within the Galaxy:

    \begin{enumerate}[a.]
        \item We identified that about three-fourths of our sample have sustained churning, whereas the remaining roughly one-fourth is either undisturbed or subject to blurring.
       
        \item The stars in the most metal-rich HC group predominantly formed in the inner regions of the Galaxy and subsequently migrated outward to larger Galactocentric radii (over 90\%).
    
        \item The most metal-poor stars, in contrast, appear to have originated at larger Galactocentric distances and migrated inward towards the Galaxy's central regions (approximately 43\%), while the largest portion seem to have retained their \rbirth~(around 48\%). It is important to note, however, that this group may still contain a small number of thick disc interlopers with metallicities that overlap with the thin disc. Despite this possible contamination, the general patterns in their motion remain consistent within this subset.
    
        \item Stars with intermediate metallicities exhibit a transitional behaviour, with some migrating outward and others inward. In other words, as median metallicities decrease throughout our metallicity-stratified groups, the proportion of stars migrating outward decreases, and the number of those migrating inward increases. Additionally, the number of stars with orbits consistent with their \rbirth\ also rises. This smooth transition is consistent throughout the groups. 
        
        \item A significant portion of stars with intermediate metallicities remain close to their \rbirth, suggesting a relatively stable positional relationship with their formation sites. This aligns with the expectation that stars with slightly lower metallicities than the Sun were likely formed near the Sun's current Galactocentric distance. Since we are observing stars in the Sun's vicinity, it is natural to find stars with these features in this region.

    \end{enumerate}

    \item There is a significant age gap between stars influenced by inward and outward churning. For instance, in Group 1 (the second most metal-rich group), this gap is approximately 5 Gyr, and it tends to persist or widen in other metallicity-stratified groups. Affected stars that have kept their orbital radii, on the other hand, exhibit intermediate ages. This age pattern remains consistent across all HC groups.

    \item Due to the overall old ages of our blurred/undisturbed group, we doubt that most of these stars are undisturbed since we believe that time is an important factor that dictates the interactions between the stars and the Galactic structures. In other words, we reckon that most stars seem to have undergone blurring.

    \item Metal-rich stars, which are usually formed in the inner regions of the Galaxy, seem to take less time to suffer from inward and outward churning. Despite the age gap, there's a clear trend across all HC groups showing that the ages of the stars suffering from both inward and outward churning increase, but the gap between them remains. This age rise with decreasing metallicity also happens with blurred/undisturbed stars.

    \item Although eccentricity is expected to play a relevant role for churned stars, we did not see any strong relation between eccentricity and type of movement (we do see that stars orbiting at lower \rgui\ have slightly larger eccentricities). The eccentricity for all the metallicity- and motion-stratified groups of stars in the sample is somewhat low, with median values ranging from 0.13 to 0.15. There is no observable trend across our HC groups.

    \item The angular momentum ($L$) analysis results provide valuable insights into the processes that shape stellar motion within our sample. Notably, \lz\ revealed distinct trends, contrasting with the lack of significant patterns in \lr\ and \lphi. Across the sample, \lz\ consistently increases as metallicity decreases, irrespective of stellar motion type. This trend may reflect the chemical enrichment history of the MW, as $L_z$ traces patterns within the Galactic disc. Further, age differences between motion-stratified groups strengthen this interpretation. Outward-churned stars, for instance, tend to have higher $L_z$ and are systematically older within each metallicity-defined group than their blurred/undisturbed and inward-churned counterparts. Inward-churned stars, conversely, are consistently the youngest, exhibiting lower $L_z$ than their blurred/undisturbed peers, who hold intermediate positions in age and $L_z$. Another perspective (or interpretation) on the behaviour of $L_z$ considers the role of Galactic structures: outward churning and elevated \lz\ might result from interactions with these structures, whereas inward-churned stars, being younger, appear to `lose' \lz\ as they move inward. These findings underscore the possible influence of Galactic structures on stellar motion types, particularly concerning $L_z$. However, the degree to which these ages align with the timescales required for inward or outward churning remains an open question.

    \item The results observed in our analysis of the action components -- \jr, \jp, \jz\ -- corroborate findings from other parameters and reinforce the interpretations discussed. Specifically, \jr, which quantifies radial orbital excursions as an integrated quantity, provides key insights into radial orbital stability. Our analysis suggests that stars in Group 6 may have stabilised their orbits earlier. A reasonable hypothesis is their proximity at birth to the OLR radius. This resonance, known to influence orbital dynamics, could have accelerated the damping of radial oscillations, allowing these stars to achieve orbital stability earlier compared to stars in the other groups. The behaviour of \jp\ aligns with \lz, as anticipated. Lastly, \jz\ remains generally constant across all metallicity-stratified groups, except for the two most metal-poor ones, which show higher vertical action. This may reflect either the remaining interlopers from the thick disc or perturbations due to the pericentric passage of Sagittarius in the outer disc.

    \item Finally, some results regarding the last two most metal-poor groups must be considered cautiously due to the interlopers that may potentially remain in the thick disc, as they may not have been completely removed. However, if all of these stars indeed belong to the Galactic thin disc, we identify at least three potential indicators that may suggest the influence of the pericentric passage of Sagittarius (which happened around 6 Gyr ago) on the outer disc. Group 4 and, especially, Group 3 consist of stars that do the following:

    \begin{enumerate}[a.]
        \item The stars reach significant heights above the Galactic plane, which suggests a potential influence from the pericentric passage of Sagittarius; however, it is important to acknowledge that outward churning could also contribute to this feature, meaning this vertical displacement may be from a combination of both mechanisms.

        \item The stars exhibit angular momentum and action components (notably \jz) that differ markedly from other metallicity-stratified groups.
        
        \item Groups 3 and 4 display a higher proportion of stars not associated with the disc, as evidenced by the preliminary chemo-dynamic analysis. Although we did not explore these outliers in detail in the main body of the paper, their increased frequency in Groups 3 and 4 may be an indication of a possible disturbance in the outer disc. Nevertheless, it is hard to differentiate them from the typical stars of the halo (usually metal-poor) at this stage.
    \end{enumerate}

    \item We conclude our study by providing estimates for the Sun's birth radius, with the most likely value being $7.08 \pm 0.24$ kpc, with a 3$\sigma$ range spanning from 6.46 to 7.81 kpc. These results are in agreement with previous studies, further validating the robustness of our methodology.
   
\end{enumerate}

Our findings provide important insights into the migration patterns of stars across the Galaxy, revealing how metallicity, age, angular momentum, and action trace the history of their motions. By using a streamlined GAM approach to extend chemical enrichment models, we provide the means to deepen understanding of the MW's dynamic structure and the complex forces influencing its stellar populations.

%-------------------------------------------------------------------
\section{Data availability} \label{sec:data_availability}
The catalogue is only available in electronic form at the CDS via anonymous ftp to \url{cdsarc.u-strasbg.fr} (130.79.128.5) or via \url{https://cdsarc.cds.unistra.fr/viz-bin/cat/J/A+A/696/A205}. Additional parameters can be provided upon reasonable request; please contact the corresponding author for such inquiries.

%-------------------------------------------------------------------
\begin{acknowledgements}

The authors thank the anonymous referee for the kind suggestions that helped to improve this work. M.~L.~L.~Dantas acknowledges the support of Agencia Nacional de Investigación y Desarrollo (ANID), Chile, through Fondecyt Postdoctorado grant 3240344. M.~L.~L.~Dantas and P.~B.~Tissera also acknowledge ANID Basal Project FB210003. M.~L.~L.~Dantas further appreciates the support from the National Science Centre, Poland, under project 2020/38/E/ST9/00395. R.~Smiljanic acknowledges funding from the National Science Centre, Poland, under project 2019/34/E/ST9/00133. R. S. de Souza acknowledges the support from the São Paulo Research Foundation under project 2024/05315-4. P.~B.~Tissera thanks Fondecyt Regular 2024/1240465. M.~L.~L.~Dantas and R.~Smiljanic thank Victor P. Debattista and L. Beraldo e Silva for their insightful discussions and invaluable suggestions, which greatly improved this work. M.~L.~L.~Dantas also acknowledges A.~R.~da Silva, R.~Giribaldi, H.~Perottoni, D.~de~Brito~Silva, and P.~Jofré for the valuable discussions. M.~L.~L.~Dantas expresses gratitude to M.~Bilicki and W.~Hellwing for the opportunity at the Center for Theoretical Physics, PAS. M.~L.~L.~Dantas is especially grateful to Miuchinha for her unwavering companionship, love, and support, despite the hardships of life.

This work made use of the following online platforms: \texttt{slack} (\url{https://slack.com/}), \texttt{github} (\url{https://github.com/}), and \texttt{overleaf} (\url{https://www.overleaf.com/}). Additionally, this work used the following \textsc{python} packages: \textsc{matplotlib} \citep{Hunter2007}, \textsc{numpy} \citep{Harris2020}, \textsc{pandas} \citep{mckinney-proc-scipy-2010}, and \textsc{astropy} \citep{Astropy2013, Astropy2018, Astropy2022}. We also benefited from \textsc{topcat} \citep{Taylor2005}. The main palette for the figures was created using \url{https://colorbrewer2.org} (credits: Cynthia Brewer, Mark Harrower, and The Pennsylvania State University), with special cases handled using the colour picker at \url{https://htmlcolorcodes.com/}.

Based on data products from observations made with ESO Telescopes at the La Silla Paranal Observatory under programme ID 188.B-3002. These data products have been processed by the Cambridge Astronomy Survey Unit (CASU) at the Institute of Astronomy, University of Cambridge, and by the FLAMES/UVES reduction team at INAF/Osservatorio Astrofisico di Arcetri. These data have been obtained from the \textit{Gaia}-ESO Survey Data Archive, prepared and hosted by the Wide Field Astronomy Unit, Institute for Astronomy, University of Edinburgh, which is funded by the UK Science and Technology Facilities Council. This work was partly supported by the European Union FP7 programme through ERC grant number 320360 and by the Leverhulme Trust through grant RPG-2012-541. We acknowledge the support from INAF and Ministero dell' Istruzione, dell' Universit\`a' e della Ricerca (MIUR) in the form of the grant "Premiale VLT 2012". 

The results presented here benefit from discussions held during the \textit{Gaia}-ESO workshops and conferences supported by the ESF (European Science Foundation) through the GREAT Research Network Programme. This publication makes use of data products from the Wide-field Infrared Survey Explorer, which is a joint project of the University of California, Los Angeles, and the Jet Propulsion Laboratory/California Institute of Technology, funded by the National Aeronautics and Space Administration. This work has made use of data from the European Space Agency (ESA) mission {\it Gaia} (\url{https://www.cosmos.esa.int/gaia}), processed by the {\it Gaia} Data Processing and Analysis Consortium (DPAC, \url{https://www.cosmos.esa.int/web/gaia/dpac/consortium}). Funding for the DPAC has been provided by national institutions, in particular the institutions participating in the {\it Gaia} Multilateral Agreement.
\end{acknowledgements}

%-------------------------------------------------------------------
\bibliographystyle{aa}        % style aa.bst
\bibliography{paper} 
%-------------------------------------------------------------------
%-------------------------------------------------------------------
\appendix
%-------------------------------------------------------------------
\section{Model using the \textsc{mgcv} package} \label{appendix:model}

\begin{figure}[h]
    \centering
    \includegraphics[width=\linewidth, trim={0 0 15cm 0}, clip]{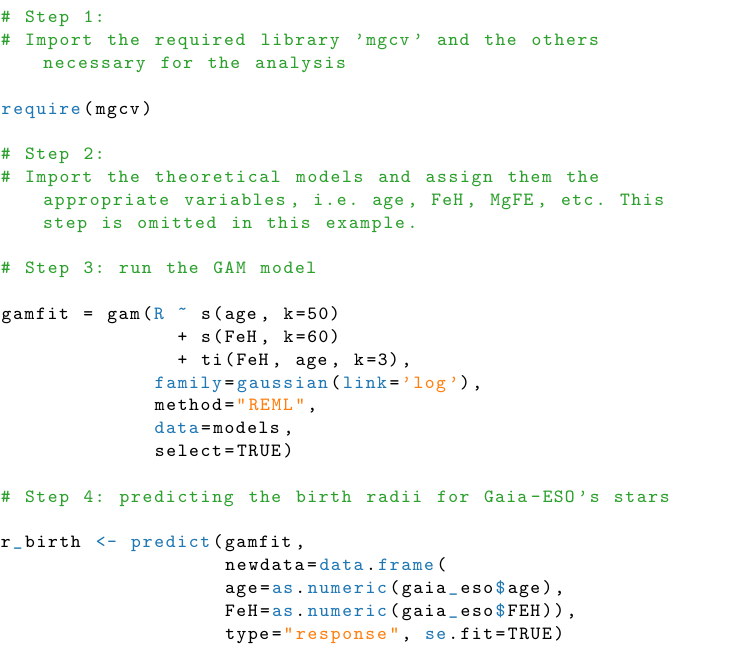}
\end{figure}

\noindent We remind the reader that \texttt{age} here refers to $t$ in Equation \ref{eq:ages}. The choice of the smooth parameter $k$, which sets the basis dimension for penalised regression smoothers in the \textsc{mgcv} package, plays a critical role in balancing computational efficiency and model flexibility. According to the \textsc{mgcv} documentation\footnote{\url{http://cran.nexr.com/web/packages/mgcv/index.html}}, $k$ determines the upper limit of the degrees of freedom associated with a smooth term, with $k-1$ (or $k$) representing this limit after accounting for the identifiability constraint.

The effective degrees of freedom, however, are controlled during fitting by methods such as generalised cross-validation (\texttt{GCV}), akaike information criterion (\texttt{AIC}), or restricted maximum likelihood (\texttt{REML}, as previously described and used in the current paper). This means that the exact choice of $k$ is not overly critical as long as it is sufficiently large to capture the underlying data pattern while remaining computationally efficient. As noted in the documentation, choosing a higher \(k\) value can help prevent underfitting, but excessively high values may result in overfitting or unnecessary computational cost.

To ensure that the choice of $k$ is appropriate, the following general guideline is recommended:
\begin{enumerate}
    \item Fit the model and extract deviance residuals.
    \item For each smooth term, fit an equivalent single smooth to the residuals using a higher $k$ value to check for residual patterns that might suggest a need for increased $k$.
\end{enumerate}

In our analysis, we selected $k=50$ for the age term, $k=60$ for the [Fe/H] term, and $k=3$ for the interaction term \texttt{ti(FeH, age)}. These values were determined based on the guidelines above and additional testing. Higher values of $k$ for the interaction term were tested but did not yield any improvement in model fit or performance, confirming that the choice of $k=3$ for the interaction term was sufficient.

Additionally, we used both individual and combined parameters in our model to assess their contribution to the accuracy of the estimated $R_{\rm b}$. The individual parameters \feh\ and $t$ were first tested independently, and then combined in the interaction term \texttt{ti(FeH, age)}. This term represents a tensor product smooth that captures the interaction between \feh\ and $t$, enabling the GAM to model their combined, non-linear effect on $R_{\rm b}$ flexibly, without enforcing the same level of smoothness for each variable. While the individual parameters each contributed to the model fit, the interaction term significantly enhanced the model's ability to capture the relationship between these variables, allowing it to more precisely recover the original chemical enrichment trends described in \citet{Magrini2009}.

\begin{figure}
    \centering
    \includegraphics[width=\linewidth]{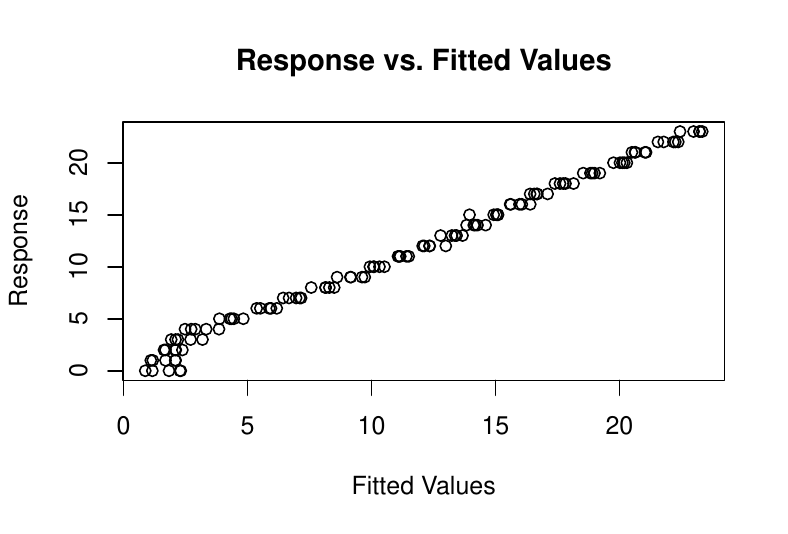}
    \caption{Response variable versus fitted variable via GAM. The response variable is $R_{\rm b}$. This plot was created by running \texttt{gam.check} command in the \textsc{mgcv} package.}
    \label{fig:gam_response_vs_fitted}
\end{figure}

The scatterplot of the response variable versus the fitted values from the GAM in Fig. \ref{fig:gam_response_vs_fitted} (generated through the \texttt{gam.check} command in the \textsc{mgcv} package) demonstrates the model's overall effectiveness in reproducing observed data. Across most radii, the points align closely along the 45-degree line, indicating a strong agreement between the fitted and observed values. The consistent spread of points suggests homoscedasticity, further validating the model's adequacy. However, minor deviations are evident at very small radii (below 2.5 kpc), where the model exhibits degeneracy. This behaviour likely stems from the limited resolution of the original chemical enrichment models and the sensitivity of the logarithmic link function (\texttt{link=`log'}) to values approaching zero, which can introduce instability. Despite these limitations, the GAM performs exceptionally well across the broader range of $t$ and \feh, reinforcing its robustness and supporting the conclusion that these stars were likely formed at even smaller radii. Far from compromising our findings, this slight overestimation of $R_{\rm b}$ at near-zero radii strengthens our key result: stars have migrated from the innermost regions of the Galaxy to the solar neighbourhood. Figure \ref{fig:gam_response_vs_fitted} also provides insight into these minor discrepancies observed in Fig. \ref{fig:comparison_feh}, tracing their origin to resolution limitations.

%-------------------------------------------------------------------
\section{Comparison between generalised additive models for [Fe/H] and [Mg/H]} \label{appendix:comparison_feh_mgh}

\begin{figure*}
    \centering
    \includegraphics[width=\linewidth]{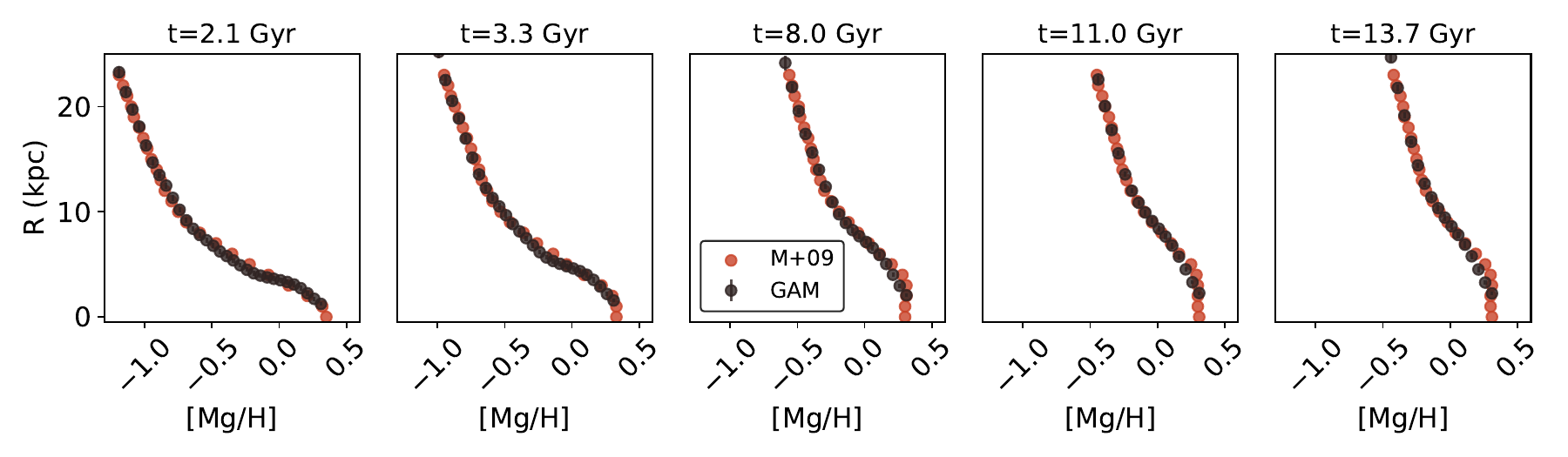}
    \caption{Similar to Fig. \ref{fig:comparison_feh} but with a model considering \mgh, instead of \feh.}
    \label{fig:comparison_mgh}
\end{figure*}

\begin{figure*}
    \centering
    \includegraphics[width=\linewidth]{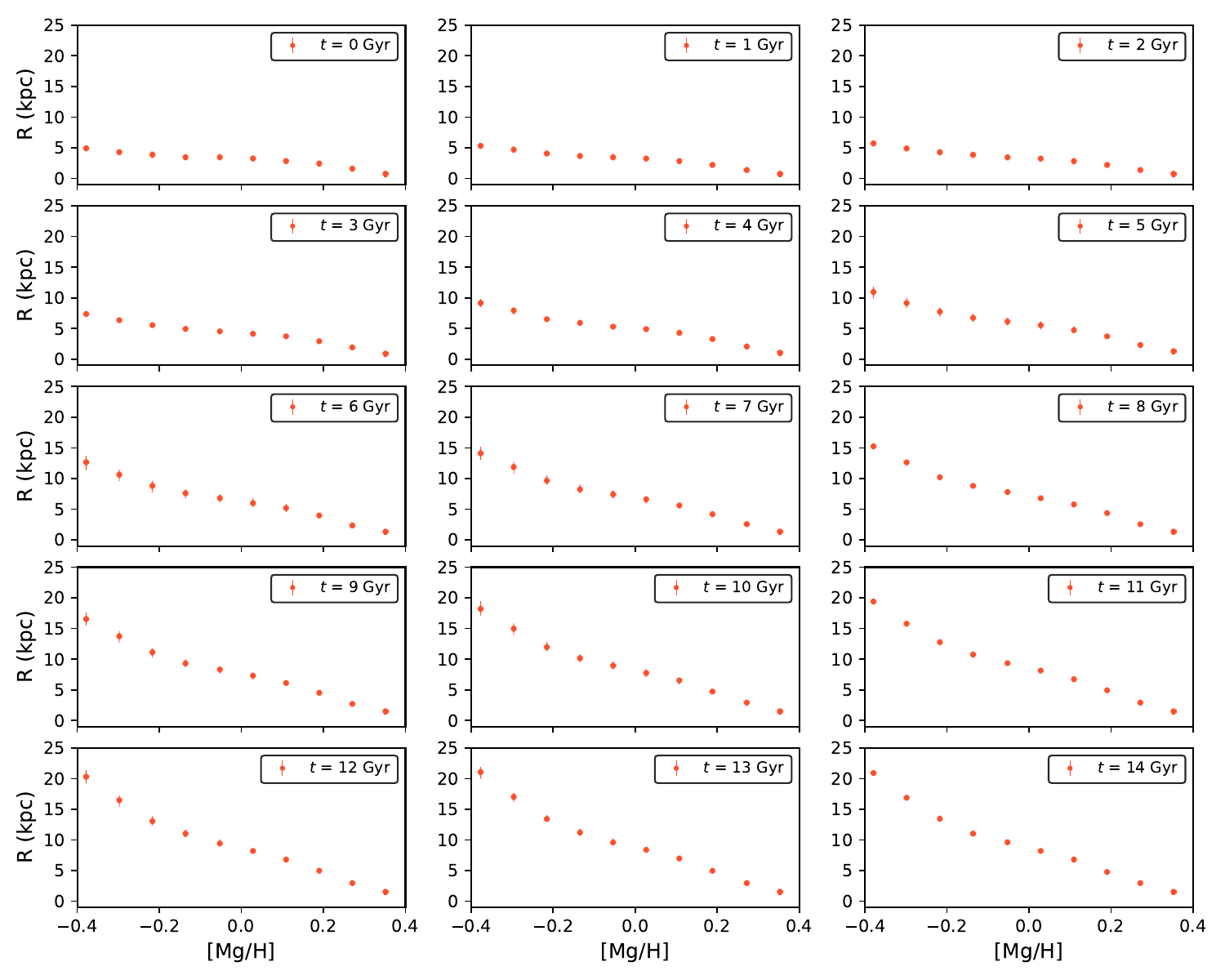}
    \caption{Similar to Fig. \ref{fig:new_grids_split_feh}, but instead of depicting \feh~in the $x$-axis, it depicts \mgh.}
    \label{fig:new_grids_split_mgh}
\end{figure*}

\begin{figure*}
    \centering
    \includegraphics[width=\linewidth]{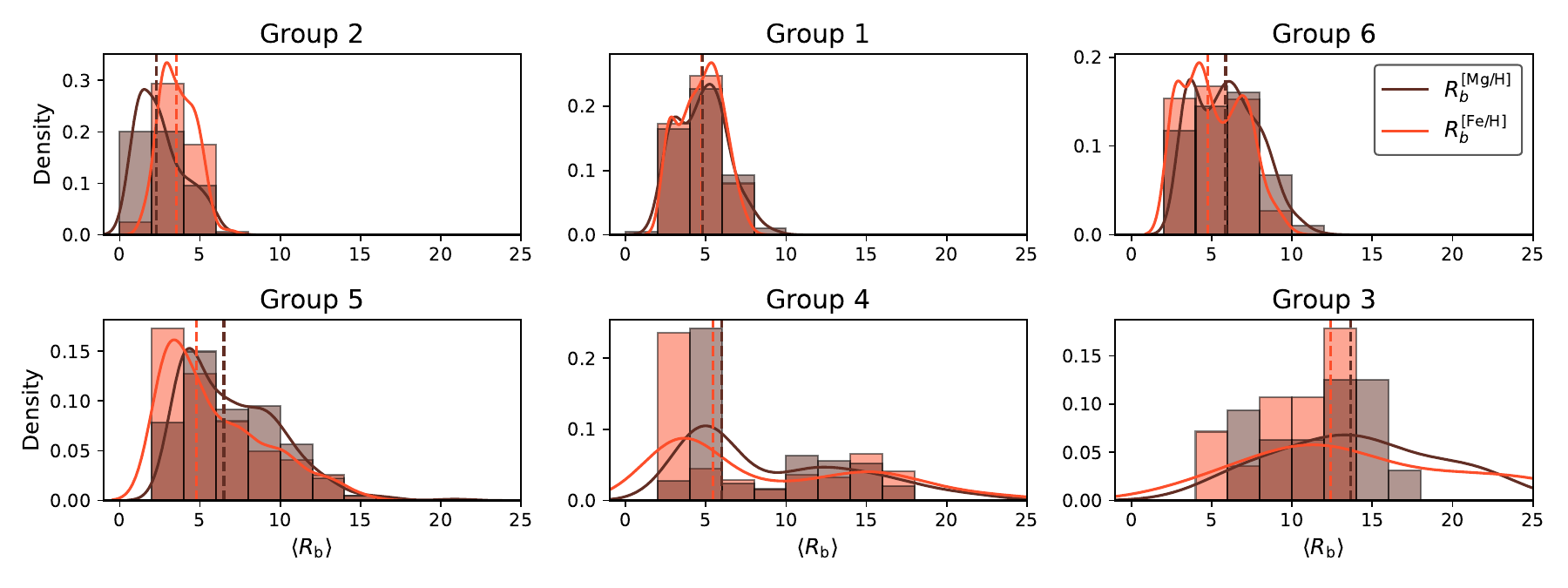}
    \caption{Histograms and Gaussian kernel density plots for the estimated \rbirth~using either \feh~(orange colours) or \mgh~(brown colours). The dashed vertical line with the corresponding colours. The order of the subplots depicts the median estimations for each groups, again, with decreasing $\langle \rm{\feh} \rangle$ for each subgroup.}
    \label{fig:kde_rbs}
\end{figure*}

\begin{figure*}
    \centering
    \includegraphics[width=\linewidth]{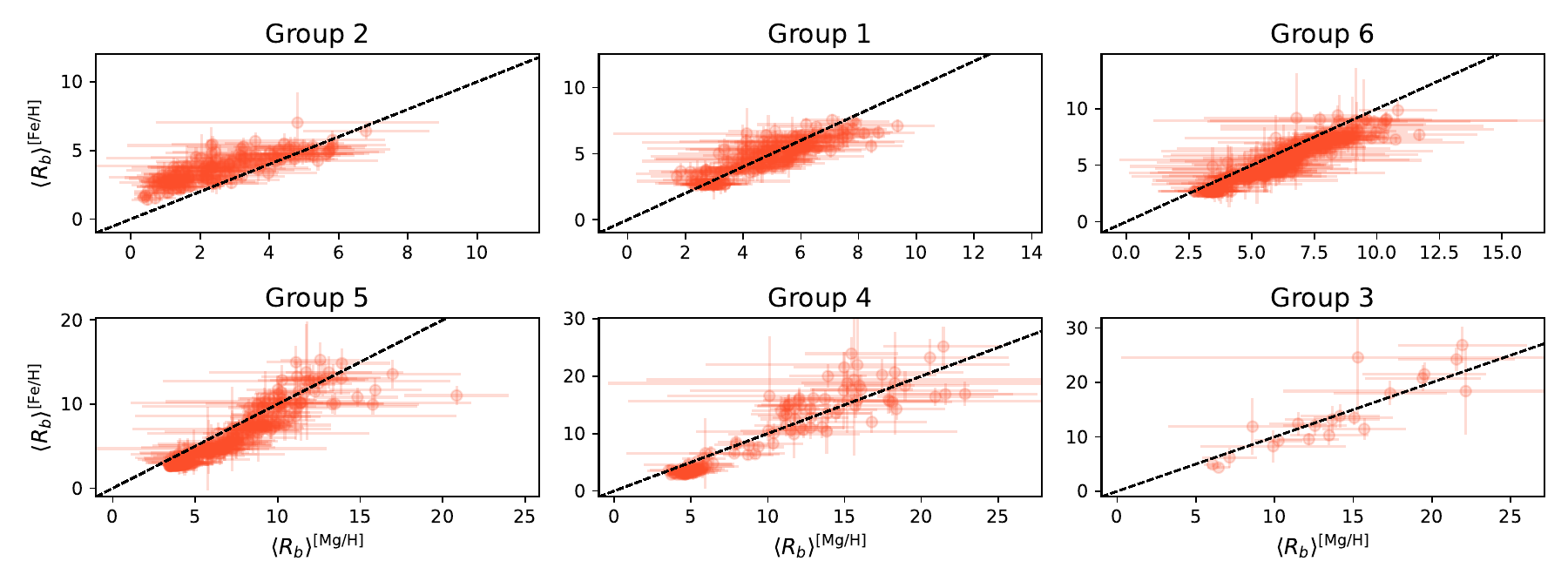}
    \caption{Comparison between the \rbirth\ estimated using either \feh~or \mgh~in a GAM. We also display a 1-to-1 dashed line to serve as a visual reference.}
    \label{fig:rbs_1x1}
\end{figure*}

In this appendix, we present a supplementary analysis showing the results from the GAM regression using \mgh\ and compare them with the results obtained using \feh. We chose to run the GAM with \mgh\ instead of \mgfe\ to fully remove the influence of Fe in this auxiliary model. Figures \ref{fig:comparison_mgh} and \ref{fig:new_grids_split_mgh} are analogous to Figs. \ref{fig:comparison_feh} and \ref{fig:new_grids_split_feh} in the main text, but use \mgh\ as an independent variable in the GAM instead of \feh. Figures \ref{fig:kde_rbs} and \ref{fig:rbs_1x1} compare the \rbirth~estimates obtained using the main GAM (i.e. with \feh~as an independent variable) and the alternative GAM using \mgh.

Although the distributions shown in Fig. \ref{fig:kde_rbs} are not identical, they are very similar for all groups. Figure \ref{fig:rbs_1x1} shows that the \rbirth\ estimates are consistent within the error bars, with very few outliers, primarily in Groups 4 and 3. These groups consist mostly of the oldest stars, which the original theoretical models by \citet{Magrini2009} struggle to fully cover, as discussed in Section \ref{subsec:challenges}. It is worth emphasising once again that the original models for \mgfe~(shown in Fig. \ref{fig:MagriniModels}) depict a large plateau for $R\gtrsim 7.5$ kpc, which can be particularly problematic when estimating the \rbirth~for stars formed above this threshold. This limitation is reflected in the significant errors for larger \rbirth\ at when using \mgfe\ as the estimator, which further supports our preference for using \feh\ over \mgfe.

It is worth mentioning that similar issues in retrieving $R_{\rm b}$ for very low radii are also present in the model using \mgfe. This discussion was previously developed in Appendix \ref{appendix:model}.

%-------------------------------------------------------------------
\section{Additional supporting material} \label{appendix:additionalmaterial}

This section provides additional supporting material for the dynamic analysis of our sample.

% -----------------
\subsection{Preliminary chemo-dynamic analysis} \label{appendix_subsec:prelim_chemodynamic}

\begin{figure*}
   \centering
   \includegraphics[width=\linewidth]{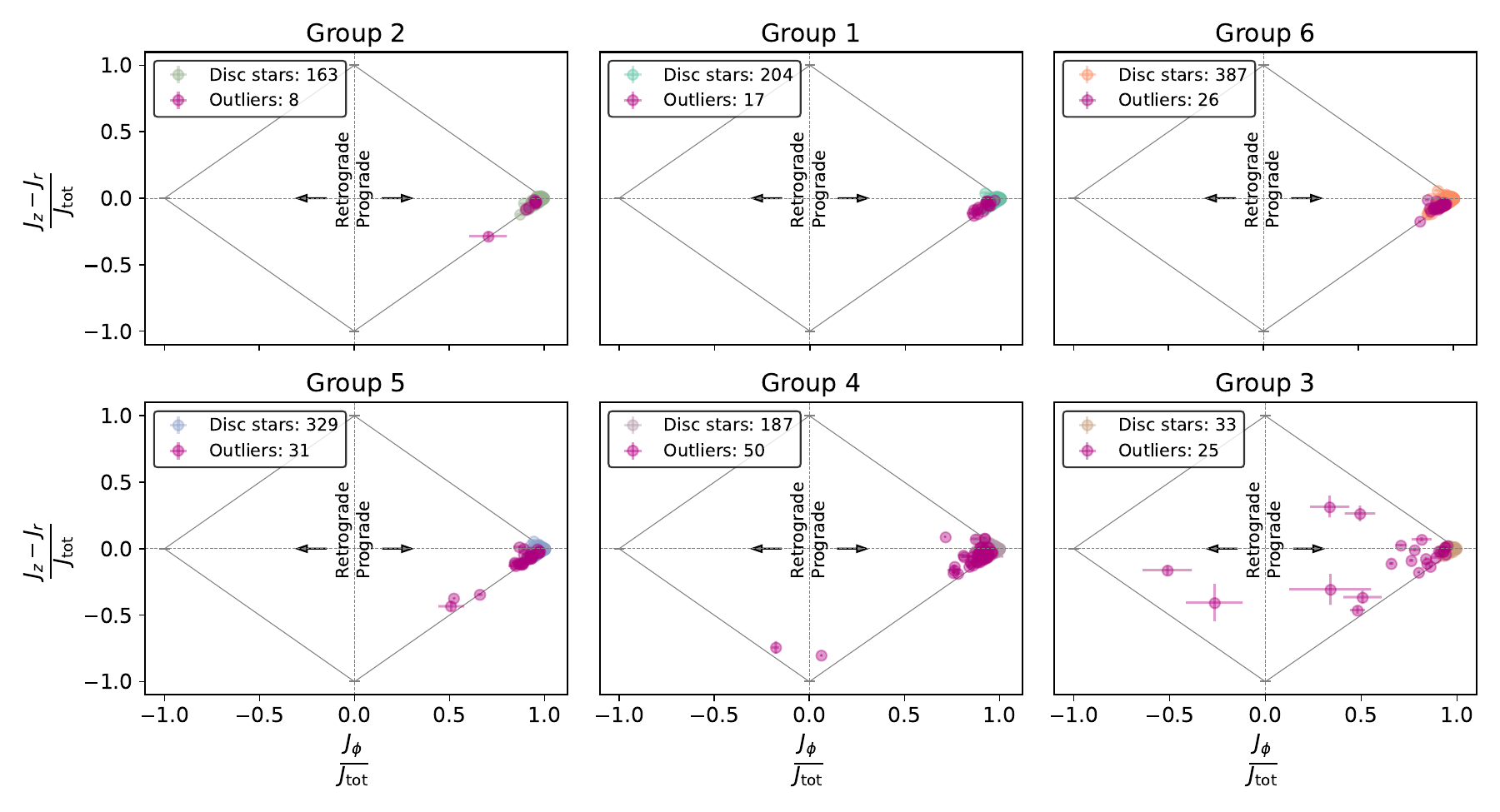}
   \caption{Action maps for the six groups retrieved from the HC, ordered by descending median metallicity $\langle \rm{\feh} \rangle$. Stars identified as belonging to the disc using the Toomre Diagram cluster around the right corner [$(J_z - J_r)/J_{\rm tot}=0$ and $J_{\phi}/J_{\rm tot}=1$]. In contrast, stars classified as outliers sometimes appear to be from the disc but increasingly occupy other regions of the action maps as the group metallicity decreases, even reaching retrograde orbits. The colour scheme for all groups remains consistent throughout this paper, with outliers always shown in magenta. }
   \label{fig:action_map}
\end{figure*}

Figure \ref{fig:action_map} presents the action map for the entire sample, with disc interlopers highlighted in magenta. This figure, along with \ref{fig:toomre} and \ref{fig:lindblad}, reinforces our criteria for identifying stars associated with the Galactic disc.

% -----------------
\subsection{Potential blurred thin disc stars classified as halo stars}
\label{appendix_subsec:disc_or_halo}

\begin{figure*}
    \centering
    \includegraphics[width=\linewidth]{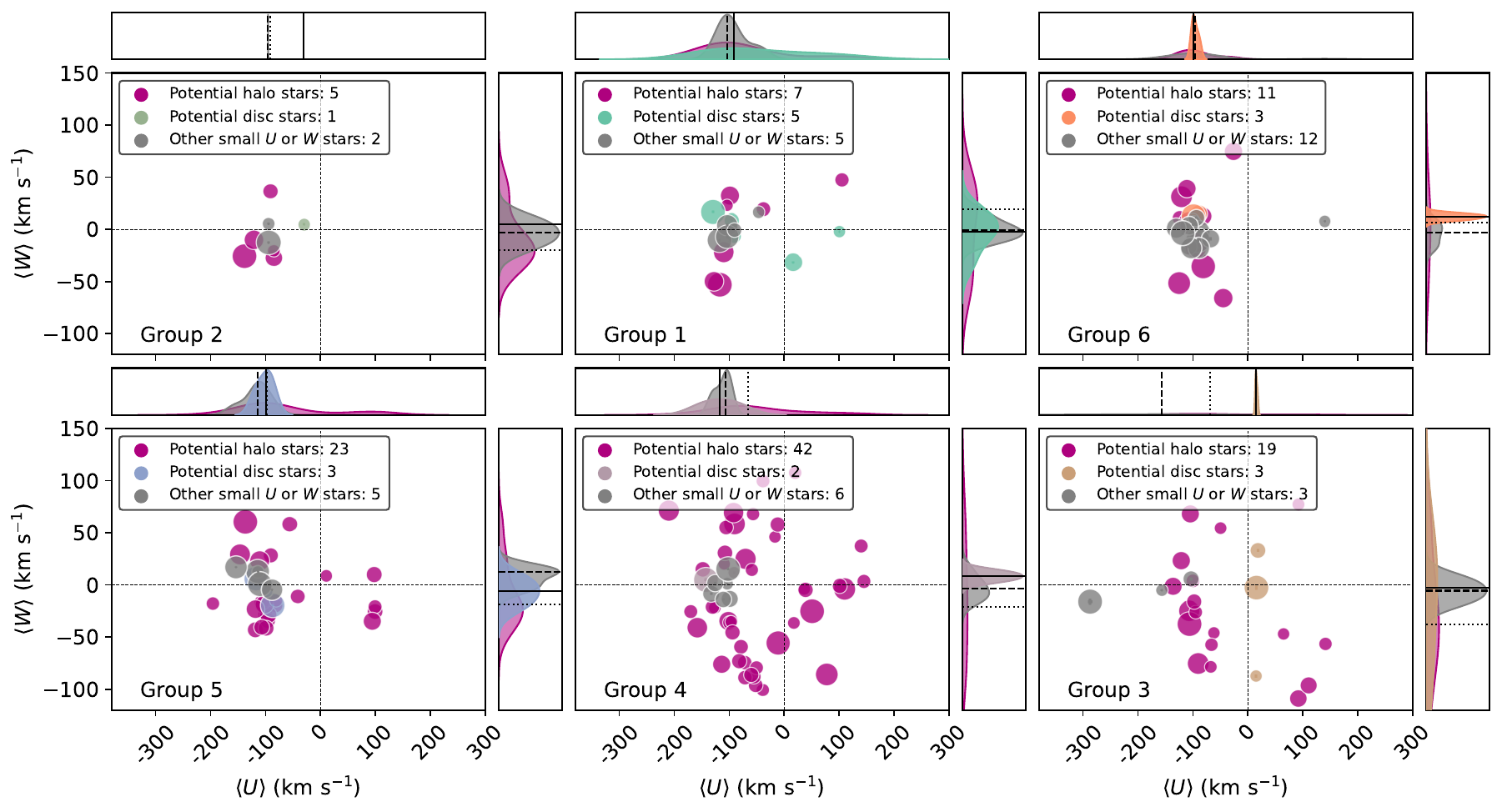}
    \caption{$W$ and $U$ velocities of the stars removed from the main analysis using the Toomre diagram. Stars classified as most likely belonging to the Galactic halo are coloured in magenta, in accordance with previous figures, while those coloured according to the sample groups are potential blurred disc stars. The marker size is proportional to $|\Delta R|$. The adjacent plots show the distribution of both velocities, with median values indicated by the dotted line for potential halo stars, the solid line for potential blurred disc stars, and the dashed line for stars with low $W$ or $U$. Vertical and horizontal dashed lines at zero velocity provide visual reference points for the reader. Unlike similar figures, we omit the annotated medians in the adjacent distribution plots to avoid information overload.}
    \label{fig:disc_or_halo}
\end{figure*}

In Fig. \ref{fig:disc_or_halo}, we present the $W$ and $U$ velocity components used in the Toomre diagram (Fig. \ref{fig:toomre}) for stars excluded from the main sample, which we classified as halo stars. We employed a detailed approach to determine whether there are blurred disc stars among these halo stars by considering two additional criteria: (i) whether their current \rgui\ aligns with their estimated \rbirth; (iii) if they are within or below the $2\sigma$ thin-thick separation detailed in Sect. \ref{subsubsec:thin_thick_disc}; and (ii) if either the $W$ or $U$ component exhibits small values (between -20 and 20 km s$^{-1}$).

The first criterion assesses whether these stars maintain their orbital radii within the estimated errors, consistent with the behaviour expected of blurred stars, which should retain their $L$ throughout their orbits. The second criterion is to remove thick disc stars, using the same criterion described in Sect. \ref{subsubsec:thin_thick_disc}. The third, though somewhat arbitrary, aims to identify stars with very small velocities in one component and correspondingly high velocities in another, which may indicate they are blurred disc stars.

Based on these criteria, we find that approximately 10.8\% of our halo stars could potentially be classified as blurred thin disc stars. However, this represents only about 1.2\% of the total sample of 1460 stars, which is why these stars were not further analysed in the main body of the paper.

A more challenging scenario arises when some halo stars that might also be blurred thin disc stars do not meet the first criterion, possibly due to a combination of changing parameters that conserve their $L$. In such cases, assessment becomes difficult. Nonetheless, we observe that other stars exhibit small $W$ or $U$ velocities, which could indicate they are blurred thin disc stars. By focusing on stars with low $W$ or $U$ velocities, we identify an additional 33 stars displaying these characteristics. Combining these 33 with the previous 17, we find that $\sim$ 3.4\% of the stars fit this profile, still a small fraction of our full sample. In sum, distinguishing between halo and blurred disc stars is challenging due to the complex interplay of changing orbital parameters.

\begin{figure*}
    \centering
    \includegraphics[width=\linewidth]{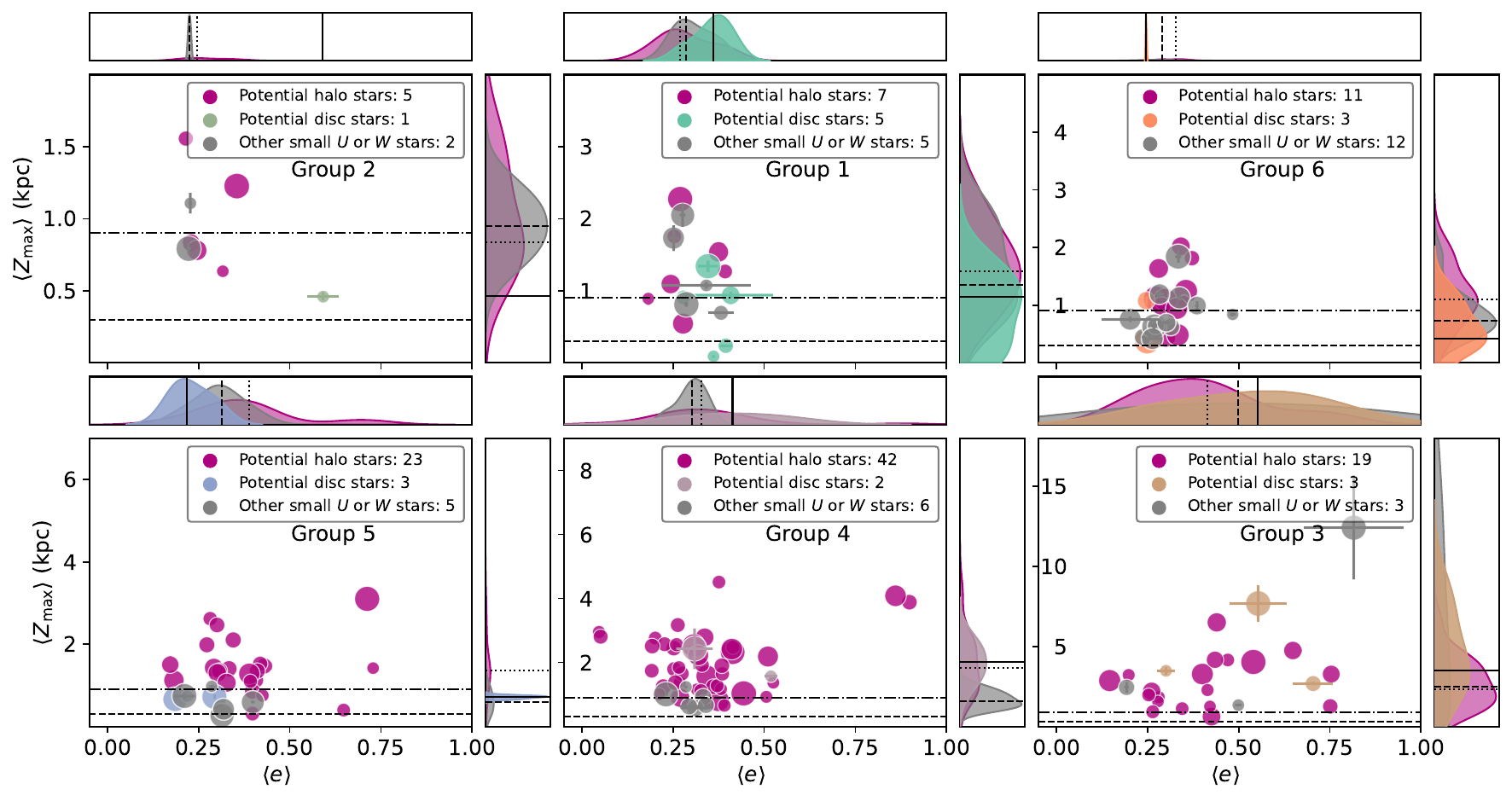}
    \caption{The \zmax\ and \eccentricity\ of the stars removed from the main analysis using the Toomre diagram are shown here. Stars classified as most likely belonging to the Galactic halo are coloured magenta, consistent with previous figures, while stars coloured according to the sample groups represent potential blurred disc stars. The marker size is proportional to $|\Delta R|$. The adjacent plots display the distributions of \zmax\ and \eccentricity, with median values indicated by a dotted line for potential halo stars, a solid line for potential disc stars, and the dashed line for stars with low $W$ or $U$. As in Fig. \ref{fig:zmax_age}, the scale heights for the thin and thick discs are shown as dotted and dot-dashed lines, respectively \citep[i.e. 300 and 900 pc;][]{McMillan2017}. We note that while the $x$-axis is the same across all subplots, the $y$-axis scales differ to better depict the distribution of each subplot individually. Unlike similar figures, we omit the annotated medians in the adjacent distribution plots to avoid information overload.}
    \label{fig:disc_or_halo_ecc_zmax}
\end{figure*}

We further investigate the properties of these stars in Fig. \ref{fig:disc_or_halo_ecc_zmax}, where we plot \zmax\ against \eccentricity. It is evident that \eccentricity\ is generally higher across all groups, even surpassing that of the likely halo stars, with the exceptions of groups 5 and 6. For all groups, except the two most metal-poor (groups 3 and 4), the potential thin disc stars exhibit median \zmax\ values that are either within or closer to the scale heights of their halo counterparts, shown in magenta. Notably, for the most metal-poor groups, \zmax\ values are elevated, which may be the result of interactions with the Sagittarius satellite galaxy. Additionally, for these two metal-poor groups, \eccentricity\ is significantly higher compared to the other groups. Given these features, it is likely that these potential thin disc stars are indeed blurred members of the disc, which we do not account for in the main analysis of the paper.

% -----------------
\subsection{Heliocentric distances} \label{appendix_subsec:helio_dist}

\begin{figure*}
    \centering
    \includegraphics[width=\linewidth]{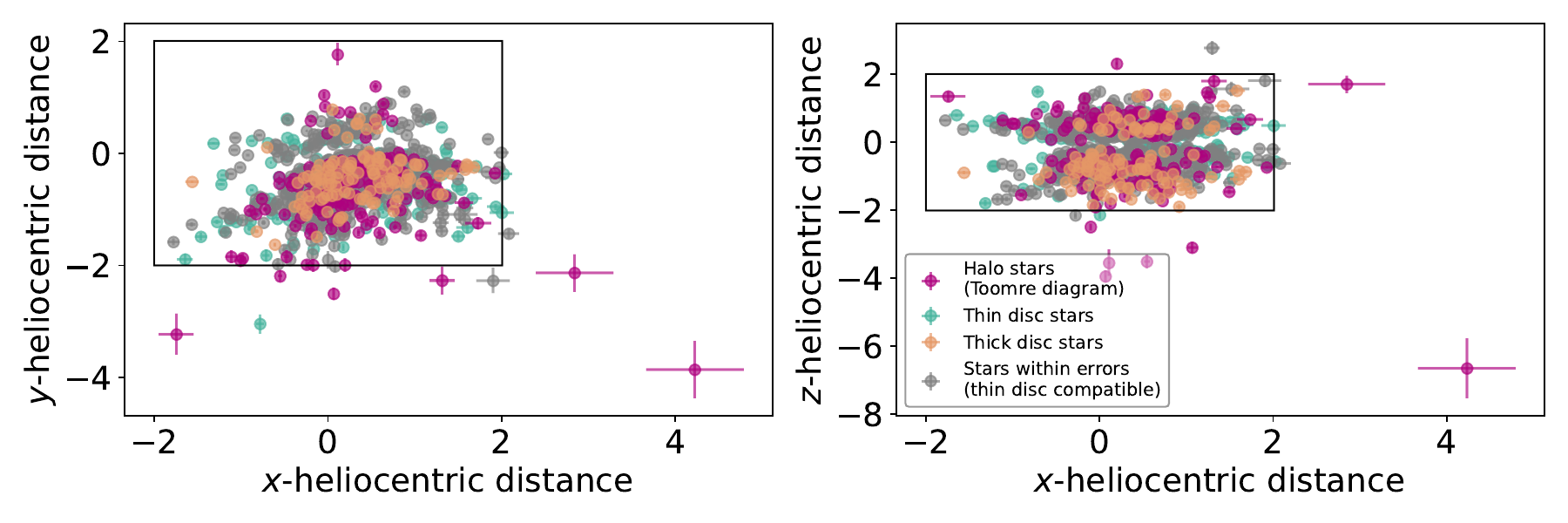}
    \caption{Heliocentric distance projections for the stars of our sample. This figure features two panels depicting the spatial distribution of stars within our observational sample in relation to the Sun. The left panel shows the projection onto the $xy$ Cartesian plane and the right panel onto the $xz$ plane. Distances are measured in kpc. Highlighted in each panel is a 2 kpc square, illustrating the region where the vast majority (> 99\%) of our disc stars are located, effectively encompassed within a 2 kpc cube. Stars are coloured according to their classes described in Sect. \ref{subsubsec:disc_or_halo} and \ref{subsubsec:thin_thick_disc}.}
    \label{fig:xyz_distances}
\end{figure*}

Figure \ref{fig:xyz_distances} illustrates the heliocentric distances for the stars in our sample, with outliers, likely not part of the Galactic disc, highlighted in magenta. Also, stars classified according to our thin-thick criterion are coloured accordingly. It is noteworthy that nearly all the analysed stars are within 2 kpc of the Sun.

% -----------------
\subsection{A few different perspectives on \rbirth\ versus \rgui, \rapo, and \rperi} \label{appendix_subsec:rb_other_rs}

\begin{figure*}
    \centering
    \includegraphics[width=\linewidth]{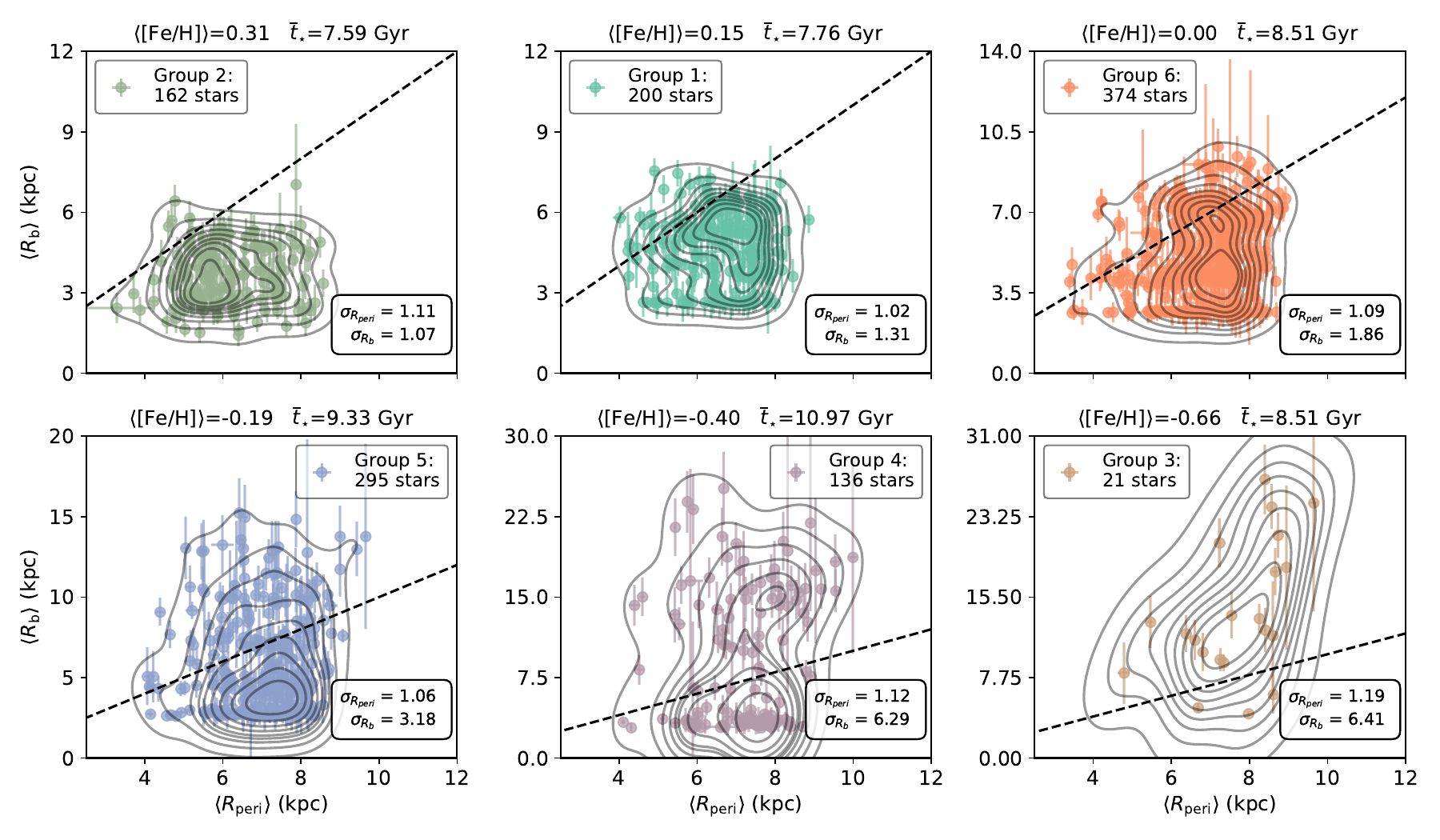}
    \caption{Similar to Fig. \ref{fig:rb_rg_groups} but contrasting \rperi~instead of \rgui.}
    \label{fig:rb_rperi_groups}
\end{figure*}

\begin{figure*}
    \centering
    \includegraphics[width=\linewidth]{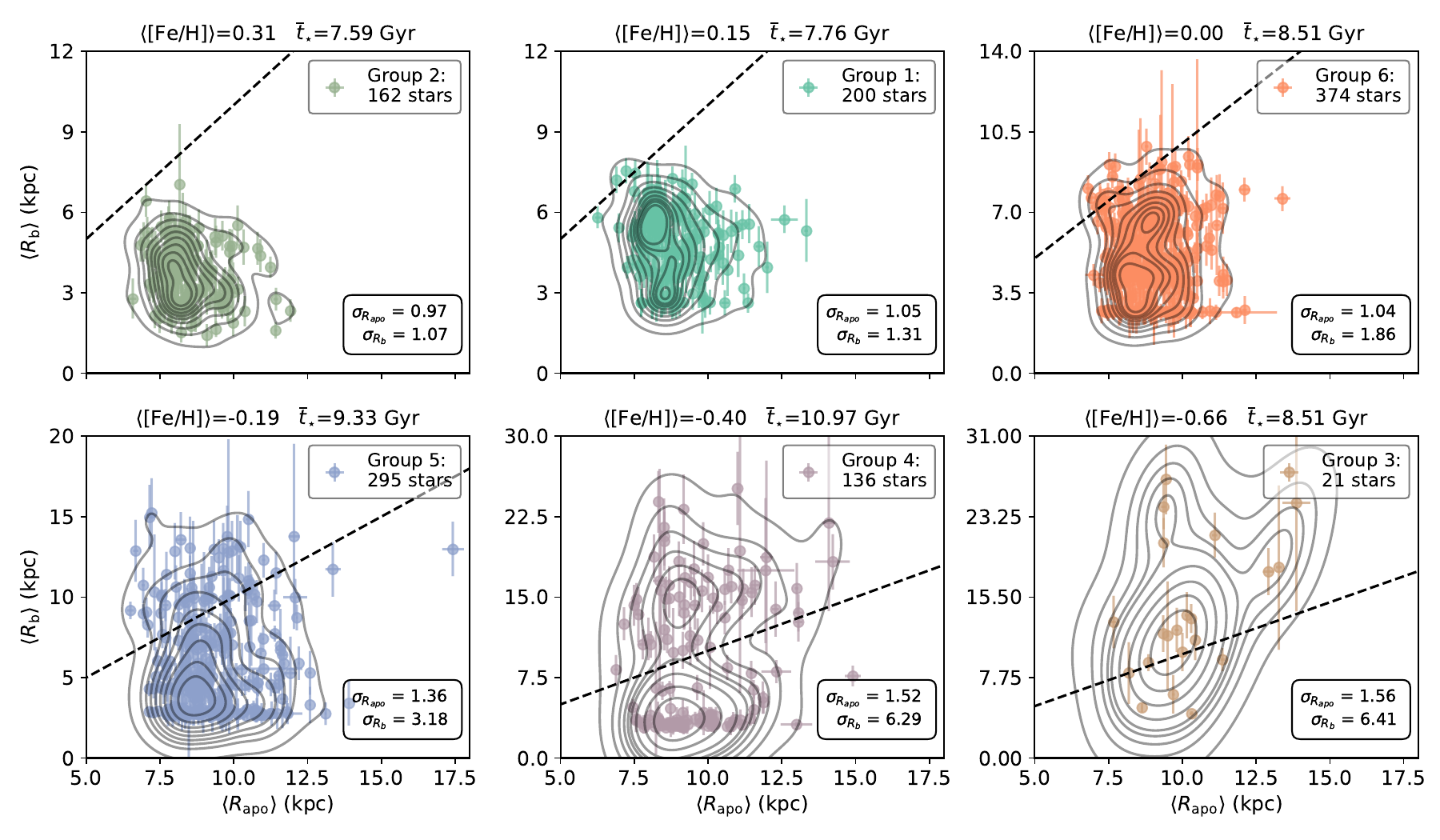}
    \caption{Similar to Fig. \ref{fig:rb_rg_groups} but contrasting \rapo~instead of \rgui. In this case, we changed the range of \rapo, to better depict all the stars.}
    \label{fig:rb_rpapo_groups}
\end{figure*}

\begin{figure*}
    \centering
    \includegraphics[width=\linewidth]{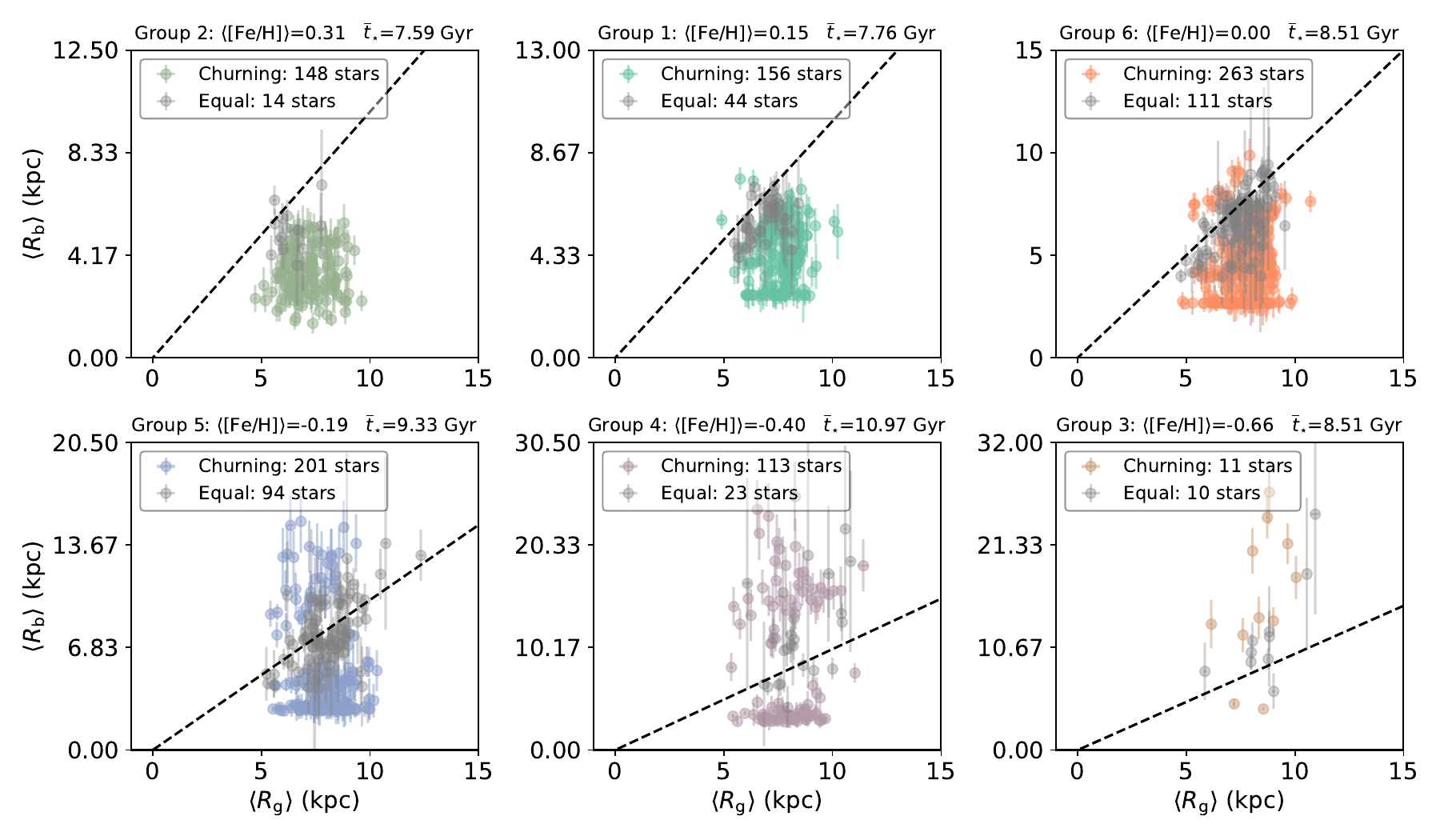}
    \caption{Similar to Fig. \ref{fig:rb_rg_groups} but with the movement classification. Stars with unchanged orbital radii are marked in grey, whereas those consistent with churning keep their colour scheme used throughout this paper.}
    \label{fig:rb_rg_churblur}
\end{figure*}

Figures \ref{fig:rb_rperi_groups} to \ref{fig:rb_rg_churblur} present similar plots with various highlights. Figs. \ref{fig:rb_rg_churblur} omits the 2D-Gaussian kernel densities to enhance the visibility of individual markers. Figures \ref{fig:rb_rperi_groups} and \ref{fig:rb_rpapo_groups} are analogous to Fig. \ref{fig:rb_rg_groups}, but depict the pericentric (\rperi) and apocentric radii (\rapo) instead of the guiding radius. Both figures display the standard deviation of the variables in each subplot (for each group), which increases with decreasing metallicity.

Figure \ref{fig:rb_rg_churblur} is similar to Fig. \ref{fig:rb_rg_groups}, but each subplot highlights stars with unchanged orbital radii in grey, while stars influenced by churning (radial migration) retain the colour palette used throughout this paper, regardless of the direction of their motion (inward or outward).

\begin{figure*}
    \centering
    \includegraphics[width=\linewidth]{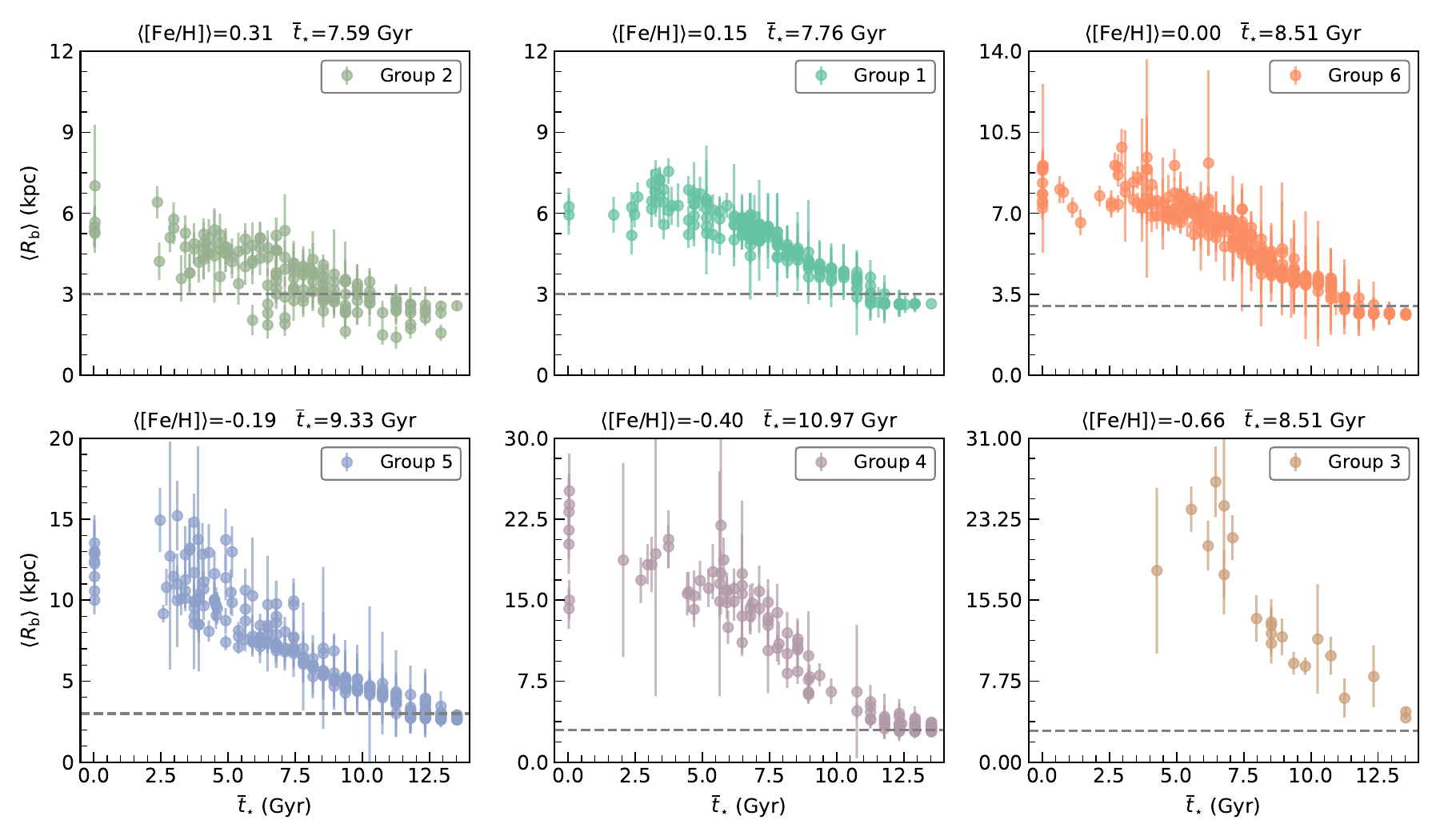}
    \caption{Scatterplot of \rbirth\ versus stellar age ($\overline{t}_{\star}$) for all groups in our sample. A grey horizontal dashed line at 3 kpc highlights the approximate \rbirth\ limit for stars that are not super-metal-rich. Group 2, consisting of super-metal-rich stars, is notably not constrained by this limitation.}
    \label{fig:rb_age}
\end{figure*}

Figure \ref{fig:rb_age} shows the relationship between \rbirth\ and stellar age ($\overline{t}_{\star}$) for all groups in our sample, demonstrating a general trend of decreasing \rbirth\ with increasing $\overline{t}_{\star}$. A dashed line at 3 kpc marks the approximate \rbirth\ threshold, which primarily impacts older stars that are not super-metal-rich. While this boundary constrains the \rbirth\ distribution of most groups, it does not apply to Group 2. Many stars in Group 2 fall below this threshold, indicating their independence from this limit due to their super-metal-rich nature. This apparent threshold arises from the combined effects of the original chemical enrichment models for the MW \citep[as adopted in this paper;][]{Magrini2009}, which do not predict the formation of non-super-metal-rich stars in the innermost regions of the MW, and the GAM methodology. These trends are further illustrated in Figs. \ref{fig:comparison_feh}, \ref{fig:new_grids_split_feh}, and \ref{fig:comparison_mgh}.

\begin{figure*}
    \centering
    \includegraphics[width=0.49\linewidth]{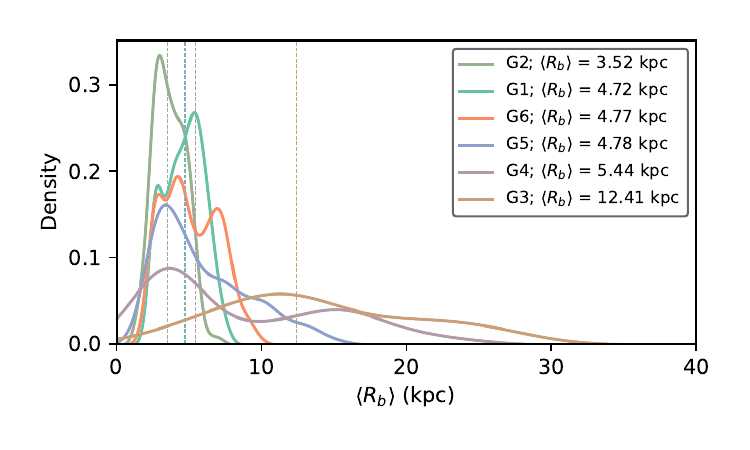}
    \includegraphics[width=0.49\linewidth]{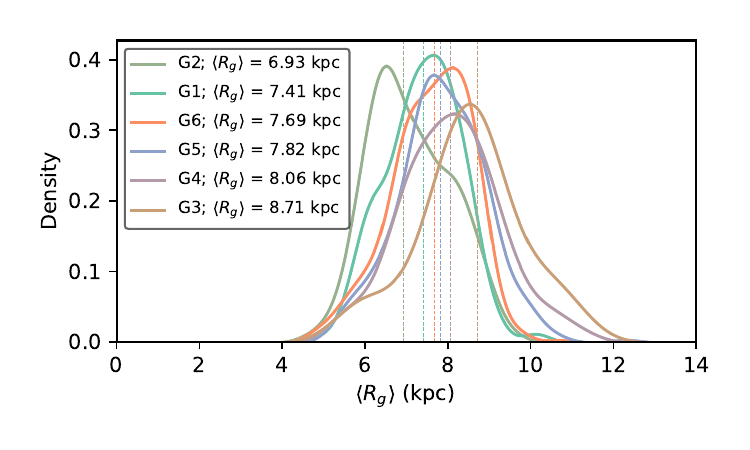}
    \caption{Distribution of \rbirth\ and \rgui\ respectively for the disc stars of our sample stratified by the HC groups. Vertical dashed lines depict the median estimates for each HC-based group of our sample.}
    \label{fig:rg_rb_kdes}
\end{figure*}

Finally, Fig. \ref{fig:rg_rb_kdes} illustrates the distributions of \rbirth\ and \rgui\ for our disc sample. It is evident that as metallicity decreases, stars are formed at increasingly larger Galactocentric distances, as expected. Notably, some stars from the most metal-poor groups (4 and 5) exhibit significantly elevated \rbirth\ estimates. This observation has been discussed in the main body of the paper and is likely attributed either to some potential thick disc intruders or to limitations within the original chemical enrichment models, which cannot be fully addressed by the GAM alone. Conversely, the distribution of \rgui\ aligns with expectations, as all stars are currently situated in the solar vicinity, making it reasonable for their \rgui\ values to be close to the Sun's Galactocentric distance.

% -----------------
\subsection{Angular momentum in Cartesian and cylindrical coordinates} \label{appendix_subsec:ang_moment}

\begin{figure*}
    \centering
    \includegraphics[width=\linewidth]{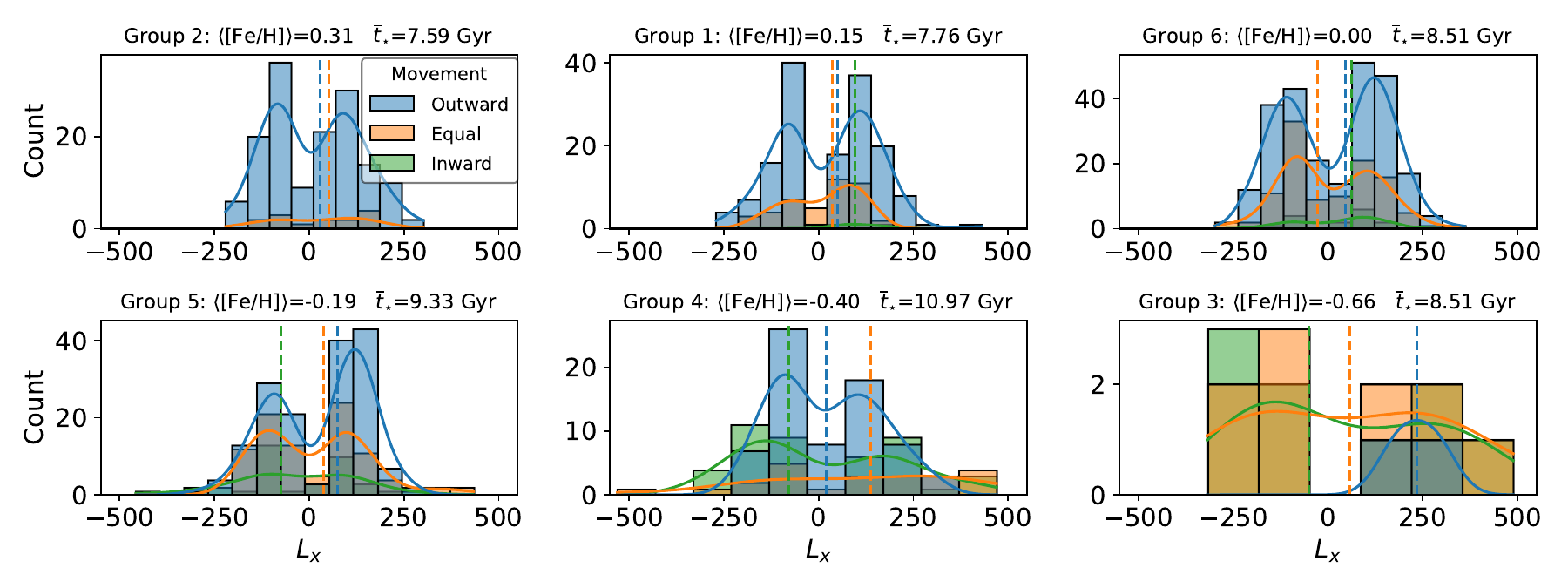}
    \caption{Distribution of $L_x$ for the stars in our sample. Each panel represents one of the HC metallicity-stratified groups, arranged in order of decreasing $\langle \rm{\feh} \rangle$. Within each panel, the $L_x$ distributions for blurred/undisturbed stars (marked as `Equal'), outward-migrating stars, and inward-migrating stars are shown in orange, blue, and green, respectively. Gaussian kernel density estimates are overlaid to smooth the binning effects. Vertical dashed lines indicate the median values for each distribution in the corresponding colours.}
    \label{fig:lx}
\end{figure*}

\begin{figure*}
    \centering
    \includegraphics[width=\linewidth]{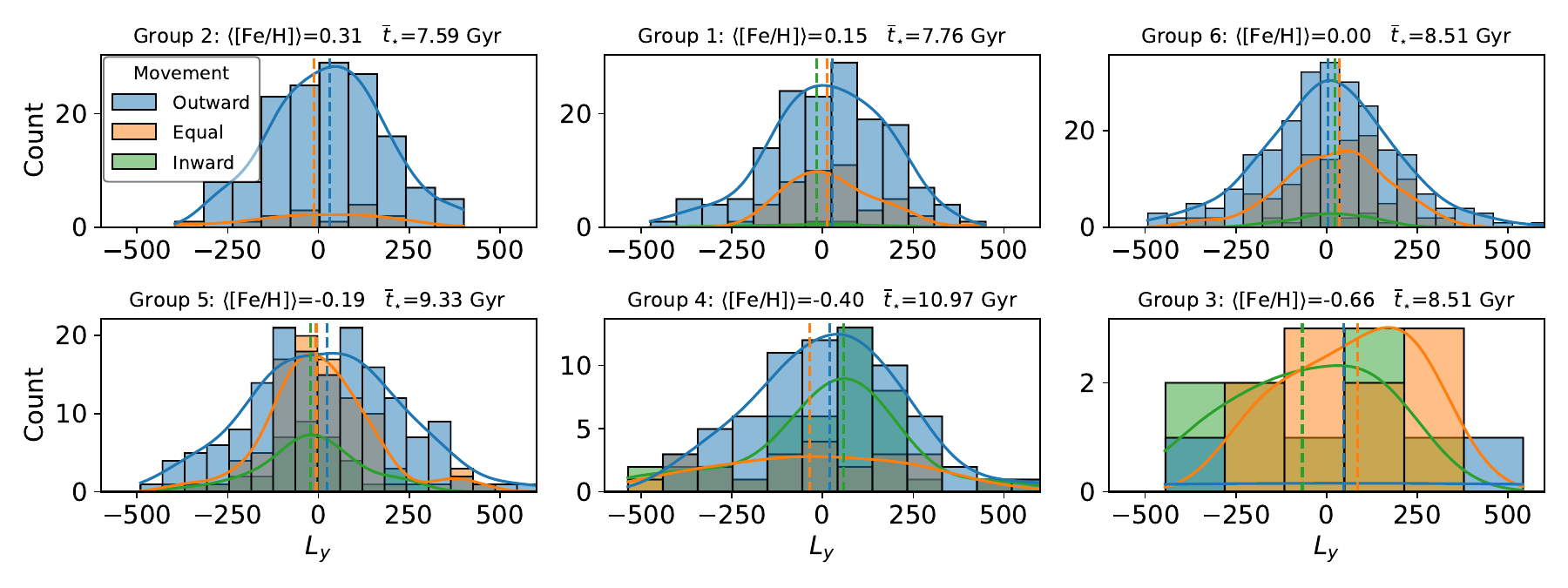}
    \caption{Same as Fig. \ref{fig:lx} but for $\langle L_y \rangle$.}
    \label{fig:ly}
\end{figure*}

\begin{figure*}
    \centering
    \includegraphics[width=\linewidth]{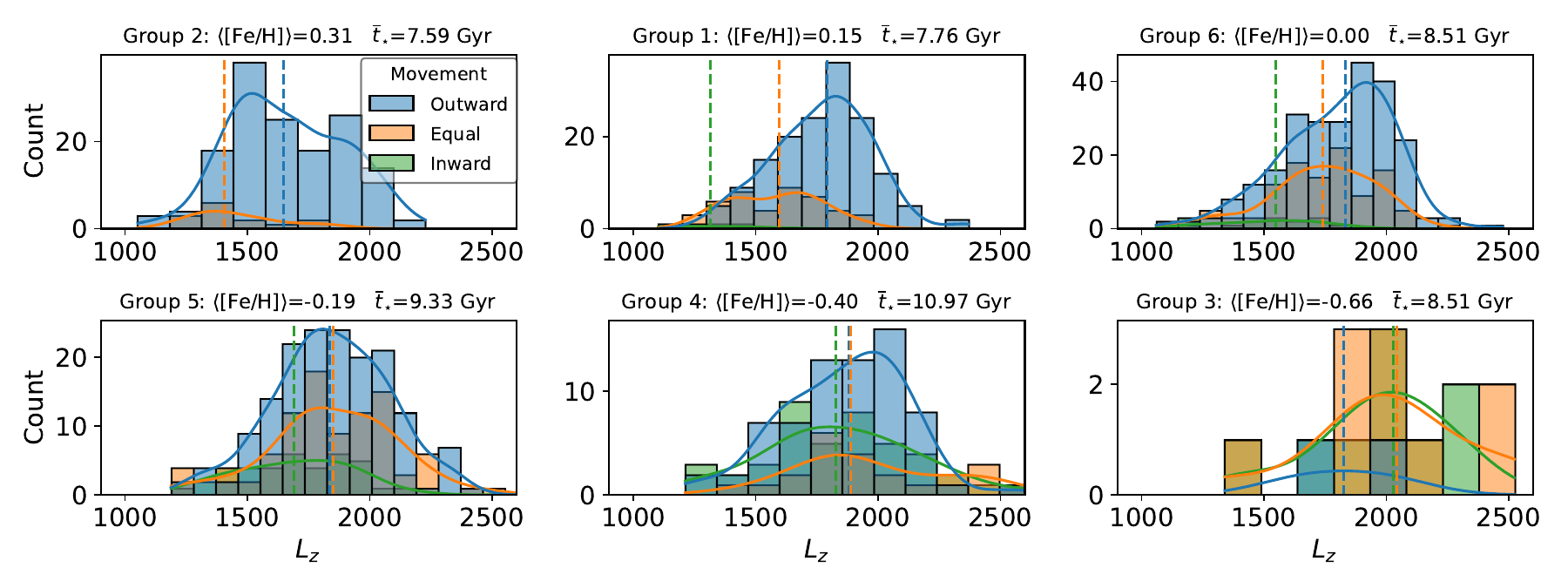}
    \caption{Same as Figs. \ref{fig:lx} and \ref{fig:ly} but for $\langle L_z \rangle$.}
    \label{fig:lz}
\end{figure*}

\begin{figure*}
    \centering
    \includegraphics[width=\linewidth]{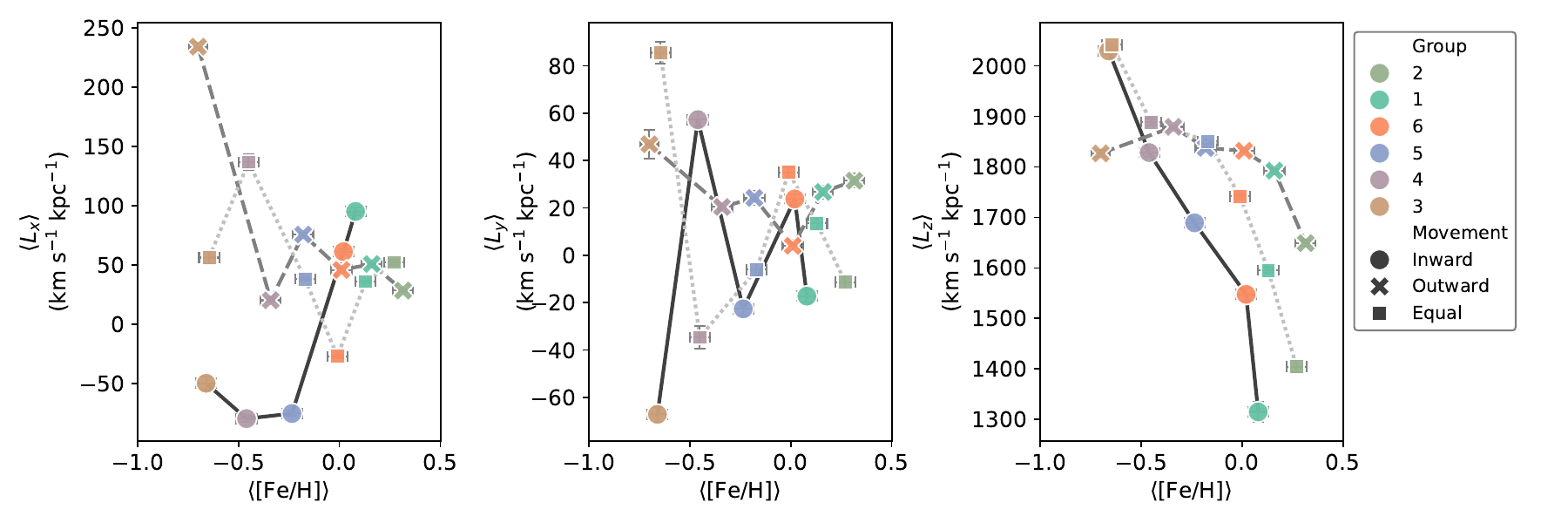}
    \caption{Same as Fig. \ref{fig:ang_mom_move_cyl} but in Cartesian coordinates.}
    \label{fig:ang_mom_move_xyz}
\end{figure*}

Figures \ref{fig:lx}, \ref{fig:ly}, and \ref{fig:lz} show the distributions of the angular momenta in the $x$, $y$, and $z$ directions respectively. Fig. \ref{fig:ang_mom_move_xyz} is similar to Fig. \ref{fig:ang_mom_move_cyl}, but in Cartesian coordinates instead of cylindrical ones. The Cartesian system is the standard output for angular momenta from \textsc{galpy} \citep{Bovy2015}. In the main text, we chose to analyse the angular momenta in cylindrical coordinates due to their ease of interpretation. Nevertheless, the results for the $x$ and $y$ directions closely resemble those of the $r$ and $\phi$ directions, respectively. The steps to convert the components of the angular momentum are outlined below:

\begin{enumerate}
    \item We first used the \texttt{galpy.util.coords.XYZ\_to\_galcencyl} method to transform the heliocentric positions (depicted in Fig. \ref{fig:xyz_distances}) into Galactocentric cylindrical coordinates in the $r$, $\phi$, and $z$ directions.

    \item We then proceed with the conversion using the standard formulae:
    
    \begin{equation} \label{eq:ang_moment_conversion}
        \begin{aligned}
        & L_r =   ~~~L_x \cos(\phi) + L_y \sin(\phi)  \\
        & L_{\phi} = -L_x \sin(\phi) + L_y \cos(\phi)  \\
        & L_z \quad \text{remains the same.} 
        \end{aligned}
    \end{equation}

    \item We finally estimated the new uncertainties using the standard error propagation rule, given by:

    \begin{equation}
        \begin{aligned}
        & \sigma_{L_r}    = [\cos^2(\phi) \sigma_{L_x}^2 + \sin^2(\phi) \sigma_{L_y}^2]^{1/2} \\
        & \sigma_{L_\phi} = [\sin^2(\phi) \sigma_{L_x}^2 + \cos^2(\phi) \sigma_{L_y}^2]^{1/2} \\
        & \sigma_{L_z} \quad \text{remains the same.}
        \end{aligned}
    \end{equation}
\end{enumerate}

\end{document}